\documentclass{aa} 
\usepackage{graphicx}
\usepackage{txfonts}

\usepackage{xcolor}
\usepackage{siunitx}
\usepackage{scalerel}
\usepackage{float}
\usepackage{placeins}
\usepackage{soul}

\newcommand\HI{H\protect\scaleto{$I$}{1.2ex}}
\newcommand\radmsquare{rad\,m\textsuperscript{-2}}

\newcommand{\angstackminsampling}{15.0}

\newcommand{\angstackinnerradiusbeams}{2.5}
\newcommand{\angstackouterradiusbeams}{7.5}
\newcommand{\beampixratioangular}{5.0}

\newcommand{\rmminnpix}{10}
\newcommand{\phyres}{2}
\newcommand{\phypixsize}{0.4}
\newcommand{\phyrefdistance}{20}

\usepackage{hyperref}
\usepackage{subcaption}

\defcitealias{2020A&A...639A.112K}{K20}
\defcitealias{2021A&A...649A..94M}{MC21}

\hypersetup{
    colorlinks=true,        
    linkcolor=blue,         
    citecolor=blue,         
    filecolor=magenta,      
    urlcolor=blue,          
}
\usepackage{booktabs} 

\begin{document}

   \title{CHANG-ES}

   \subtitle{XXXIX. Magnetic field structure in edge-on galaxies: Stacking Stokes parameters}

    \author{M. Stein
          \inst{1}
          \and
          R. Beck
          \inst{2}
          \and
          B. Adebahr
          \inst{1}      
          \and
          R.-J. Dettmar
          \inst{1}
          \and
          C. Mele
          \inst{1}        
          \and
          S. Taziaux
          \inst{1,3}
          \and
          P. Kamphuis
          \inst{1}          
          \and 
          J. English
          \inst{4}
          \and
          T. Wiegert
          \inst{5}
          \and
          J. Stil
          \inst{6}
          \and
          V. Heesen
          \inst{7}        
          \and
          C. Riseley
          \inst{1}
          \and 
          J. Irwin
          \inst{8}
          \and
          N. B. Skeggs
          \inst{8}
          \and
          R. Henriksen
          \inst{8}
          }     

   \institute{Ruhr University Bochum, Faculty of Physics and Astronomy, Astronomical Institute (AIRUB), 44780 Bochum, Germany \\
              \email{mstein@astro.ruhr-uni-bochum.de}
        \and
        Max-Planck-Institut für Radioastronomie, Auf dem Hügel 69, 53121 Bonn, Germany
        \and
        CSIRO Space and Astronomy, PO Box 1130, Bentley WA 6102, Australia
        \and
        University of Manitoba, Dept of Physics and Astronomy, Winnipeg, Manitoba R3T 2N2, Canada
        \and
        Dpto. Astronomía Extragaláctica, Instituto de Astrofísica de  Andalucía (IAA-CSIC), Glorieta de la Astronomía s/n, 18008 Granada, Spain
        \and
        Department of Physics and Astronomy, The University of Calgary 2500 University Drive NW Calgary AB T2N 1N4, Canada
        \and
        Hamburger Sternwarte, University of Hamburg, Gojenbergsweg 112, 21029 Hamburg, Germany
        \and
        Department of Physics, Engineering Physics \& Astronomy, Queen’s University, Kingston, ON K7L 3N6, Canada}

   \date{Received 4 December 2025 / Accepted 2 May 2026}
 
  \abstract
   {Galactic magnetic fields regulate star formation and cosmic-ray (CR) transport. Understanding their three-dimensional structure is key to constraining galactic CR transport.}
   {By stacking the linearly polarised signal from a sample of star-forming edge-on galaxies, we aim to analyse the average magnetic field structure of star-forming galaxies.}
   {Using synthetic data, we explored the validity of stacking Stokes $Q$ and $U$ spectra to infer the intrinsic polarisation characteristics of star-forming galaxies. Before stacking, we aligned, scaled, convolved, and reprojected $C$-band (6\,GHz, $\lambda=0.05$\,m) Stokes $Q$ and $U$ cubes of 27 star-forming edge-on galaxies. For the stacked cubes, we performed an RM-synthesis and discussed the derived polarised intensity (PI), polarisation angle ($\chi_0$), and RM maps.}
   {Synthetic data tests demonstrate that stacking Stokes $Q$ and $U$ spectra is valid for tightly constrained underlying distributions of PI, $\chi_0$, and RM.  We identify the spread of the underlying RM distribution ($\sigma_\mathrm{RM}$) as a critical parameter and establish a threshold of  $\sigma_\mathrm{RM}^\mathrm{thresh}=180$\,rad\,m\textsuperscript{-2} ($\sigma_\mathrm{RM}^\mathrm{thresh}\,\lambda^2=0.45$\,rad). For conditions that represent star-forming galaxies, stacking introduces a systematic uncertainty of $\delta_\mathrm{RM}^\mathrm{sys}=90$\,rad\,m\textsuperscript{-2} and underestimates the recovered PI. Stacking results reveal an X-shaped pattern, and polarised emission is detected up to 7\,kpc above the galactic disc. We find higher PI on the approaching side of galaxies and a decrease in PI in the galactic halo of $\sim\!60\%$\ near the galaxy's minor axis. For an X-shaped halo field, this aligns with the fitted opening angle of these galaxies and suggests that there is energy equipartition between CRs and the magnetic field rather than a uniform CR density. A global RM pattern, as reported in a previous study, cannot be confirmed.}
   {We present stacking of Stokes $Q$ and $U$ cubes as an effective tool to recover faint polarised emission in the halo of nearby galaxies, if the underlying distributions of PI, $\chi_0$, and RM are tightly constrained.}


   \maketitle

\section{Introduction}
\label{sec:intro}
In the context of galaxy evolution, magnetic fields (B fields) are understood as a moderator of star-formation activity \citep{2018NatAs...2...83T}. Additionally, galactic B fields govern the transport of highly energetic, charged particles, namely cosmic rays (CRs), which further increases their importance in galactic feedback processes. Radio continuum emission, as a tracer of cosmic ray electrons (CREs) and magnetic fields, has therefore become an essential tool for deepening our understanding of CR transport in galaxies and revealing the multi-scale nature of galactic magnetic fields \citep[see][for reviews]{2021Ap&SS.366..117H,2015A&ARv..24....4B}. At large scales, modelling the B field in the Milky Way (MW) and in external galaxies remains a challenging task, with a variety of approaches already available \citep[e.g.][]{2012ApJ...761L..11J, 2019ApJ...877...76K,2014A&A...561A.100F,2019A&A...623A.113S,2022A&A...658A.101H,2024ApJ...970...95U}. The observed X-shaped B field in the MW halo and in the haloes of edge-on late-type galaxies is seen as an opportunity to gain insight into the process of halo magnetisation and to better understand the role of magnetic fields in disc-halo interaction processes \citep[see Sect. 1 of][and references therein, for a more detailed discussion on possible formation scenarios of the X shape]{2025A&A...696A.112S}.

To create a rich database for studying the radio emission of edge-on late-type galaxies, the Continuum Halos in Nearby Galaxies: An EVLA\footnote{Now known as the Karl G. Jansky VLA.} Survey (CHANG-ES) \citep{2012AJ....144...43I} collected data for 35 nearby galaxies in multiple array configurations at the $L$ band (1.5\,GHz, $\lambda_L=0.2$\,m) and $C$ band (6\,GHz, $\lambda_C=0.05$\,m)\footnote{See \url{https://projects.canfar.net/changes/} for the project website and data release information.}. CHANG-ES data have been used to study the CRE transport from the galactic disc into the halo, as well as the observed B field geometry of individual galaxies or small subsamples in great detail \citep[e.g.][]{2019A&A...632A..11M,2020A&A...639A.111S,2022MNRAS.509..658H, 2023A&A...670A.158S}. In addition to these results obtained with the CHANG-ES $L$ band and $C$ band data, the first case studies of NGC~4217 \citep{2024A&A...691A.273H} and NGC~3556 \citep{2025ApJ...978....5X} reveal the potential offered by the newly obtained CHANG-ES $S$ band (3\,GHz) data to analyse the CR transport and the B field geometries of these systems at this intermediate frequency. 

Complementary to these case studies or analyses of small subsamples, \citet{2018A&A...611A..72K} and \citet{2025A&A...699A.243H} analyse the total intensity radio halo scale heights of the whole CHANG-ES sample and relate them to star-formation properties in the galactic disc. Further, \citet{2025A&A...696A.112S} analyses the observed X shape in the polarisation angle morphology and presents a relation between the X-shape opening angle and the star-formation rate surface density in the disc. While these `large' sample studies (relative to case studies)  still analyse the signal of each galaxy individually, \citet[][hereafter \citetalias{2020A&A...639A.112K}]{2020A&A...639A.112K} present an alternative approach. By stacking the weighted Stokes $Q$ and $U$ images (after rescaling and aligning the individual galaxies), \citetalias{2020A&A...639A.112K} present a stack of 28 CHANG-ES galaxies for polarised intensity and polarisation angle and show that the X-shaped polarisation pattern that has been observed in individual galaxies also remains in this stack. This finding strengthens the idea that this X shape is a common feature in radio haloes of star-forming late-type galaxies.

In this paper, we continue the work of \citetalias{2020A&A...639A.112K} not only by stacking the Stokes $Q$ and Stokes $U$ images of galaxies but also using the spectral information that is embedded in the datasets. Therefore, we stack Stokes $Q$ and Stokes $U$ cubes\footnote{Throughout this paper, Stokes $Q$ and Stokes $U$ cubes will be abbreviated as $Q$ and $U$ cubes.} and perform an RM synthesis on these stacked cubes. This allows us to create a stack of polarised intensity and polarisation angles similar to that presented in \citetalias[][Fig.~1]{2020A&A...639A.112K} but with the benefit of de-rotating the observed polarisation angles for the Faraday rotation in the line of sight. \citep{2005A&A...441.1217B,2009IAUS..259..591H}. This new approach further allows us to search for possible common features in the morphology of the observed rotation measure signal (see Sect. \ref{sec:method} for a detailed description of the data processing procedure).

This study uses the following basic principles and definitions. The polarisation angle ($\chi$, pol. ang.) of the linear polarised radio emission is derived from the  on-sky measured Stokes parameters $Q$ and $U$:
\begin{equation}
    \chi = \frac{1}{2} \arctan\left({\frac{U}{Q}}\right).
    \label{eq:lin_pol_ang}
\end{equation}
As we do not account for circular polarisation, we compute the polarised intensity (PI) from the linear polarisation components alone:
\begin{equation}
    \mathrm{PI} = \sqrt{Q^2 + U^2}.
    \label{eq:lin_pol_int}
\end{equation}

The wavelength- ($\lambda$) dependent change in observed pol. ang., caused by Faraday rotation, is characterised by the rotation measure (RM) of the line of sight (LoS):
\begin{equation}
    \chi(\lambda) = \chi_0 + \mathrm{RM}\, \lambda^2,
    \label{eq:rm_foreground}
\end{equation}
 where $\chi_0$ is the undisturbed pol. ang. of the emission angle and RM traces the LoS integral of the total electron density ($n_e$) times the B field component that is parallel to the LoS ($B_\parallel$):
\begin{equation}
    \mathrm{RM} = 0.81 \int_{\text{LoS}} \left( \frac{n_e}{\mathrm{cm}^{-3}} \right) \left( \frac{B_\parallel}{\mu\mathrm{G}} \right) \left( \frac{dr}{\mathrm{pc}} \right) \ \mathrm{rad}\,\mathrm{m}^{-2}.
\label{eq:rm_def}
\end{equation}
A positive RM indicates that $B_\parallel$ is pointing towards the observer \citep{2005A&A...441.1217B}. Furthermore, \citet{2005A&A...441.1217B} point out that generally there is a difference between the Faraday depth $\phi(r)$ and the RM, which only vanishes in the case of a single synchrotron source located behind a purely rotating Faraday screen.

As the physical conditions along a given LoS can change drastically, one typically distinguishes multiple RM contributors \citep[e.g.][]{1977A&A....61..771K}:
\begin{equation}
    \mathrm{RM} = \mathrm{RM_{MW}} + \mathrm{RM_{IG}} + \mathrm{RM_{host}} .
    \label{eq:rm_component_general}
\end{equation}
Here, $\mathrm{RM_{MW}}$ indicates the RM contribution of the Milky Way, $\mathrm{RM_{host}}$ is the RM that is caused by the host galaxy itself, and $\mathrm{RM_{IG}}$ summarises the contributions that occur between the host and the MW. Depending on the studied object, $\mathrm{RM_{IG}}$ can contain RM contributions of the circumgalactic medium (CGM), the intergalactic medium (IGM), the intracluster medium (ICM), or cosmic filaments. As our analysis only consists of nearby galaxies  ($D\leq42\,\mathrm{Mpc}$), the effect of large-scale structure components (e.g. filaments, intervening clusters) is limited. In addition, \citet{2023A&A...670L..23H} estimate the RM contribution of the CGM in a sample of nearby galaxies to $\mathrm{RM_{CGM}}=3.7\,\mathrm{rad\,m^{-2}}$, and \citet{2020MNRAS.495.2607O} report an upper limit of $\mathrm{RM_{IGM}}<1.9\,\mathrm{rad\,m^{-2}}$ for the IGM contribution. Both components are subdominant compared to the RM contributions of the host galaxy. Therefore, in this study, we  consider the observed RMs to contain only two components:
\begin{equation}
     \mathrm{RM} = \mathrm{RM_{MW}} + \mathrm{RM_{host}}.
     \label{eq:rm_component_study}
\end{equation}

In addition to the basic principles described above, we point out two depolarisation effects that play a crucial role in understanding the observed polarised emission in galaxies. The depolarisation factor ($\mathrm{DP}=p/p_i$, where $p$ and $p_i$ refer to the observed and initial degree of polarisation) due to internal Faraday dispersion (IFD, caused by turbulent magnetic fields) strongly depends on the observational frequency \citep{2011MNRAS.418.2336A}:
\begin{equation}
    \mathrm{DP} = \frac{1-e^{-S}}{S}, S\propto\lambda^4.
\end{equation}
Therefore, IFD affects low-frequency emission much more severely than high-frequency emission. In addition to the DP effect due to IFD, differential Faraday rotation (DFR, caused by regular magnetic fields)\footnote{Compared to the DP due to IFD, DFR is a subdominant process in star-forming galaxies.} can cause flips in $\chi$ and RM that can lead to an incorrect result \citep[][Fig.~13]{1992A&A...265..417H}. Following \citet{1992A&A...265..417H} and \citet[][Eq.~4]{2011MNRAS.418.2336A}, this `flip' occurs if $\chi=\pm n \times (\pi/2)$. For the $L$ band and $C$ band, this corresponds to critical RM values of $\mathrm{RM}_{\mathrm{crit}}^L= \pm n \times 39$\,\radmsquare\  and $\mathrm{RM}_{\mathrm{crit}}^C=\pm n \times 628$\,\radmsquare.

An example that shows the impact of both effects can be found in \citet[][Fig.~22]{2015A&A...578A..93B}. Here, the RM values derived from the low frequency dataset ($\lambda$6cm-$\lambda$20cm) are much smaller (due to IFD) than RM values derived from a high frequency dataset ($\lambda$3cm-$\lambda$6cm) and can also show  the opposite sign of predicted RMs. In  \citetalias{2020A&A...639A.112K} as well, the authors mention a large difference between the RM values derived from $L$ band and $C$ band data and conclude that the $L$ band data are strongly affected by DP effects.

Following up on the results presented in \citetalias{2020A&A...639A.112K}, \citet[][hereafter \citetalias{2021A&A...649A..94M}]{2021A&A...649A..94M} claim to find a large-scale quadrupolar RM pattern when stacking the RM signal of 24 galaxies of the CHANG-ES sample. To perform the stacking, \citetalias{2021A&A...649A..94M} constructed synthetic RM maps of the CHANG-ES galaxies by comparing pol. ang. maps in the $L$ band and $C$ band. Converted to the Quadrant (Q) labelling described in Fig. \ref{fig:sectors}, the authors find predominantly negative RMs in QI and QIII and positive RMs in QII and QIV. The reported RM amplitude of this structure is $\sim\!15$\,rad\,m\textsuperscript{-2}.

However, here we highlight some limitations of the data processing described in \citetalias{2021A&A...649A..94M} that could strongly influence their findings. Firstly, by comparing the polarisation angle at the $L$ band ($\chi_L$) and $C$ band ($\chi_C$) the authors compare physically disjunct areas.  As IFD causes stronger depolarisation in the $L$ band than in the $C$ band, the $C$ band emission traces regions that are located much deeper in the galaxy, while the $L$ band emission only traces an outer layer. Therefore, comparing $\chi_C$ and $\chi_L$ cannot produce a physically meaningful RM measurement.

As noted above, DFR can cause flips in the derived RM value. As shown in studies of CHANG-ES galaxies using RM synthesis \citep[e.g.][]{2019A&A...632A..11M}, the observed distribution of RM values ranges from several tens up to a few hundreds \radmsquare\ and is therefore higher than the derived critical RM in the $L$ band. Accordingly, we conclude that the $L$ band pol. ang. information used by \citetalias{2021A&A...649A..94M} is unsuitable for inferring the RM structure in galactic discs.

Lastly, judging from the derived RM map (Fig.~1, left panel, \citetalias{2021A&A...649A..94M}), the detected symmetry could be influenced by the choice of the analysed region. By slightly increasing their region of interest (highlighted as a white ellipse), the distribution of RM values in their top left quadrant, which were reported to be predominantly positive, could equal out or even become negative and therefore destroy the reported RM structure. To conclude, we argue that the RM structure presented by \citetalias{2021A&A...649A..94M} is not conclusive.

Nevertheless, a global RM pattern observed in stacked samples of galaxies, if genuine, is not yet explained by current theoretical frameworks and would strongly influence the current understanding of galactic B fields \citep{2022A&A...658A.101H}. Therefore, we explore the possibility of detecting a global RM pattern derived from our stacking procedure (see Sect.~\ref{sec:method}).

This paper is structured as follows. In Sect.~\ref{sec:method}, we describe the data processing as well as the reasoning behind our sample selection. Tests on synthetic data regarding the stacking technique are presented in App.~\ref{app:stack_sys}. The results of the described data processing are presented in Sect.\ref{sec:res}, focusing on the morphology of the detected polarised emission (Sect.~\ref{sec:res_pi_pol_ang}, Sect.~\ref{sec:res_pi_asym}, and Sect.~\ref{sec:res_minor_axis}) and the search for large-scale RM patterns (Sect.~\ref{sec:res_rm_struc}). Then, we discuss implications of our findings in Sect.~\ref{sec:dis} and conclude this paper in Sect.~\ref{sec:SandO}.

\section{Methodology}
\label{sec:method}
A summarising flowchart of the data processing outlined below is displayed in Fig. \ref{fig:meth_flowchart} and the code to produce the results of this paper is publicly available\footnote{\url{https://github.com/msteinastro/changes_xxxix}}.
\subsection{Sample selection}
\label{sec:meth_sample_selection}
Starting from the original CHANG-ES galaxy sample, \citetalias{2020A&A...639A.112K}, rejected seven galaxies (NGC~2992, NGC~4244, NGC~4438, NGC~4594, NGC~4845, NGC~5084, and UGC~10288) because of the absence of any radio emission from the galaxy, the detection of radio emission solely from the galaxy’s core or jet, or the presence of background radio sources that dominate over the galaxy’s emission. We additionally excluded NGC~660 from our analysis as it is a highly disturbed galaxy, classified as a polar ring galaxy \citep{1990AJ....100.1489W}. After excluding these eight galaxies, our initial sample consisted of 27 star-forming edge-on galaxies with detected diffuse radio emission (see Table \ref{tab:fundamental_parameters}).

\subsection{Initial data reduction}
\label{sec:meth_initial_dr}
As in \citetalias{2020A&A...639A.112K}, in this study we analysed the CHANG-ES $C$ band ($6\,\mathrm{GHz}$) datasets. The calibration and initial imaging procedure were performed using the CASA routines \citep{2022PASP..134k4501C}. First, the $C$ band data in each of the 16 spectral windows were averaged, resulting in a frequency spacing of {128\,MHz}, covering a spectral range from 5\,GHz to 7\,GHz. Polarisation calibrated measurement sets from the C and D array observations were combined and imaged using a Briggs weighting of {\texttt{robust}=+2} and a pixel scale of \SI{1.5}{\arcsec}.  We further applied multi-scale CLEAN on scales of \SI{0}{\arcsec}, \SI{15}{\arcsec}, and \SI{40}{\arcsec}, with a noise threshold of $25\,\mathrm{\mu Jy\, Beam^{-1}}$ and a maximum of 2000 clean iterations. Afterwards, all individual slices were convolved to a common resolution, full width half maximum (FWHM) of the synthesised radio beam, of \SI{12}{\arcsec}\footnote{Except for NGC~5907, which was convolved to \SI{12.5}{\arcsec}.} and primary beam corrected. For some galaxies, individual spectral windows had to be excluded in the flagging process. Furthermore, the computed central frequency of spectral windows may change slightly if individual channels were flagged. Therefore, we compiled a reference frequency grid by averaging the central frequencies of the individual spectral windows for all galaxies.

\subsection{Correction for the MW foreground RM}
\label{sec:meth_MWfg}
To correct our data cubes for the RM MW foreground, we extracted the MW foreground RM at the position of each galaxy from the Galactic RM map published by \citet{2022A&A...657A..43H}. As derived in Sect. \ref{sec:intro} (Eqs. \ref{eq:rm_foreground} and  \ref{eq:rm_component_study}), we subtracted the RM contribution of the MW by first computing the observed polarisation intensity and angle cubes PI\textsubscript{obs} and $\chi$\textsubscript{obs} from the pre-processed $Q$ and $U$ cubes (see Sect. \ref{sec:meth_initial_dr}). We then de-rotated the observed pol. ang. as:
\begin{equation}
    \chi_\mathrm{{host}}=  \chi_\mathrm{{obs}} - \mathrm{RM_{MW}}\, \lambda^2.
\end{equation}
Assuming that the MW contribution is only rotating the signal and not adding to the overall emission, we constructed $Q$\textsubscript{host}- and $U$\textsubscript{host}-cubes from the de-rotated pol. ang. cube and the observed PI cube:

\begin{eqnarray}
        Q_\mathrm{{host}} &=& \mathrm{PI_{obs}}\,  \cos(2\, \chi_\mathrm{host}), \\
        U_\mathrm{{host}} &=& \mathrm{PI_{obs}}\,  \sin(2\, \chi_\mathrm{host}). 
\end{eqnarray}

\subsection{Galaxy alignment}
\label{sec:meth_GA}

To align the galaxies, we rotated them so that the galaxy major axis is parallel to the image \textit{x}-axis. Since the galaxy morphology in PI can be very complex, we applied the rotation based on the Hyperleda position angle (PA, Table \ref{tab:fundamental_parameters}) to the CHANG-ES D array $C$ band ({\texttt{robust}=0}) total intensity images \citep{2015AJ....150...81W} to check if the applied rotation resulted in a proper alignment. If not, we visually adjusted the rotation angle, so that the radio total intensity major axis of each galaxy aligns properly with the image \textit{x}-axis. The applied rotation angles are listed in Table \ref{tab:stacking_parameters}. Using the MW-foreground corrected $Q$\textsubscript{host}- and $U$\textsubscript{host}-cubes, we constructed $\chi$\textsubscript{host} and PI\textsubscript{host} according to Eqs. \ref{eq:lin_pol_ang} and \ref{eq:lin_pol_int}. To preserve the morphology of the polarisation angle cube under the applied rotation, we subtracted the rotation angle ($\omega$) from the pol. ang. cube:

\begin{equation}
    \chi_\mathrm{{host}}^\prime = \chi_\mathrm{{host}} - \omega.
\end{equation}
Then, we rotated\footnote{Rotation of the cubes was performed using the \texttt{scipy.ndimage.rotate} routine \citep{2020SciPy-NMeth}.} the  $\chi_\mathrm{{host}}^\prime$ cube and PI\textsubscript{host} cube and constructed $Q$\textsubscript{rot} and $U$\textsubscript{rot} from these rotated datasets. In Sect. \ref{sec:res} we discuss results of three different alignment strategies:
\begin{enumerate}
    \item `standard' alignment (std): Galaxies were aligned using a smallest possible rotation. We use this quasi-random strategy as baseline for our other approaches.
    \item `rotation' alignment (rot): We rotated all galaxies so that their approaching side is in the eastern (left) half of the image. This strategy enables us to check if the rotation sense of the galaxy impacts the morphology of the detected polarised emission.
    \item `double' alignment (dbl): Each galaxy entered the stack twice, once with standard alignment and with an extra $180^\circ$ rotation applied. With this approach, we synthetically increase our sample by a factor of two, which allows us to further trace the polarised emission, at the cost of introducing a point symmetry. 
\end{enumerate}

\subsection{Masking background sources}
\label{sec:meth_masking}
To mitigate the effect of background sources on our stacking experiments, we ran \texttt{PyBDSF} \citep[][using a pixel detection threshold of $5\,\sigma$ ($\sigma$ indicates the standard deviation of the background) and an island boundary threshold of $2.5\,\sigma$]{2015ascl.soft02007M} on the rotated total intensity images to detect sources in the projected vicinity of our target galaxies. Then, we removed the target galaxy from the resulting source mask and applied it to $Q$\textsubscript{rot} and $U$\textsubscript{rot} to construct $Q$\textsubscript{mask} and $U$\textsubscript{mask}. With the automatic source detection, we typically remove less than 5\% data. While this process is effective in removing isolated and compact sources in the projected vicinity of the target galaxy, background sources that overlap with the diffuse halo will remain in the data. Therefore, if necessary, we masked individual background sources manually. 
\subsection{Galaxy scaling, resolution matching, and reprojection}
\label{sec:meth_smr}
As a final step before stacking the $Q$ and $U$ cubes, the data had to be scaled to a common scale and reprojected to a reference coordinate frame. Additionally, we applied a convolution to the individual frequency slices so that all datasets have a similar resolution. Here, we implemented two different approaches: angular size stacking ($Q_\mathrm{{ang,\,regrid}}$, $U_\mathrm{{ang,\,regrid}}$) and physical size stacking ($Q_{\mathrm{phy,\,regrid}}$, $U_{\mathrm{phy,\,regrid}}$). The choice between angular and physical size scaling depends on the nature of the observed B  fields. Angular size scaling is the appropriate choice if the morphology of the B field is expected to scale proportionally with the overall size of the galaxy. Conversely, physical size scaling is preferred if the characteristic scales of the magnetic patterns remain constant in absolute units regardless of the host galaxy's total dimensions.

\begin{figure}
    \centering
    \includegraphics[width=0.65\linewidth]{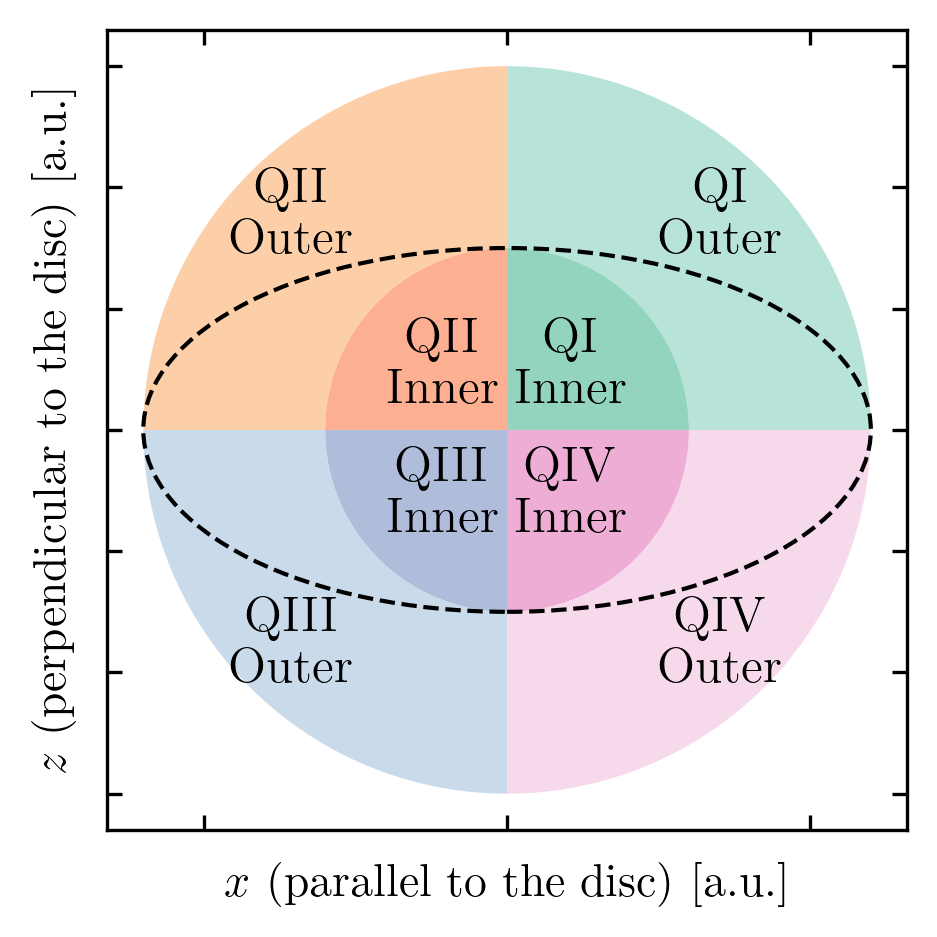}
    \caption{Sketch of the sectors that were analysed separately. Each quadrant was separated into an `inner' and an `outer' sector. The dashed ellipse indicates the galactic polarised emission.}
    \label{fig:sectors}
\end{figure}

For angular size scaling (ang), we only selected galaxies whose angular extent is large enough that the diameter of the star-forming disc (Table \ref{tab:fundamental_parameters}) is covered by at least $N^{\mathrm{Beam}}_{\mathrm{min}}=\angstackminsampling$ beams. The choice of the minimal required sampling has two effects. By reducing the minimal sampling requirement $N^{\mathrm{Beam}}_{\mathrm{min}}$, more galaxies can be used in the stacking experiment. However, increasing the size of the radio beam reduces the detected polarised signal due to beam depolarisation \citep{1998MNRAS.299..189S}. Furthermore, since we plan to compare different regions in the galaxy stack (`inner' vs. `outer', individual quadrants, see Fig. \ref{fig:sectors}), we chose $N^{\mathrm{Beam}}_{\mathrm{min}}$ so that an individual inner sector is still covered by more than a single beam. In the angular stacking, we define `inner` as $r<\angstackinnerradiusbeams\,\mathrm{B_{maj}}$ and `outer' as $\angstackinnerradiusbeams\,\mathrm{B_{maj}}\leq r <\angstackouterradiusbeams\,\mathrm{B_{maj}}$. Using this sampling requirement, we convolved all galaxies so that $d_{\mathrm{ang}}/\mathrm{B_{maj}}=\angstackminsampling$. From the galaxy sample described in Table \ref{tab:fundamental_parameters}, a total number of $N_\mathrm{ang}=21$ satisfy the resolution requirement and were included in the angular size stacking (see Table \ref{tab:stacking_parameters}). 
In the initial data reduction, the images were compiled so that the \texttt{CRPIX} keywords point to the centre of each cube slice and the \texttt{CRVAL} keywords represent the position of the target galaxy on the sky. Finally, we set the celestial \texttt{CRVAL} keywords to (0,0) for all galaxies and reproject\footnote{Image reprojection is performed using the \texttt{astropy reproject} library (\url{https://reproject.readthedocs.io/en/stable/}).} each cube so that the radio beam is sampled by \beampixratioangular\,pixels in each direction. This procedure resulted in Q- and U-cubes of matching coordinate centre, resolution, and pixel size with regard to the angular extent of each galaxy. To reduce the bias towards galaxies with high PI in the stacking routine, we introduced a flux normalisation\footnote{In this article, we use ‘flux’ as an abbreviation for flux density. In addition, we use the term ‘scaling’ for a change in size and ‘normalisation’ for a change in the PI or RM domain.} strategy where we normalised the $Q_{\mathrm{ang,\,regrid}}$ and $U_{\mathrm{ang,\,regrid}}$ cubes using the polarised flux measured on PI maps of the individual galaxies (see Table \ref{tab:stacking_parameters}, PI\textsubscript{ang}):
\begin{equation}
 \mathrm{X_{ang,FluxNorm}} = \frac{\mathrm{X_{ang,reproject}}}{\mathrm{PI_{ang}}} \quad \text{for} \quad \mathrm{X} \in {Q, U}.
 \end{equation}
 
For stacking galaxies with regard to their physical size (phy) we computed the physical resolution and pixel size for each galaxy using the distances listed in Table \ref{tab:fundamental_parameters}. In this stack, we only considered galaxies with a resolution better than \phyres\,kpc and reprojected all galaxies into a matching coordinate frame with a pixel size of \phypixsize\,kpc. For physical size scaling, a total number of $N_\mathrm{phy}=25$ (see Table \ref{tab:stacking_parameters}) galaxies were included in the stacking procedure. To account for brightness differences due to the distance $D$ of each galaxy, we normalised the fluxes by virtually moving all galaxies to a distance of \phyrefdistance\,Mpc (approximately the mean distance of the analysed sample):
\begin{equation}
        \mathrm{X_{phy,\,DistNorm}} = \mathrm{X_{phy,\,reproject}} \left(\frac{D\,\mathrm{[Mpc]}}{\phyrefdistance}\right)^2 \quad \text{for} \quad \mathrm{X} \in {Q, U}.
\end{equation}
As an alternative approach (similar to the angular scaling),  we also stacked the galaxies after normalising for their measured PI (see Table \ref{tab:stacking_parameters}, PI\textsubscript{phy}):
\begin{equation}
 \mathrm{X_{phy,FluxNorm}} = \frac{\mathrm{X_{phy,reproject}}}{\mathrm{PI_{phy}}} \quad \text{for} \quad \mathrm{X} \in {Q, U}.
 \end{equation}

\subsection{Galaxy stacking}
\label{sec:meth_stacking}
\begin{figure}
    \centering
    \includegraphics[width=0.75\linewidth]{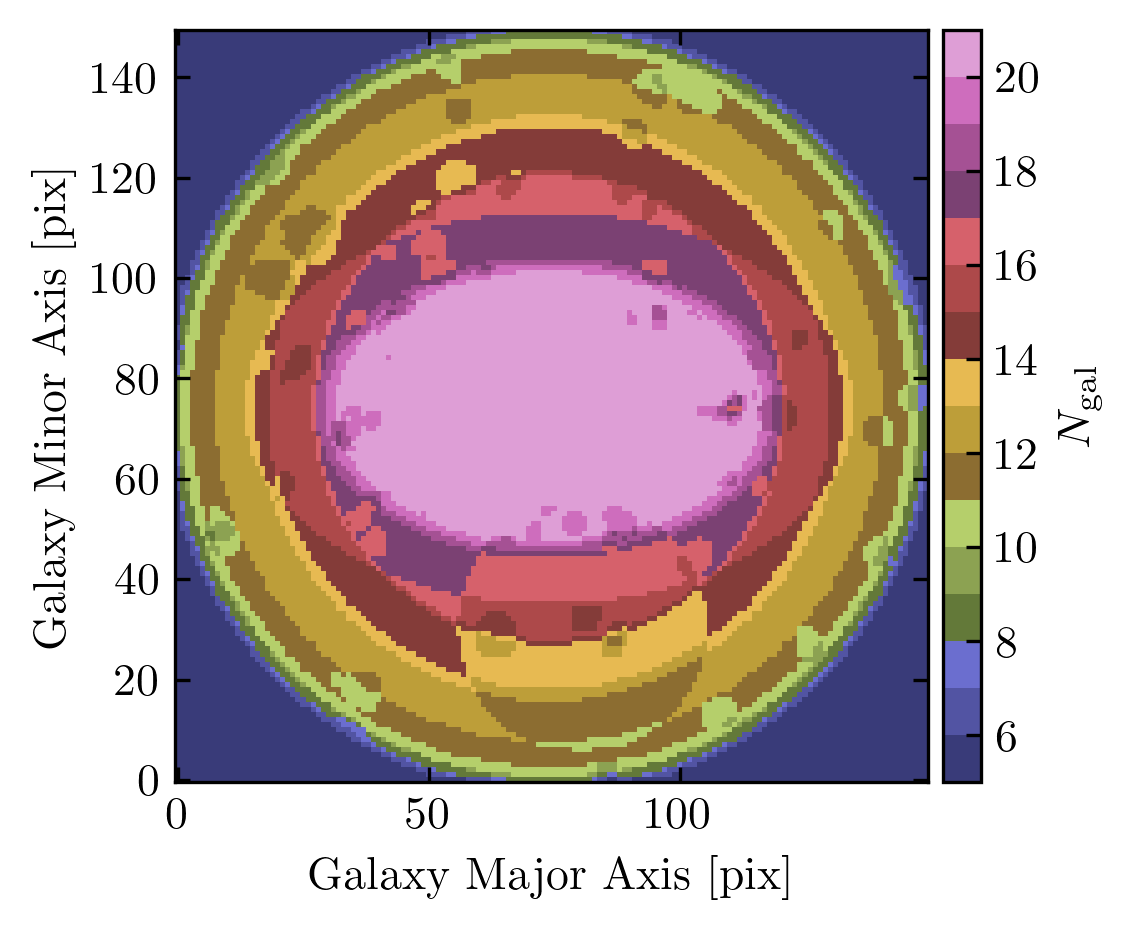}\\
    \includegraphics[width=0.75\linewidth]{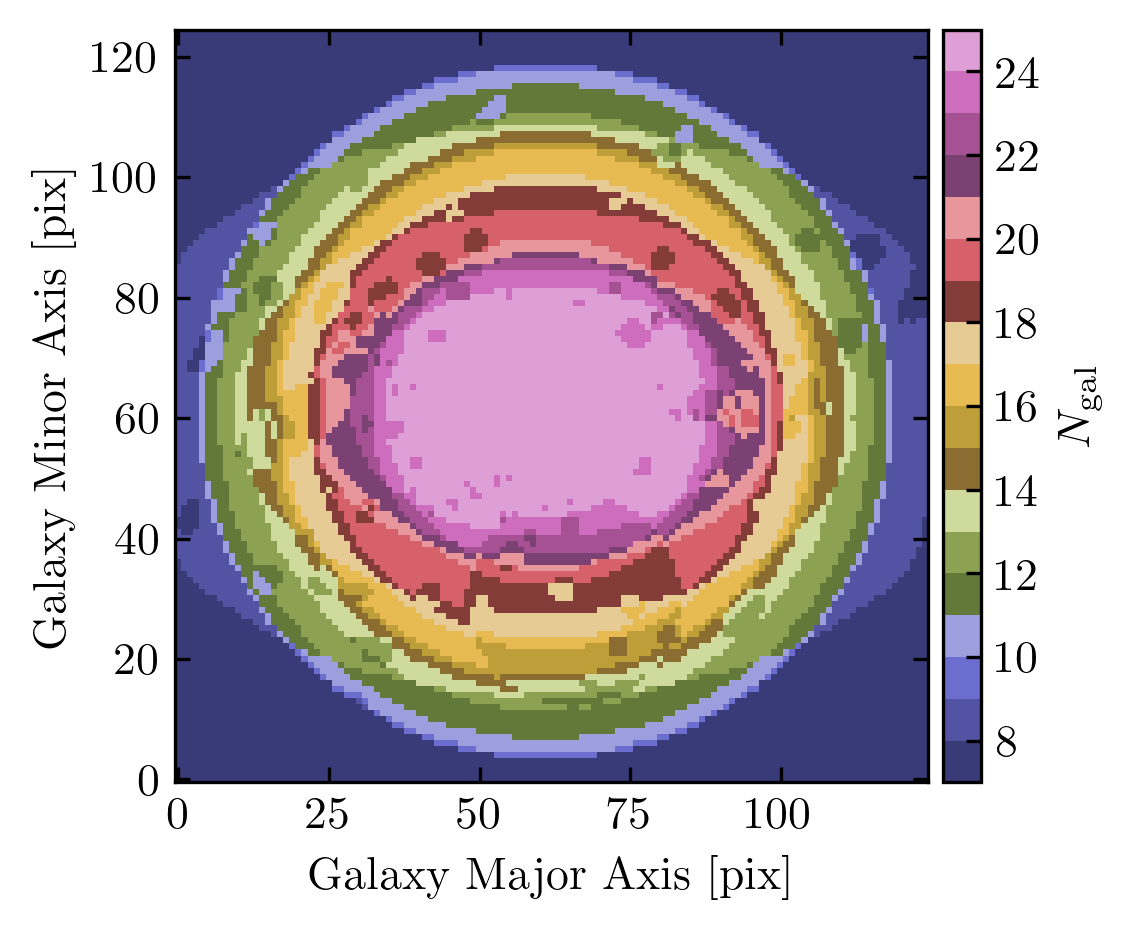}
    \caption{Heatmap of the index=0 slice, indicating the number of galaxies that contribute to each pixel, for the angular stack (top panel) and physical stack (bottom panel)  in the standard alignment. Galaxies that were originally observed with two pointings and later mosaiced show an elliptical footprint in these plots.}
    \label{fig:heatmap_ang_standard}
\end{figure}

For the angular stack and the physical stack, we place the individual frequency slices of each galaxy in the reference frequency grid and pixel-wise compute mean and median. As an example, we display the number of contributing galaxies for the angular size stack using the standard alignment in Fig. \ref{fig:heatmap_ang_standard}. Here, the effect of the size scaling as well as the masking of background sources in the individual galaxies becomes visible.

\subsection{RM synthesis}
\label{sec:meth_rm}
Finally, RM synthesis was performed on the stacking results (i.e. the $Q$ and $U$ cubes with applied MW foreground correction, masking, alignment rotation, size-scaling, and reprojection), using the three-dimensional RM synthesis implementation \texttt{rmsynth3d} of the \texttt{RM-Tools} package \citep{2020ascl.soft05003P, 2026arXiv260120092V}. We set the absolute maximum probed Faraday depth to 4000\,rad\,m\textsuperscript{-2} with a Faraday depth channel width of 100\,rad\,m\textsuperscript{-2} and activated channel weighting using the inverse variance of the background. To create the complete RM synthesis output, we further processed the resulting data products using the \texttt{rmtools\_peakfitcube} utility. 
With the peak-fitting algorithm implemented in \texttt{rmtools\_peakfitcube}, we computed estimates for PI, $\chi_0$, and RM as well as uncertainty estimates for these parameters ($\delta_\mathrm{PI}$, $\delta_{\chi_0}$, and $\delta_{\mathrm{RM}}$).

Applying RM synthesis to our dataset ($C$ band, 5-7\,GHz) resulted in a measured full width half maximum (FWHM) of the RM spread function (RMSF) of 2118\,rad\,m\textsuperscript{-2}. In the analysis of the RM distribution (Sect. \ref{sec:res_rm_struc}), we only consider RM pixels with a $5\sigma$ detection in the PI map. All polarised fluxes in Table~\ref{tab:stacking_parameters} were derived by placing an aperture that encompasses the detected polarised emission on the polarised intensity map that has been corrected for polarisation bias\footnote{As pointed out in the \texttt{RM-Tools} documentation, polarisation bias correction is only applied to pixels with a signal-to-noise ratio larger than 5.} polarised intensity map. To estimate the flux uncertainty for each measurement, the background emission was estimated in an empty sky region\footnote{This region was not corrected for polarisation bias}.

In App. \ref{sec:app_img_atals}, we show maps of total intensity, polarised intensity, RM, and RM error maps for all analysed galaxies. We only show the results for individual galaxies when applying RM synthesis to the $Q$ and $U$ cube after the initial data reduction (Sect. \ref{sec:meth_initial_dr}) without further data processing (i.e. no alignment, no masking, no scaling, and no flux normalisation has been applied).

To identify possible systematics that arise from stacking Stokes $Q$ cubes and Stokes $U$ cubes before performing RM synthesis, we performed tests based on synthetic Stokes $Q$ and $U$ spectra in App. \ref{app:stack_sys}.
Based on these tests, we derived systematic uncertainties $\delta_\mathrm{RM}^{\mathrm{sys}}$, $\delta_{\chi_0}^{\mathrm{sys}}$, and $\delta_{\mathrm{PI}}^{\mathrm{sys}}$ that contribute to the combined uncertainty estimates\footnote{ $\delta_i^{\mathrm{comb}}=\sqrt{\left(\delta_i\right)^2+\left(\delta_i^{\mathrm{sys}}\right)^2}, \mathrm{for}\ i \in[\mathrm{RM}, \chi_0,\mathrm{PI}]$.} for the parameters computed in RM synthesis. For the analyses presented in this paper, we accounted for a systematic uncertainty in RM\footnote{Appendix \ref{app:stack_sys} also provides systematic uncertainty estimates of $\chi_0$. However, the distribution of the detected polarisation angles is not quantitatively analysed in this work.} of $\delta_\mathrm{RM}^{\mathrm{sys}}=90\,\mathrm{rad\,m^{-2}}$. Furthermore, the tests showed that the recovered PI is significantly underestimated by up to 70\%.

\section{Results}
\label{sec:res}
\subsection{PI and polarisation angle morphology of stacked cubes}
\label{sec:res_pi_pol_ang}
\begin{figure}
    \centering
    \includegraphics[width=0.75\linewidth]{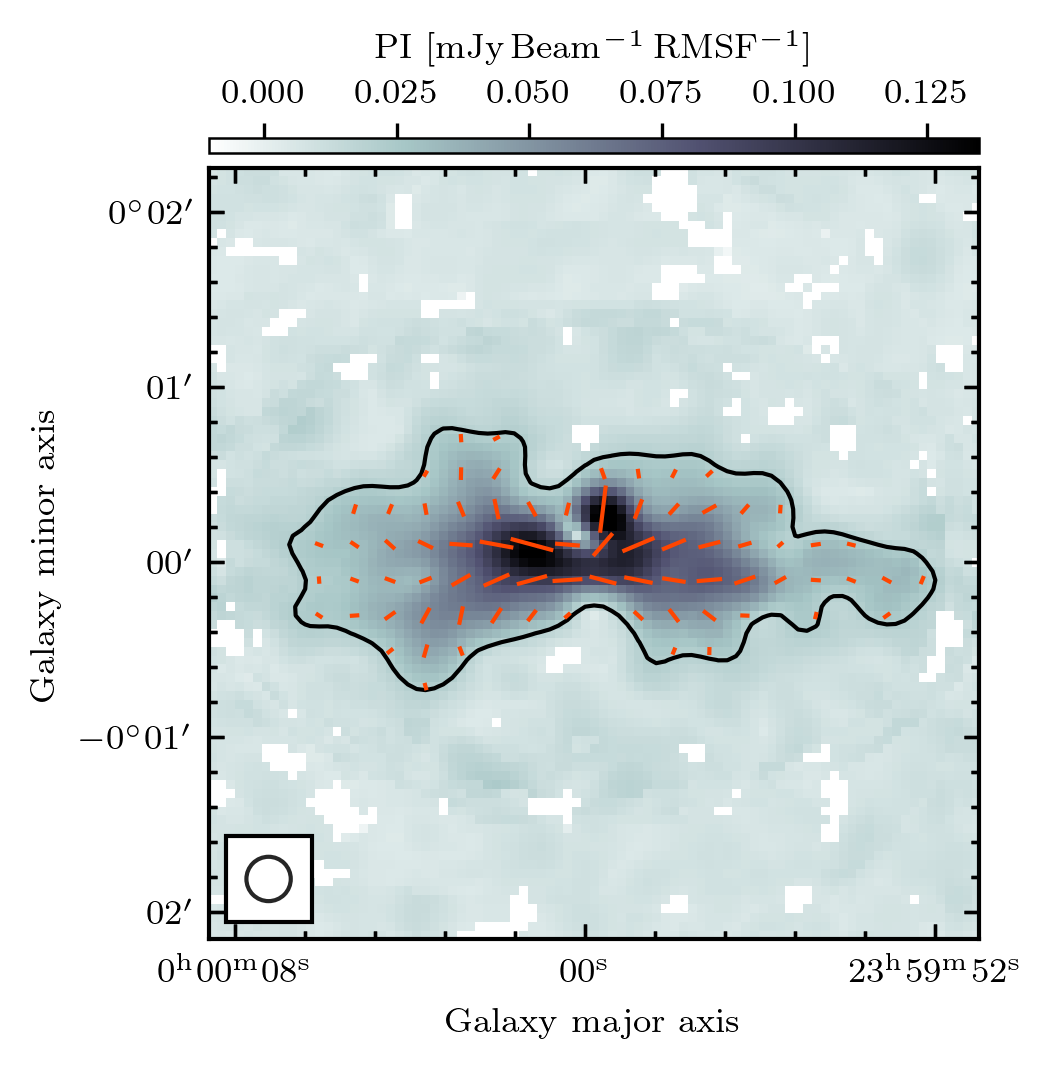} \\
    \caption{PI map with overlaid pol. ang. (orientation of the B field) information for the angular-scaled mean stack without flux normalisation. The vector length corresponds to the PI. The black contour indicates the $3\sigma$-level of the PI emission. The shape of the synthesised beam is indicated in the bottom left corner.}
    \label{fig:stack_standard_mean}
\end{figure}

Figure \ref{fig:stack_standard_mean} presents the results of the standard alignment procedure, which employs angular scaling without flux normalisation. To perform the stacking, we calculated the pixel-wise mean across the individual galaxy samples within the Stokes $Q$ and $U$ cubes. The resulting PI and $\chi_0$ information show a strong similarity to the stacked PI and $\chi$ image derived by \citetalias{2020A&A...639A.112K}. Both datasets show an X shape pol. ang. morphology and a larger $z$-extent in the outer regions of the galactic disc. However, the PI map from \citetalias{2020A&A...639A.112K} as well as the non-normalised mean stack in this study show individual high surface brightness regions that are most likely caused by individual galaxies. The high surface brightness region directly above the centre of the mean stack, the most prominent local structure, can be traced back to the radio lobe of NGC~3079 \citep{1977A&A....58..221D,2017MNRAS.464.1333I,2019AJ....158...21I} (see Fig. \ref{fig:app_img_atlas_3003_3044_3079_3432}).

\begin{figure*}
    \centering
    \begin{subfigure}[b]{0.28\linewidth}
        \centering
        \includegraphics[width=\linewidth]{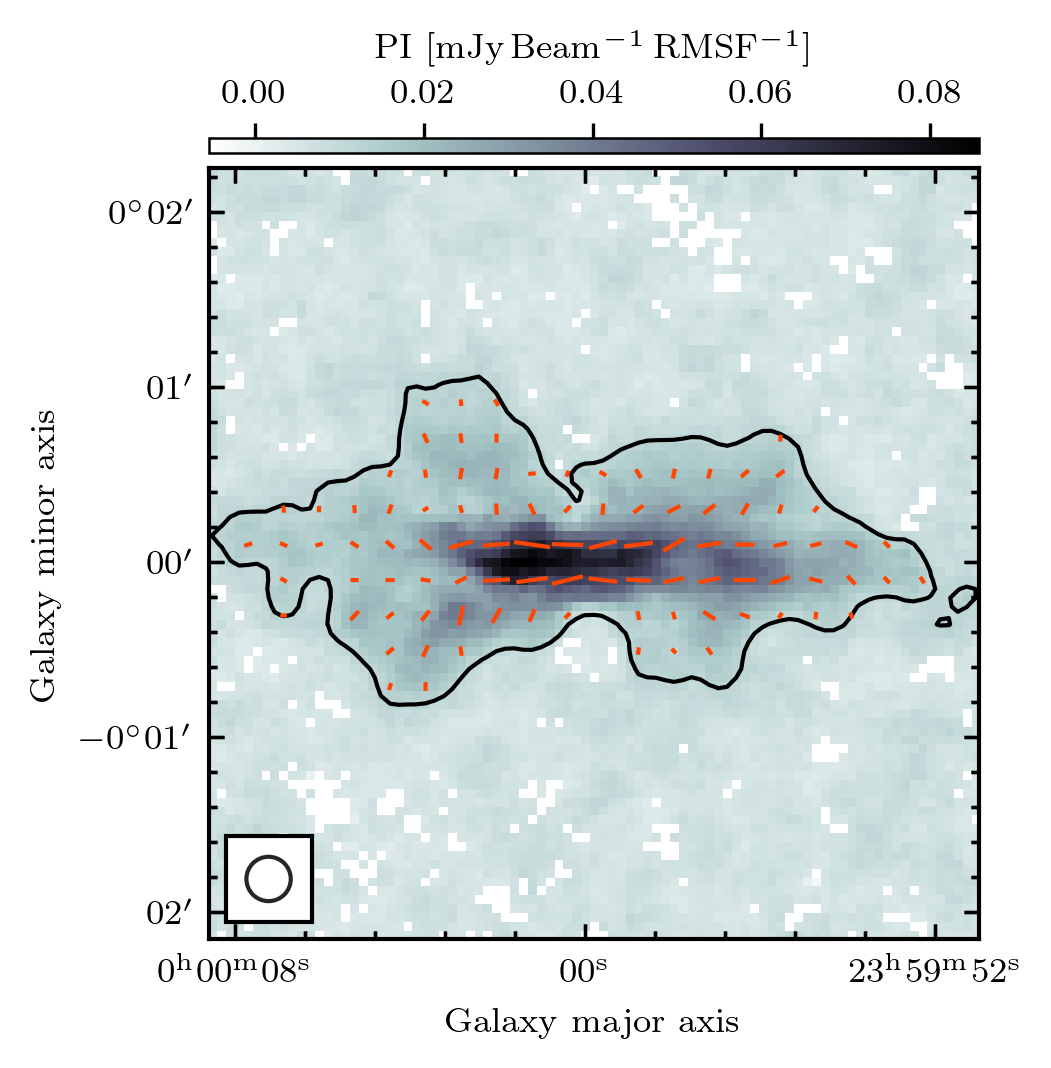}
    \end{subfigure}
    \begin{subfigure}[b]{0.28\linewidth}
        \centering
        \includegraphics[width=\linewidth]{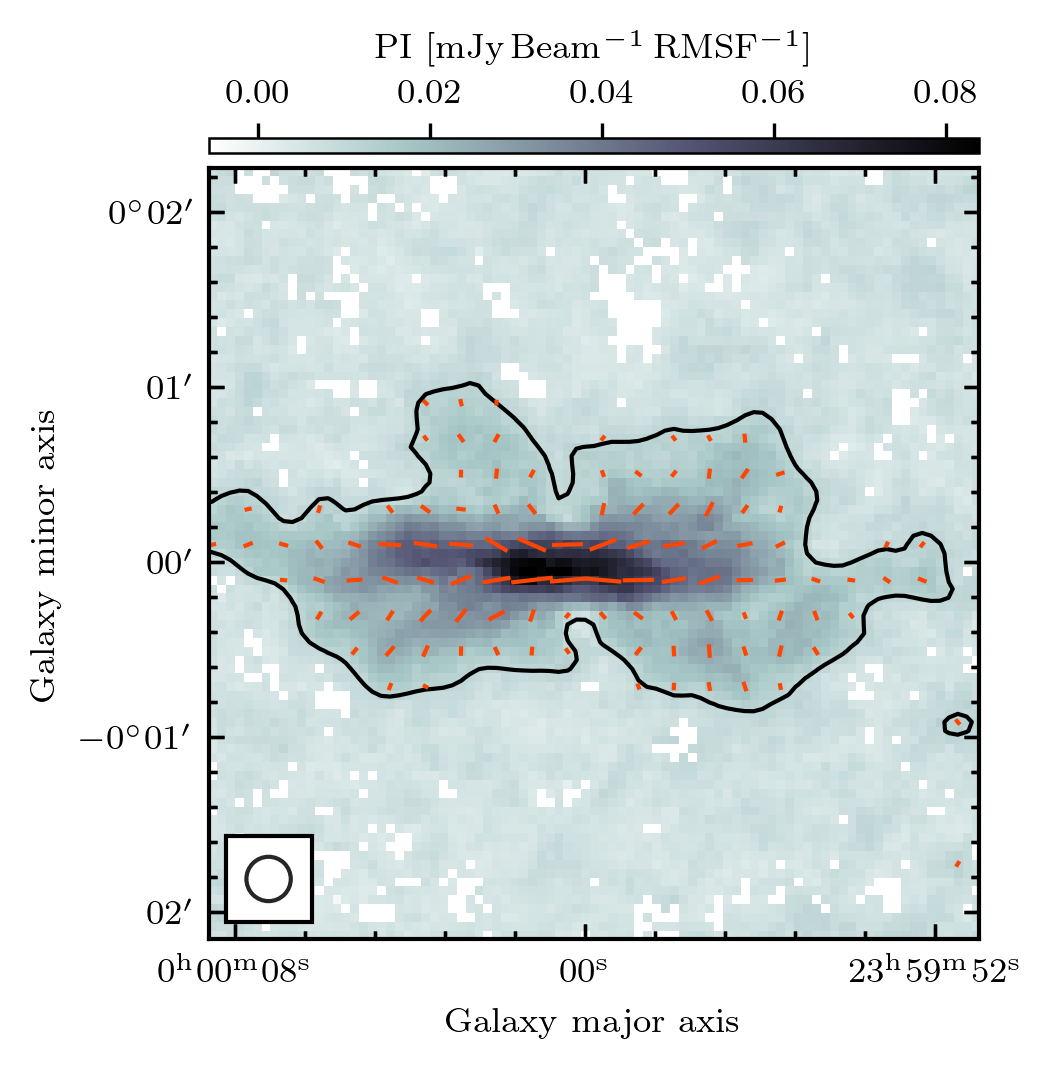}
    \end{subfigure}
    \begin{subfigure}[b]{0.28\linewidth}
        \centering
        \includegraphics[width=\linewidth]{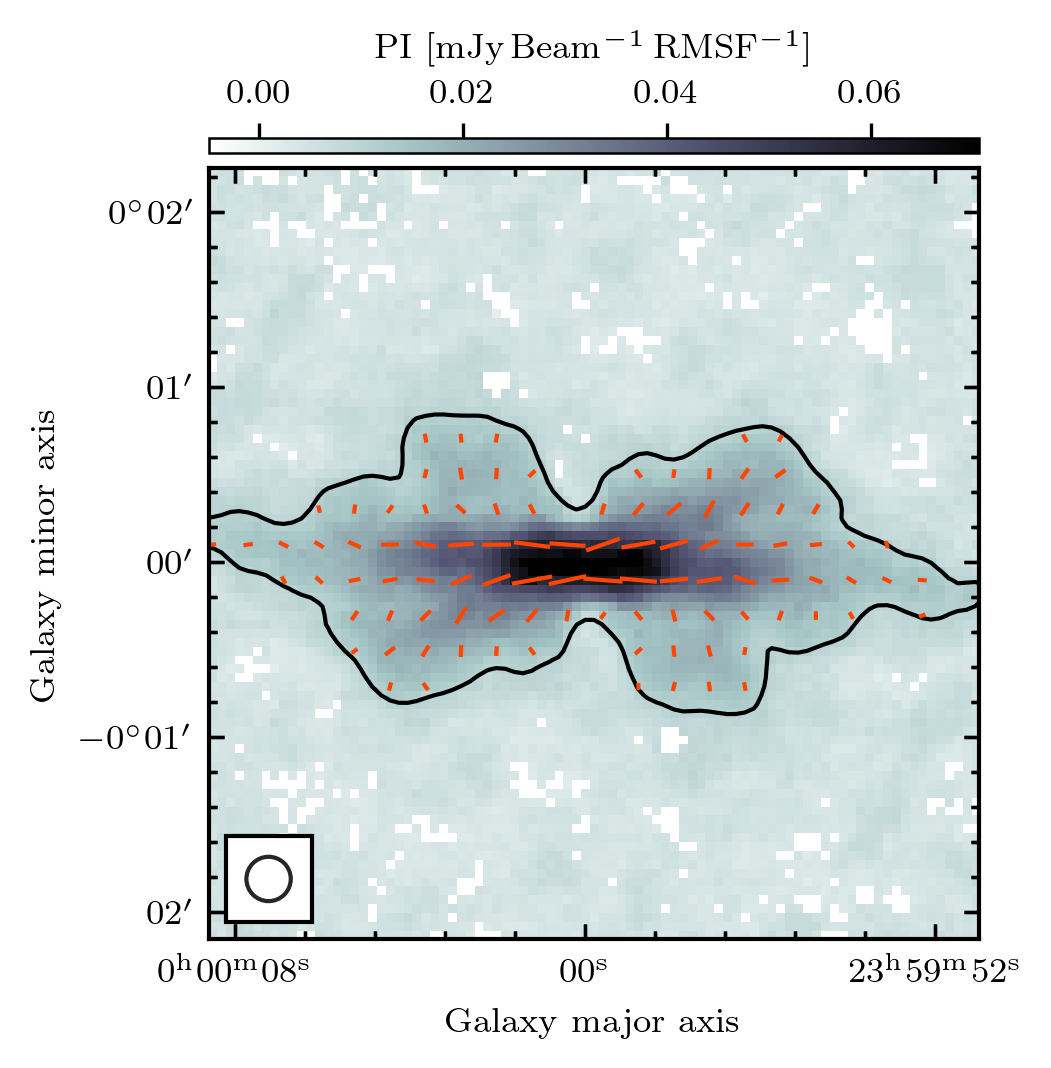}
    \end{subfigure}
    \\
    \begin{subfigure}[b]{0.28\linewidth}
        \centering
        \includegraphics[width=\linewidth]{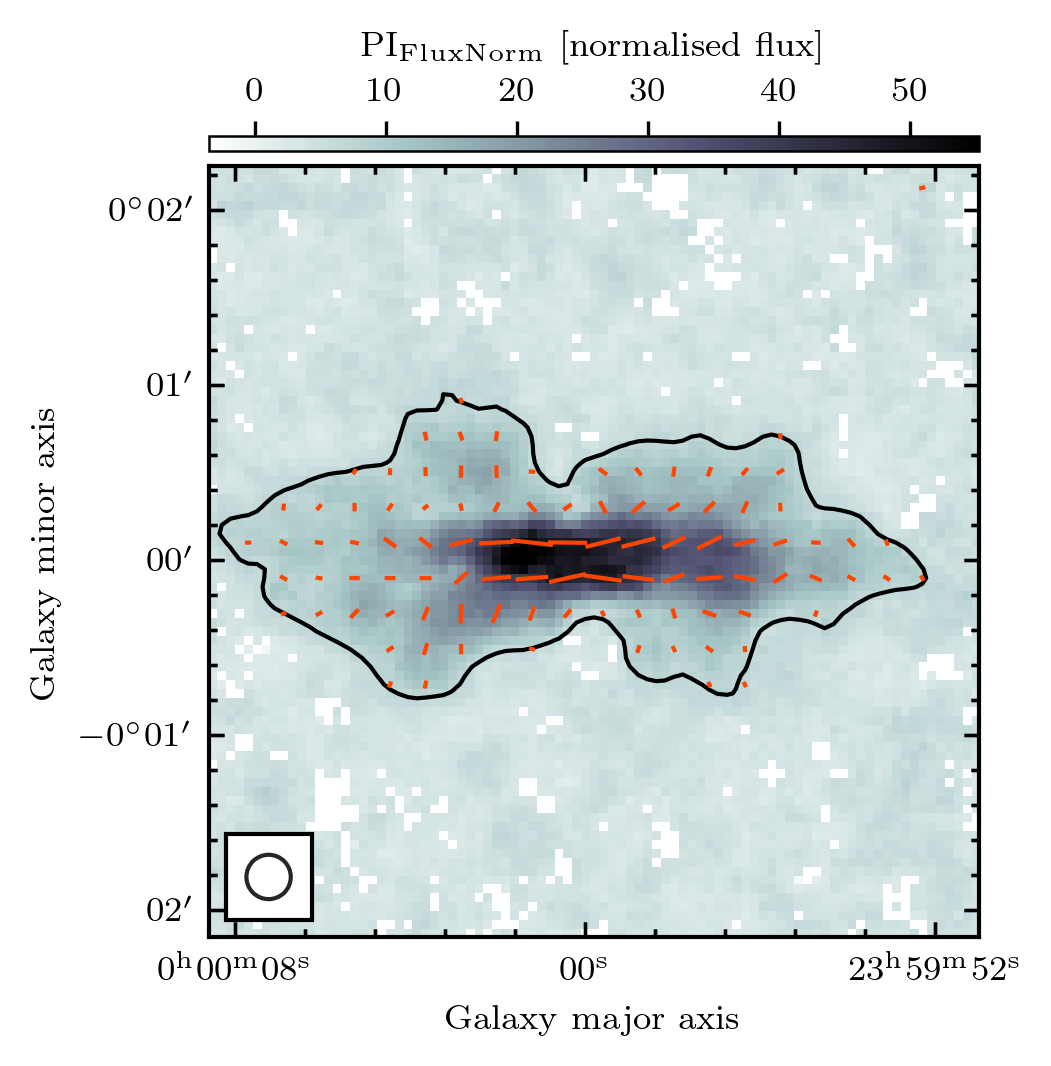}
    \end{subfigure}
    \begin{subfigure}[b]{0.28\linewidth}
        \centering
        \includegraphics[width=\linewidth]{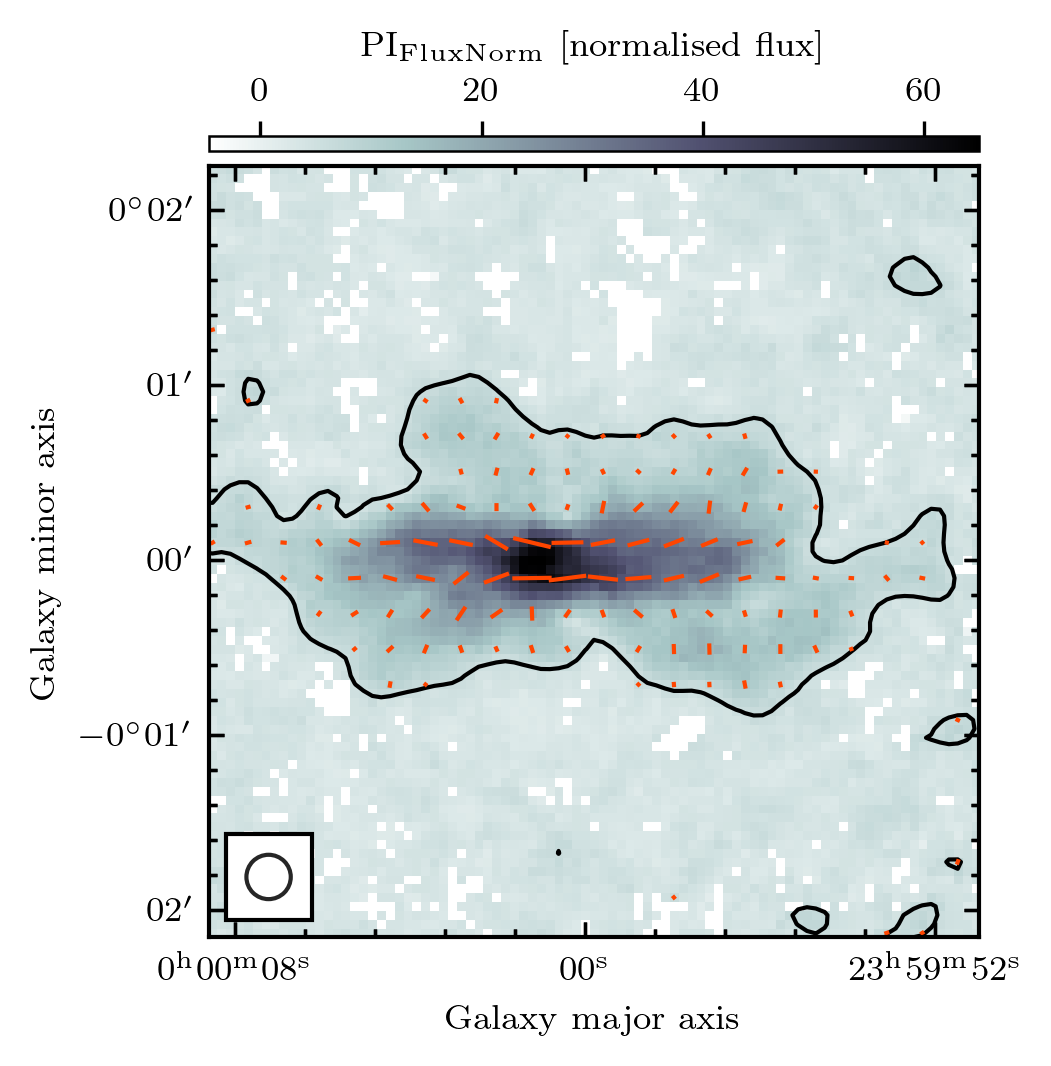}
    \end{subfigure}
    \begin{subfigure}[b]{0.28\linewidth}
        \centering
        \includegraphics[width=\linewidth]{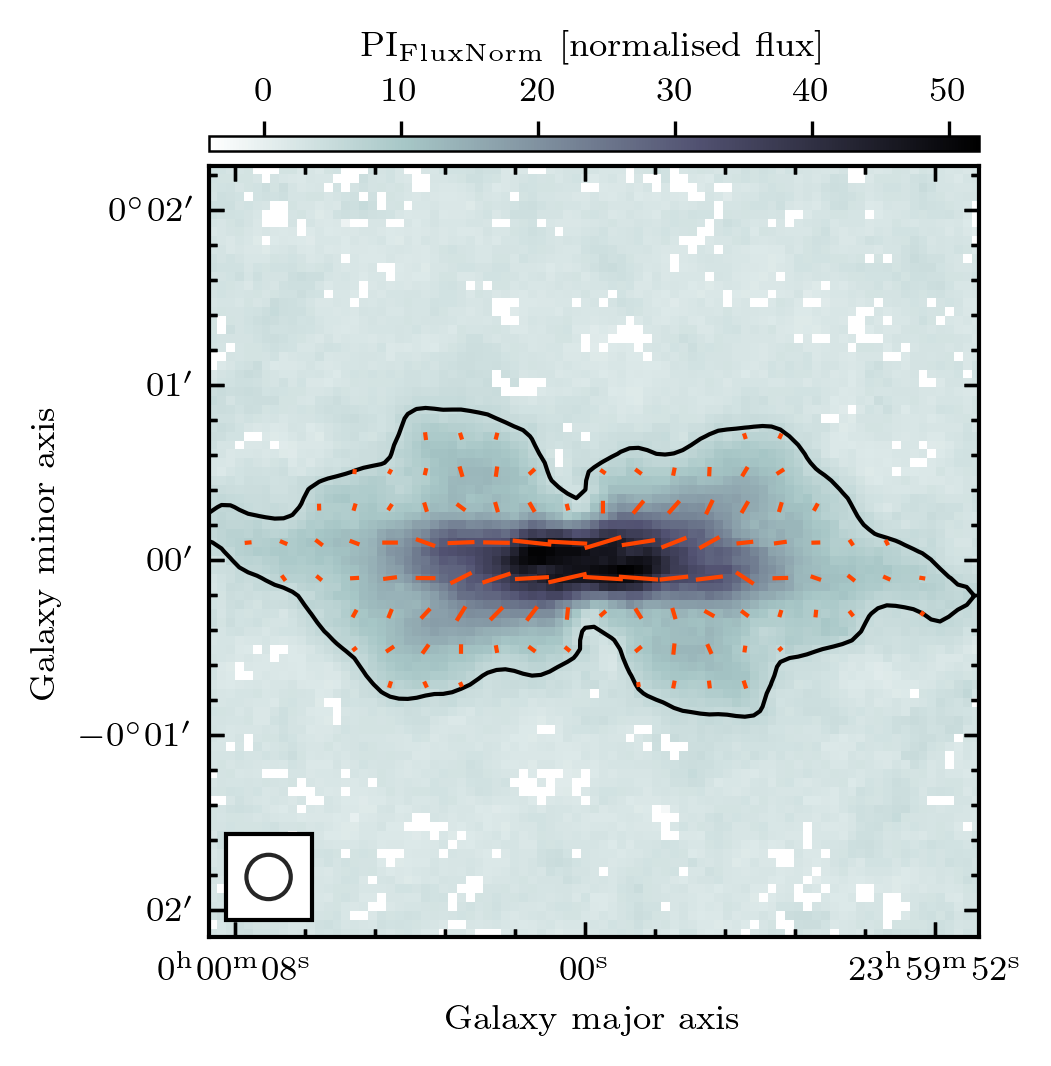}
    \end{subfigure}
    \begin{subfigure}[b]{0.28\linewidth}
        \centering
        \includegraphics[width=\linewidth]{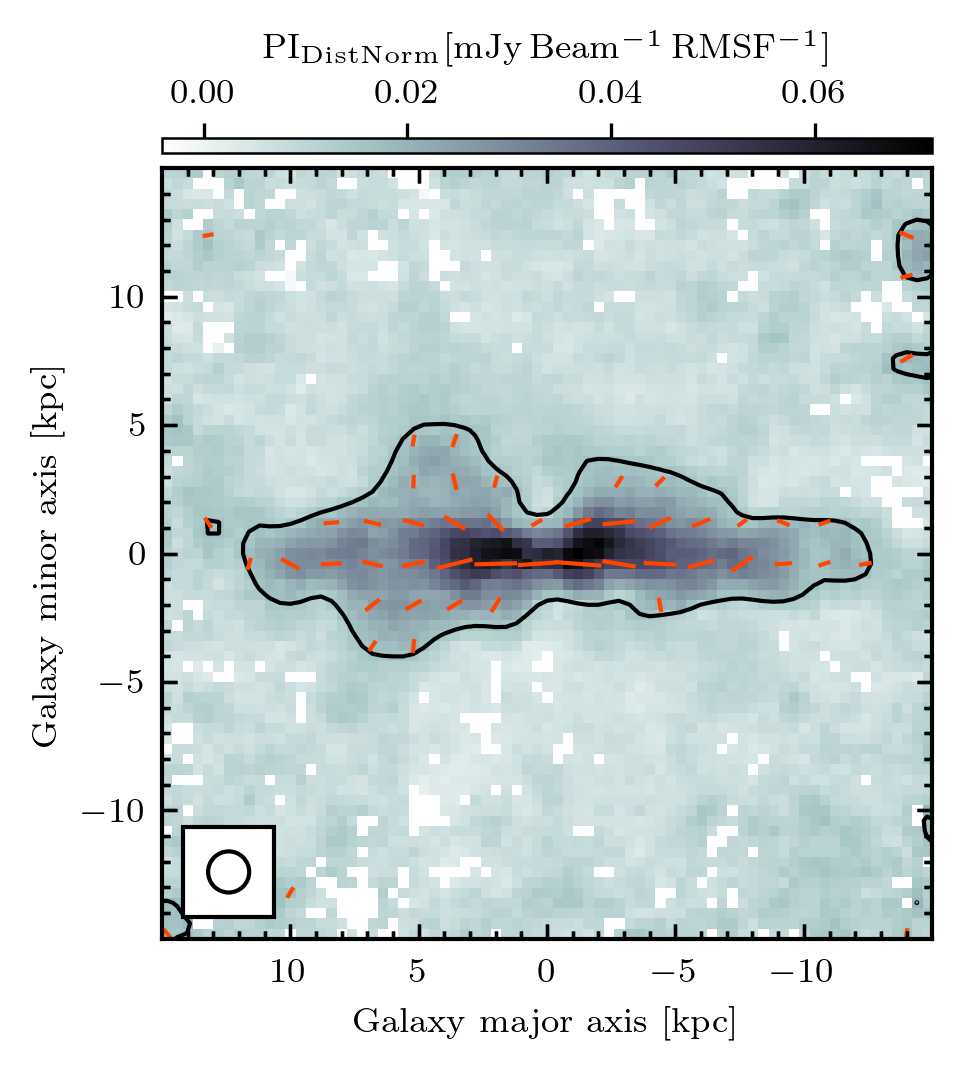}
    \end{subfigure}
    \begin{subfigure}[b]{0.28\linewidth}
        \centering
        \includegraphics[width=\linewidth]{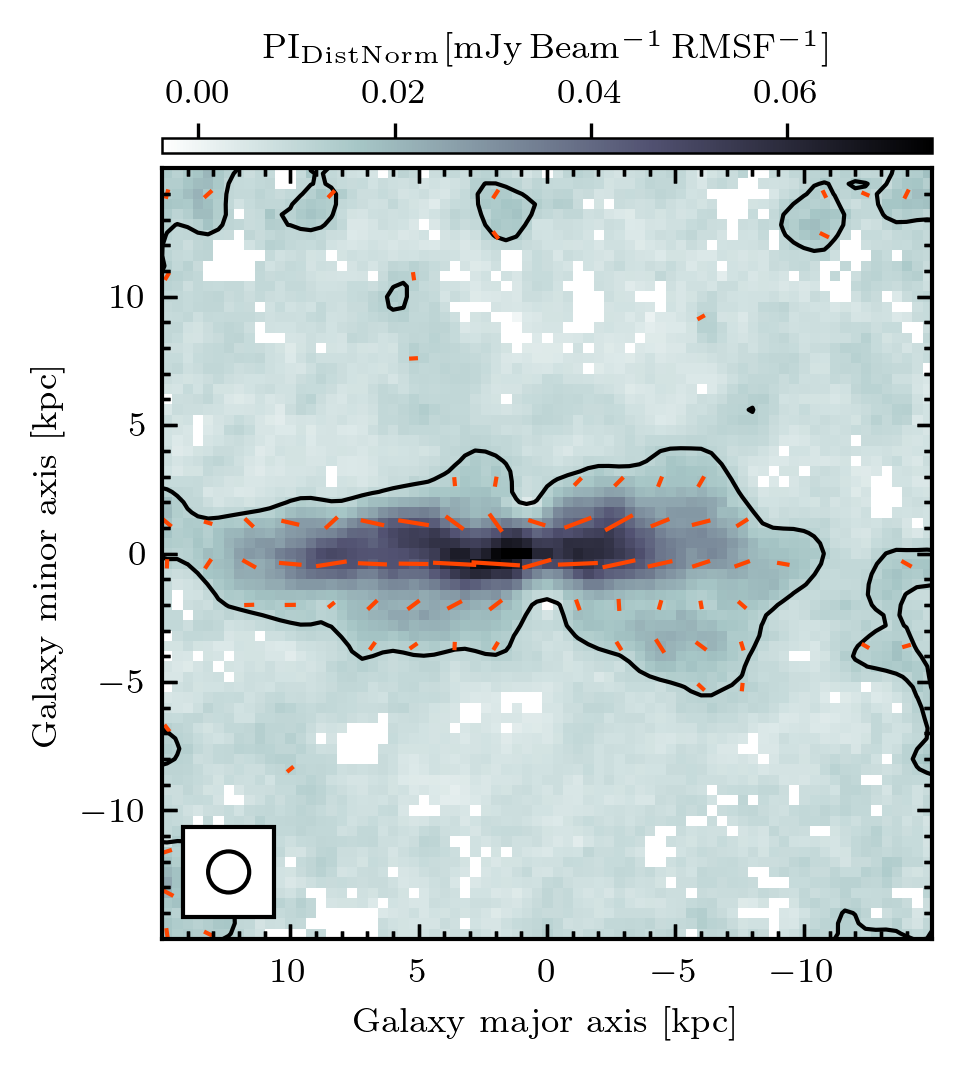}
    \end{subfigure}
    \begin{subfigure}[b]{0.28\linewidth}
        \centering
        \includegraphics[width=\linewidth]{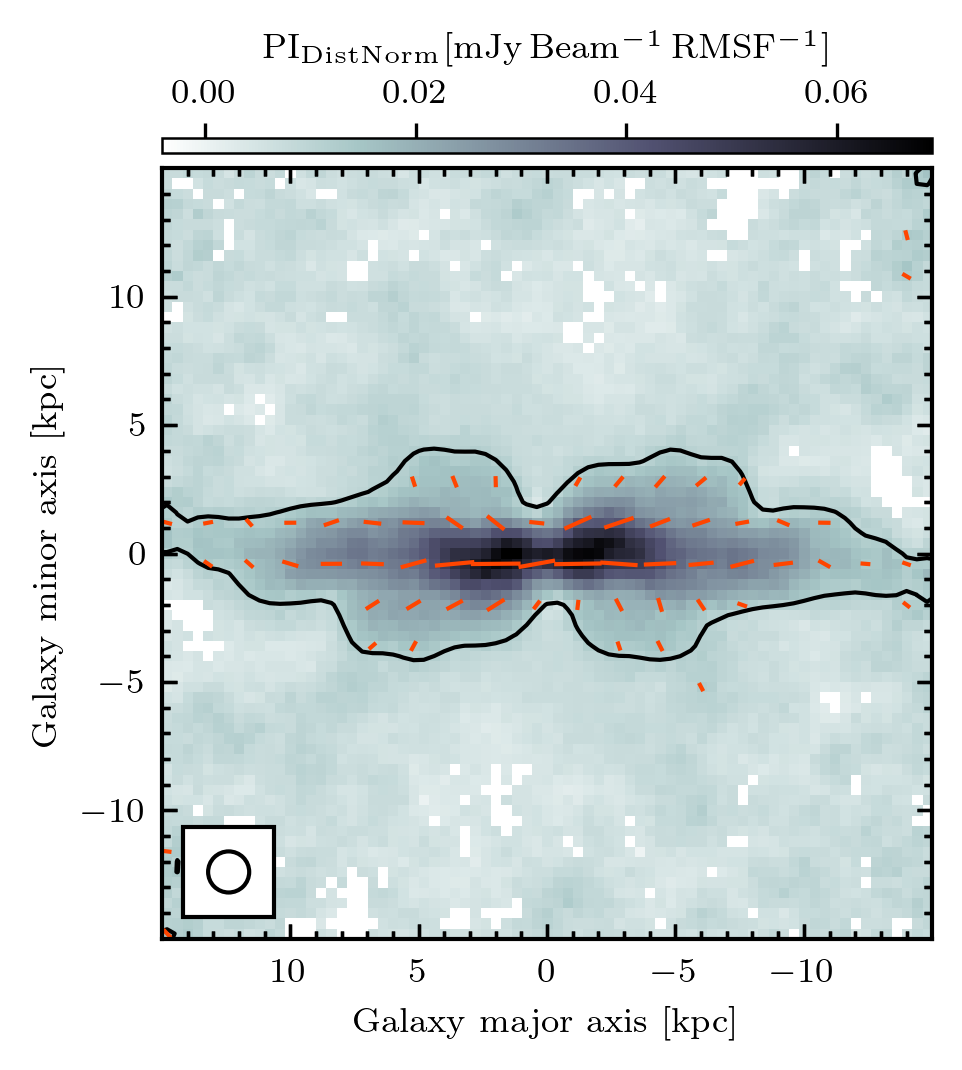}
    \end{subfigure}
    \\
    \begin{subfigure}[b]{0.28\linewidth}
        \centering
        \includegraphics[width=\linewidth]{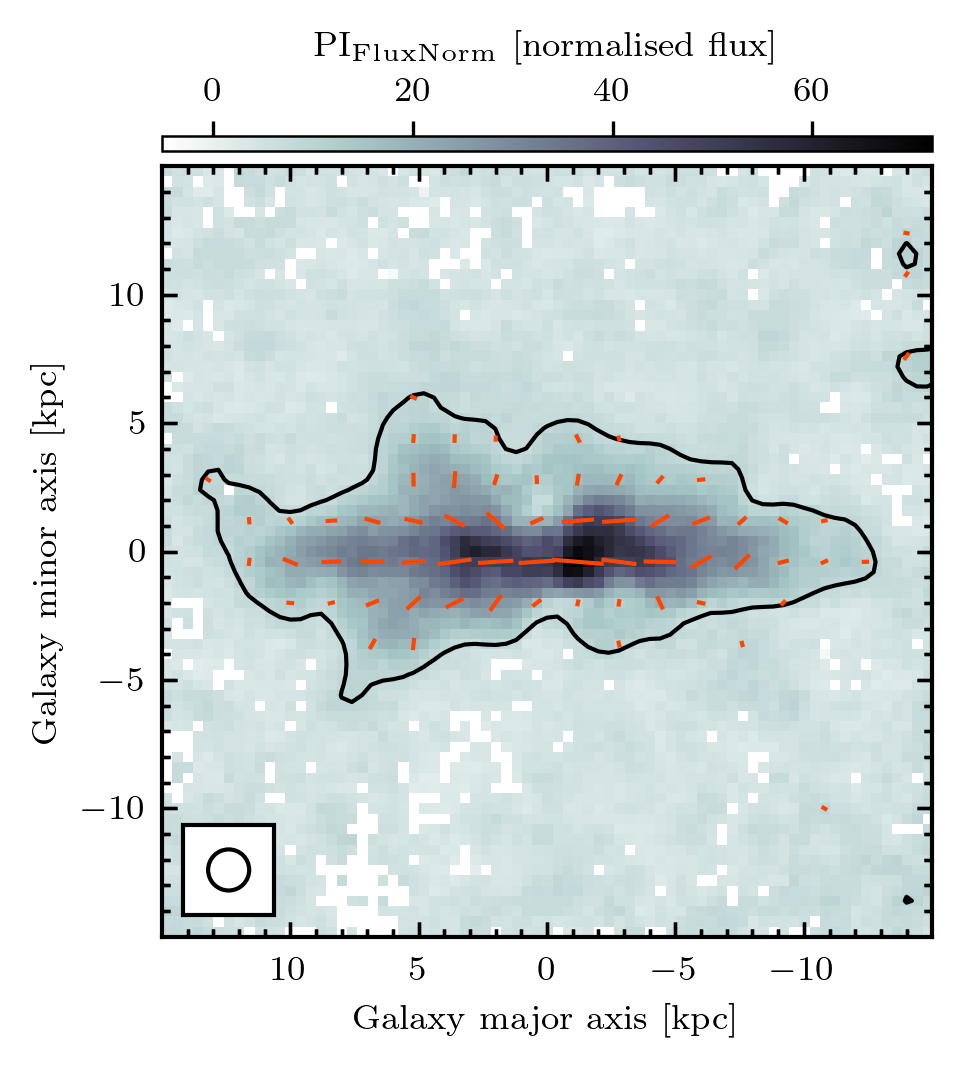}
    \end{subfigure}
    \begin{subfigure}[b]{0.28\linewidth}
        \centering
        \includegraphics[width=\linewidth]{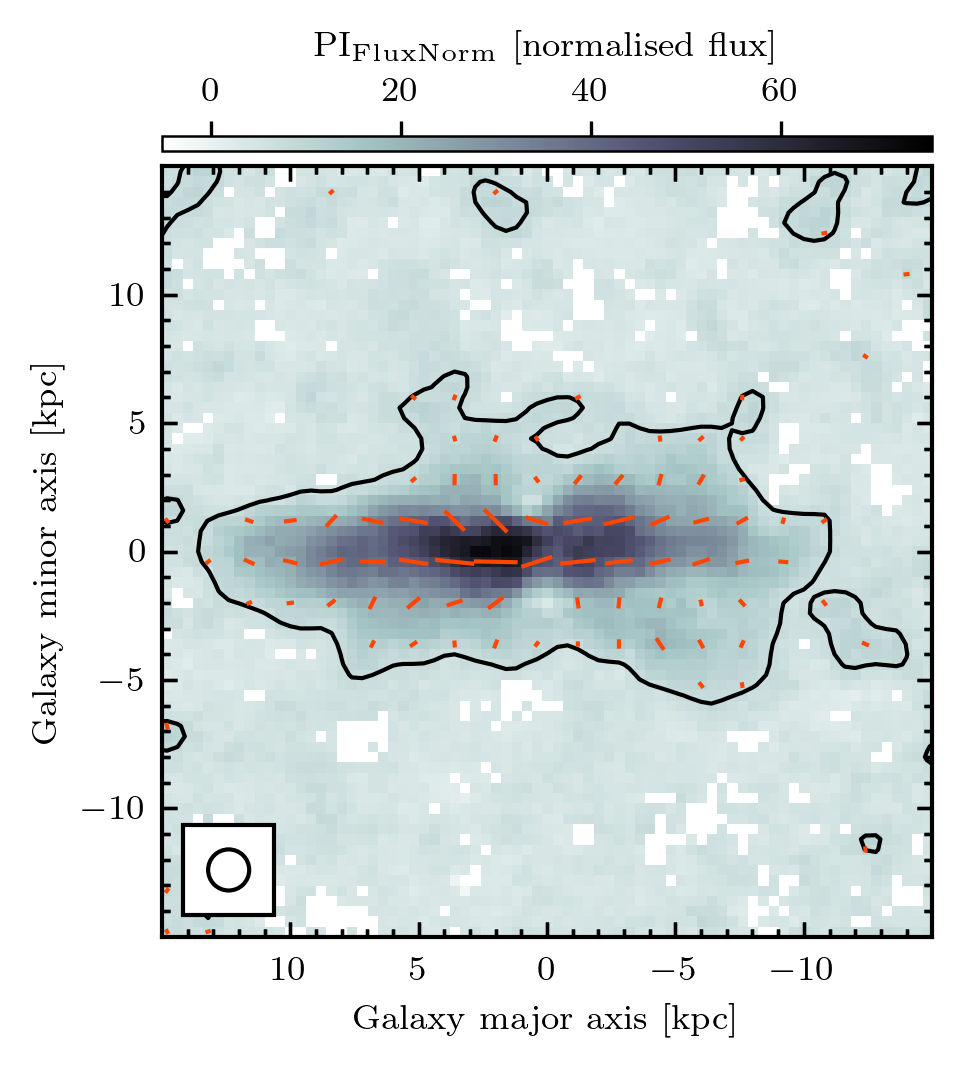}
    \end{subfigure}
    \begin{subfigure}[b]{0.28\linewidth}
        \centering
        \includegraphics[width=\linewidth]{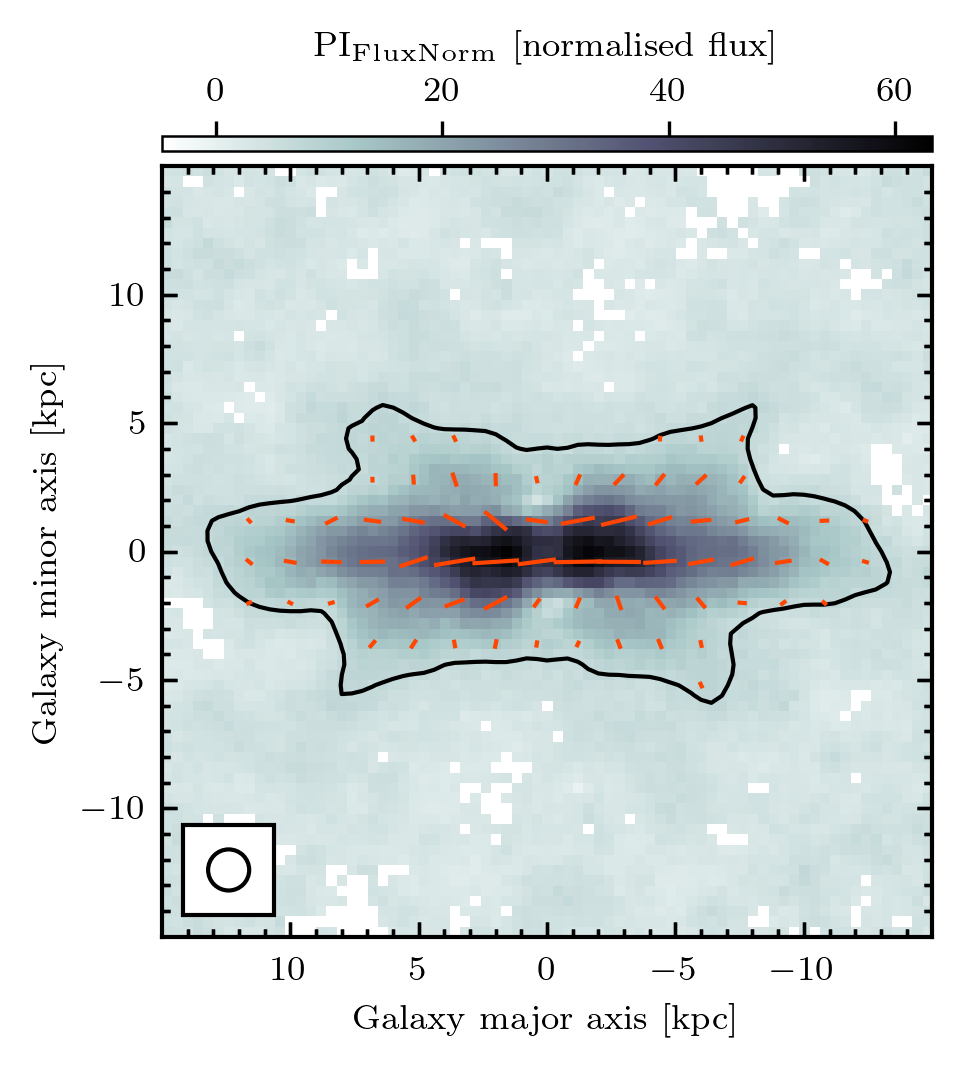}
    \end{subfigure}

    \caption{Maps of PI with overlaid pol. ang. information (orientation of the B field)  of the angular scaled (first and second row) and physical scaled (third and fourth row) median stacks for the three alignment strategies: standard (left), rotation (middle), and double (right). For angular size scaling, we show results from stacking without flux normalisation (first row) and PI flux normalised stacking results (second row). For physical size scaling, we show results from stacking with normalising the flux by the galaxy distance (third row) and the PI flux normalised stacking results (fourth row). The vector length corresponds to the PI. The black contour indicates the $3\sigma$-level of the PI emission. The shape of the synthesised beam is indicated in the bottom left corner of each panel.}
    \label{fig:pi_pa_median}
\end{figure*}

 We typically find a lower background noise level, a larger halo extent, as well as a larger $z$ extent difference between the central and outer regions, and a generally smoother halo morphology in the resulting data products when computing the median for stacking the $Q$ and $U$ cubes. Additionally, median stacks are less affected by such individual structures. Therefore, we only present the median stacked images using angular and physical scaling in Fig. \ref{fig:pi_pa_median}.
 
 Inspecting the panels of Fig. \ref{fig:pi_pa_median}, we find that the X shape and the $z$-extent difference appear in all panels. Therefore, those method-independent characteristics seem to be general features in galaxy haloes. Using each galaxy twice in the double alignment strategy (right column Fig. \ref{fig:pi_pa_median}) significantly increases the extent of the polarised emission and results in a smoother halo morphology but introduces a point mirror symmetry with regard to the galactic centre. Comparing the results from the standard and the rotation alignment strategy (left and middle columns Fig. \ref{fig:pi_pa_median}) overall the morphology of the polarised emission is similar but some details change (e.g. in the case of physical size scaling (third and forth row in Fig. \ref{fig:pi_pa_median})), the extended polarised emission channel in the second quadrant of the standard aligned stacks (left column) seems to be moved into the fourth quadrant in the rotation aligned stack (middle column). Therefore, we note that some details in the observed structure might be influenced by individual galaxies.

\begin{figure*}
\sidecaption
\centering
  \includegraphics[width=12.8cm]{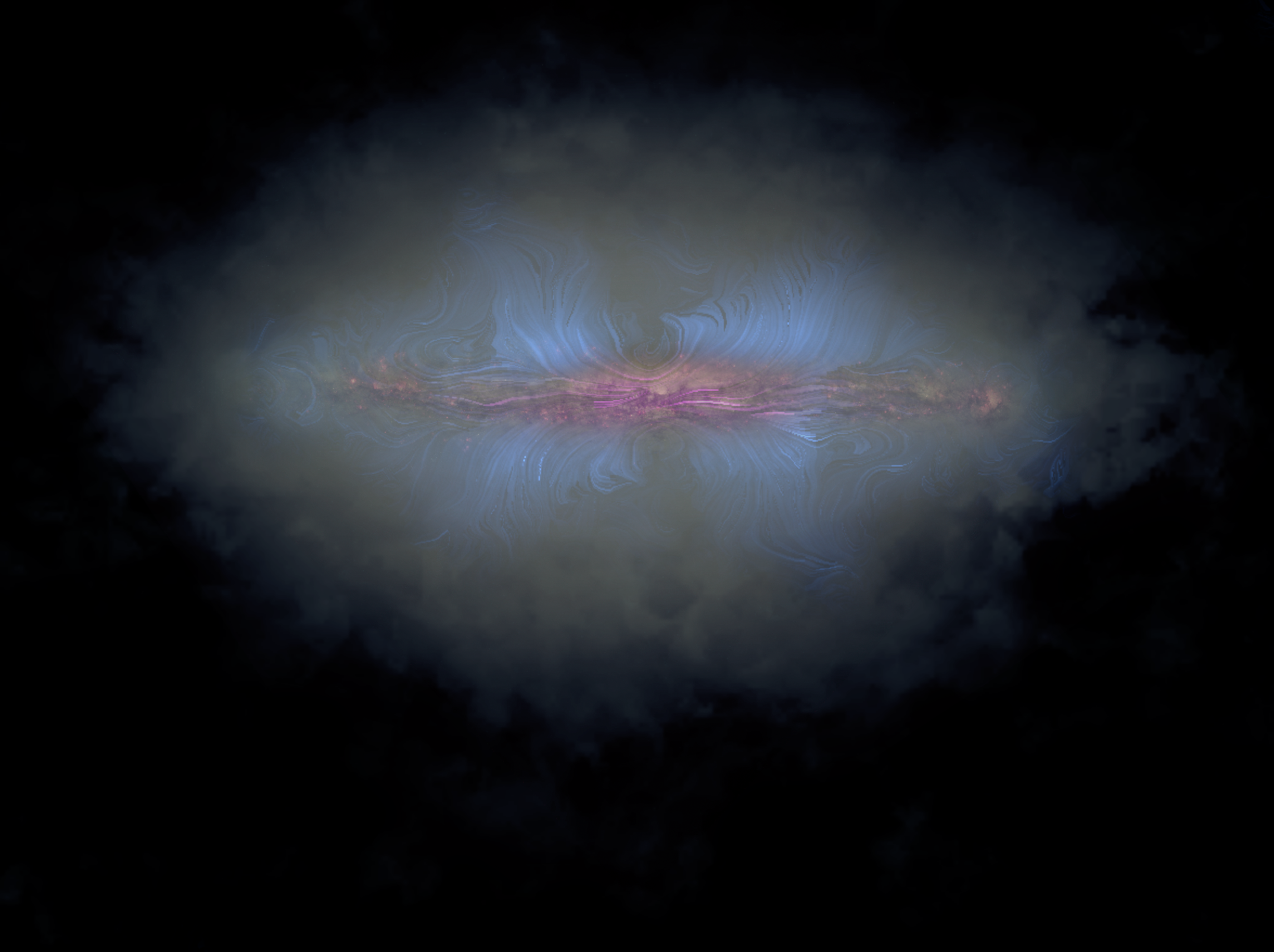}
    \caption{LIC image of the polarised emission of the dbl aligned angular scaled and PI normalised median stack of high resolution galaxies. The stack consists of 15 galaxies that are covered by at least 19 beams along the galactic disc. Polarised emission close to the disc is coloured in magenta while the halo emission is displayed in blue. The visual symmetry arises from the dbl alignment strategy, where each galaxy enters the stack twice (once rotated by 180\,deg). The underlying composite image \citep{2019Galax...7...42I} consists of the median stacked total synchrotron intensity of 30 CHANG-ES galaxies \citep{2015AJ....150...81W}, displayed in grey, and the disc of NGC 5775 constructed from Hubble Space Telescope (HST) Advanced Camera for Surveys data in filters F625W (assigned green) and F658N (assigned red). The HST image is employed to show the extent of an optical disc relative to the halo.}
    \label{fig:lic_image}
\end{figure*}

In Fig. \ref{fig:lic_image} we present a composite image of total radio emission, polarised radio emission and optical light. Here, we use the line integral convolution technique \citep[LIC,][]{cabral1993imaging}\footnote{The LIC code was adapted by Y. Stein  (Ruhr University Bochum), J. English (U. Manitoba) and A. Miskolczi (Ruhr University Bochum) from the \texttt{scipy} Line Integral Convolution \href{https://scipy-cookbook.readthedocs.io/items/LineIntegralConvolution.html}{code}.} to visualise the polarised radio emission. For the LIC imaging process, high resolution data are necessary. Therefore, we reran the data processing by selecting only galaxies that are covered by at least 19 beams and sampled each beam with $7\times7$ pixels, using angular scaling and dbl alignment. The overlay shows that we detect polarised radio emission across the whole galactic disc as well as in the galactic halo. The visual symmetry is caused by the dbl alignment strategy.

In Sect.~\ref{sec:res_pi_asym} and Sect.~\ref{sec:res_rm_struc} we discuss potential symmetries or asymmetries that can be traced in our stacking results. Therefore we exclude the dbl-alignment approach from these sections, as this approach can artificially increase or decrease symmetries as well as asymmetries in the data.

\subsection{PI asymmetry in rotation aligned galaxies}
\label{sec:res_pi_asym}
\citetalias{2020A&A...639A.112K} compare the polarised emission with the rotation sense of each galaxy and find that most galaxies show stronger polarised emission on the approaching than the receding side and thereby confirm results found by \citet{2010A&A...514A..42B} at lower frequencies (1300-1760\,MHz). Here, we repeated this experiment, but instead of analysing individual galaxies, we compared the measured PI fluxes in boxes that cover the polarised emission in the left (approaching) and the right (receding) side of the images for the std and rot alignment strategies. We used the same $q$ factor as introduced by \citetalias{2020A&A...639A.112K} where $q$ is defined as 
\begin{equation}
q=(A-R)/(A+R).
\label{eq:q_fac}
\end{equation}
$A$ and $R$ refer to the polarised flux measured on the approaching (left) and receding (right) sides of the galaxy stack. We present the derived $q$ factors for the median stacks, using angular and physical scaling in Table \ref{tab:PI_q_factors}. Additionally, we examine the effects of stacking with and without flux normalisation. As expected, we typically do not find a significant polarised flux difference between the left and right sides in the standard alignment measurements\footnote{As listed in Table \ref{tab:stacking_parameters}, 15 out of 27 galaxies require an extra rotation of $180^\circ$ so that their approaching side is on the left side.}. Here, only one realisation (physical scaling and PI normalisation) barely reached the $3\sigma$ level. In contrast, all measurements on the rotation-aligned datasets show significantly higher polarised fluxes on the approaching side, indicating that the asymmetry is connected to the rotation sense of the galaxies.

\begin{table}
    \centering
    \caption{Calculated $q$ factors of polarised emission for different alignment and flux normalisation strategies.}
    \label{tab:PI_q_factors}
    \begin{tabular}{lllr}
    \hline \hline
    Scaling & Alignment & Normalisation &  $q$\\
    \hline
ang & rot & FluxNorm &  $\mathbf{0.055 \pm 0.009}$ \\
ang & rot & noNorm      &  $\mathbf{0.041 \pm 0.008}$ \\
ang & std & FluxNorm &  $0.021 \pm 0.010$ \\
ang & std & noNorm      &  $0.018 \pm 0.011$ \\ 
phy & rot & piNorm &  $\mathbf{0.081 \pm 0.009}$ \\
phy & rot & distNorm &  $\mathbf{0.052 \pm 0.013}$ \\
phy & std & piNorm &  $\mathbf{0.037 \pm 0.012}$ \\
phy & std & distNorm &  $0.046 \pm 0.017$ \\
    \hline
    \end{tabular} 
\tablefoot{We list $q$ factors (Eq.\ref{eq:q_fac}) (comparing the left (approaching) and right (receding) side) for the standard and rotation alignment strategies in the median stacked datasets. If $q$ factors are significantly (3$\sigma$) different from zero, they are highlighted in bold.}
\end{table}
\subsection{Emission minimum along the minor axis}
\label{sec:res_minor_axis}
\begin{figure}
    \centering
    \includegraphics[width=0.75\linewidth]{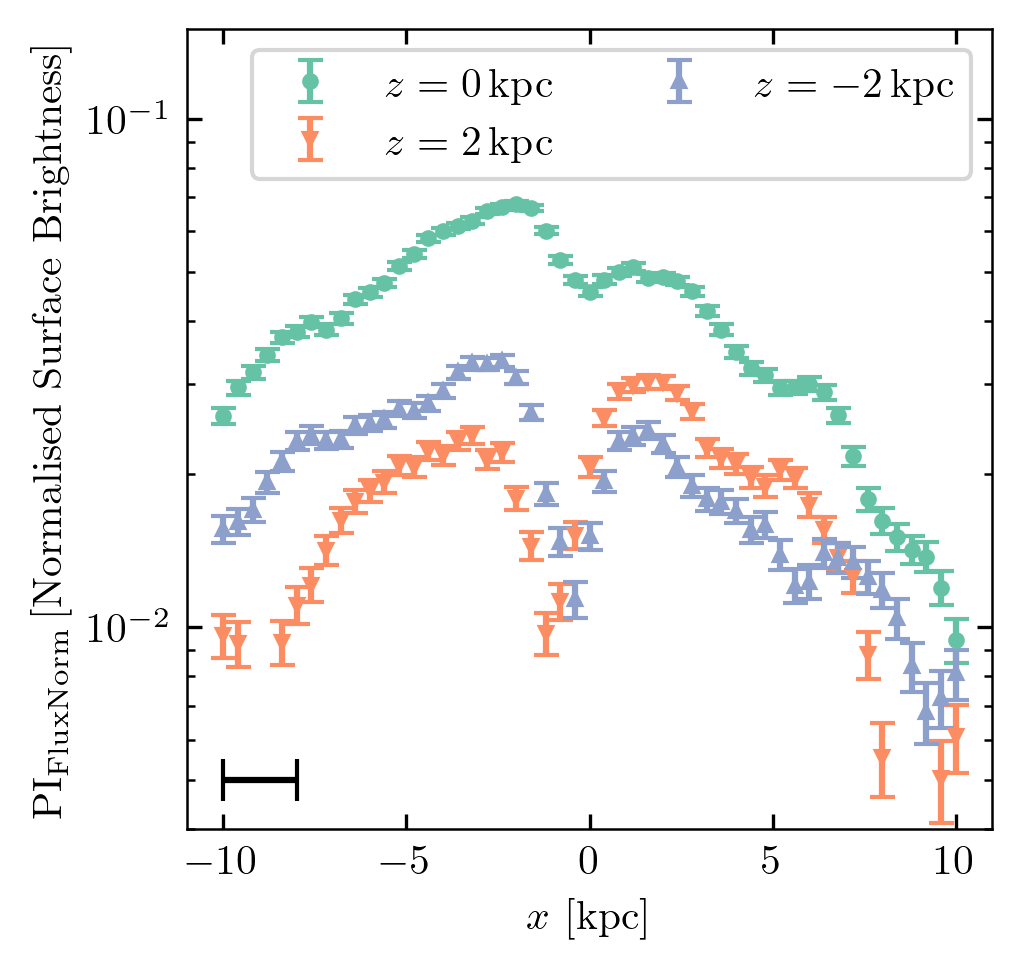}
    \caption{PI surface brightness profiles parallel to the galactic disc extracted from the stack using physical scaling, rotation alignment, and  PI flux normalisation (middle panel in the fourth row of Fig. \ref{fig:pi_pa_median}). Three profiles with vertical offsets 0\,kpc, 2\,kpc, and -2\,kpc and a width of 2\,kpc (five pixels) are shown using green, orange, and blue symbols. The beam size (2\,kpc) is indicated in the bottom left corner of the panel.}
    \label{fig:major_axis_profile}
\end{figure}

In addition to the PI asymmetry presented in the previous section, most PI maps displayed in Fig.~\ref{fig:pi_pa_median} show reduced emission along the minor axis of the galaxy. To further quantify this `minor-axis PI-deficit', Fig.~\ref{fig:major_axis_profile} shows surface brightness profiles that were measured on the PI map of the physically scaled, rotation-aligned stack using PI flux normalisation. In total, three profiles are shown, where the first profile is centred on the galactic disc ($z=0$), and two additional profiles are offset by one beam (2\,kpc) to the top and bottom. Two effects are visible. Firstly, all three profiles show a local minimum close to the minor axis ($x=0$). Here, the flux is more strongly reduced for the halo profiles ($\sim\!60\%$) than the profile that is centred on the disc ($\sim\!30\%$)\footnote{This effect is not only visible for this specific stacking approach result, but we only show results for one routine to keep the paper concise.}.

Secondly, Fig.~\ref{fig:major_axis_profile} shows that the PI asymmetry presented in the previous section mainly results from an asymmetry in the galactic disc. For the disc profile (orange data points), the mean surface brightness on the approaching side ($x\in[-5,0]$\,kpc) is $\sim40\%$  stronger than on the receding side ($x\in[0,5]$\,kpc). In contrast to that, the profiles with an offset to the disc show mixed results. Here, the profile at $z=-2$\,kpc follows the trend of the central profile but the profile at $z=2$\,kpc shows an inverse trend, with a slightly higher mean surface brightness on the receding side.

\subsection{Large-scale RM structures}
\label{sec:res_rm_struc}
 In this section, we analyse the RM maps produced by our stacking technique and discuss various systematic effects that could influence our results. We present the results of two approaches. In Sect.~\ref{sec:res_rm_struc_cubes}, we present RM maps that result from performing RM synthesis on the stacked $Q$ and $U$ cubes. As an alternative, we present results of stacking RM values of individual galaxies (similar to \citetalias{2021A&A...649A..94M}) in Sect.~\ref{sec:res_rm_struc_rm_maps}. To keep the paper concise, we only present the results of the angular-scaled stack in Sect.~\ref{sec:res_rm_struc_rm_maps}.

\begin{figure}
    \centering
    \includegraphics[width=0.75\linewidth]{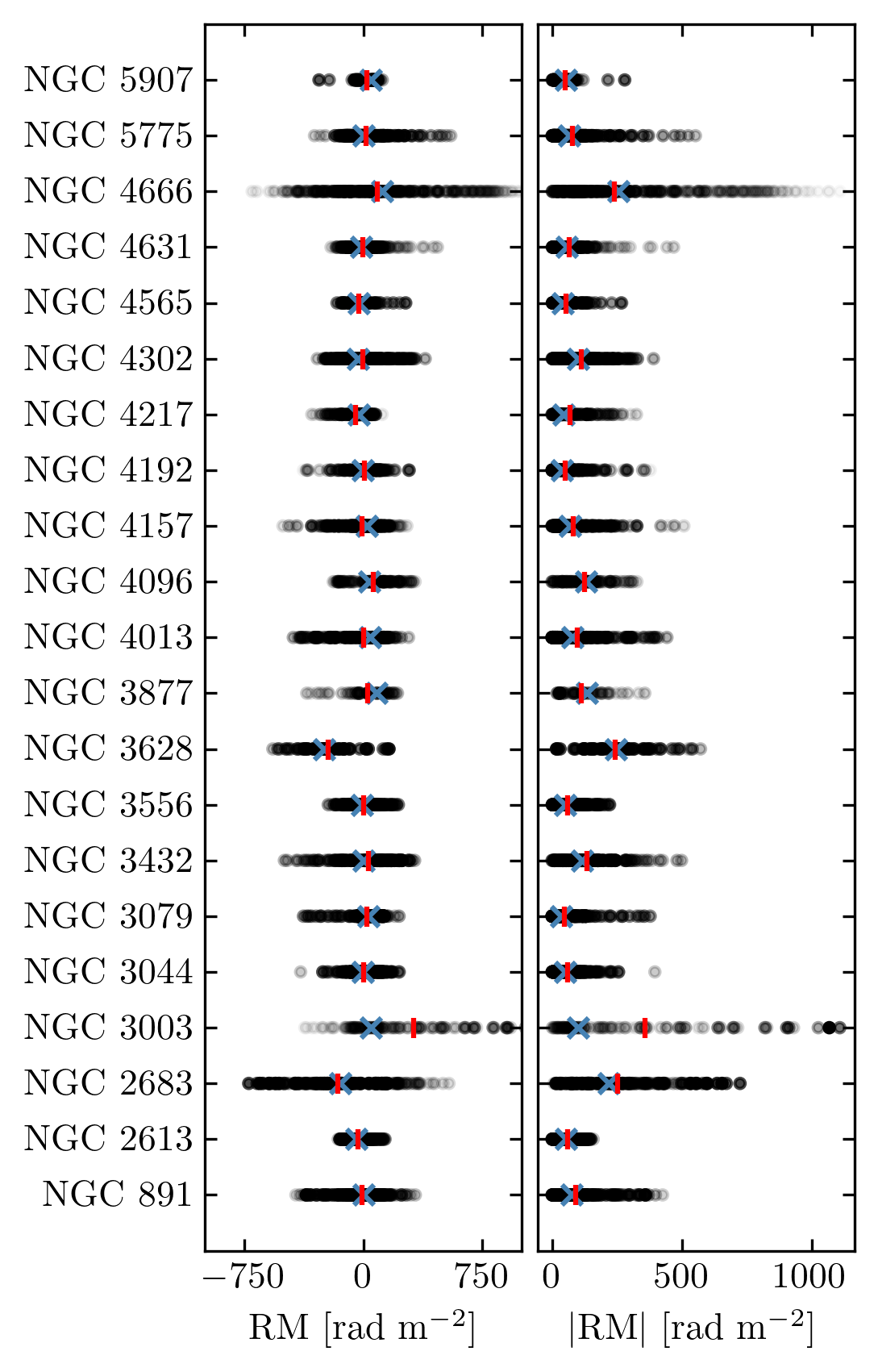}
    \caption{Distribution of RM and absolute RM values for individual galaxies that are included in the angular scaled stack. The black dots represent the RM or  |RM| distribution, weighted by $\delta_\mathrm{RM}$ per pixel (dark colours indicate a relatively low $\delta_\mathrm{RM}$). The blue x and the red vertical line indicate the median and weighted mean (weighted by $1/\delta_\mathrm{RM}^2$) of the distributions.}
    \label{fig:rm_dist_individual}
\end{figure}

To prepare the analysis of the stacked RM information, we compared the RM distributions for individual galaxies that are part of the angular scaling stack (running RM synthesis on the datasets just before performing the stacking; see Sect. \ref{sec:meth_smr}).
First, for each galaxy, we calculated the weighted mean (weighted by $(1/\delta_\mathrm{RM})^2$) and median of its MW foreground corrected RM and |RM| distributions (Fig. \ref{fig:rm_dist_individual}). In Table \ref{tab:rm_sample_stats}, we subsequently computed the mean and standard deviation of the distributions of these individual galaxy means and medians.

\begin{table}
    \centering
    \caption{Arithmetic mean ($\mu$) and standard deviation ($\sigma$) of RM values across the galaxies displayed in Fig. \ref{fig:rm_dist_individual}.}
    \label{tab:rm_sample_stats}
    \begin{tabular}{l|rrrr}
    \hline \hline
        & $\mathrm{RM_{mean}}$ & $\mathrm{RM_{med}}$ & $\mathrm{|RM|_{mean}}$ & $\mathrm{|RM|_{med}}$\\
        \hline
        $\mu$ [\radmsquare]     & 4 & -5 & 117 &96\\
        $\sigma$ [\radmsquare]  & 97& 75 & 82  &66\\
        \hline
    \end{tabular}
\end{table}

\subsubsection{RM structures derived from Q and $U$ cubes}
\label{sec:res_rm_struc_cubes}

\begin{figure}
    \centering
    \includegraphics[width=1\linewidth]{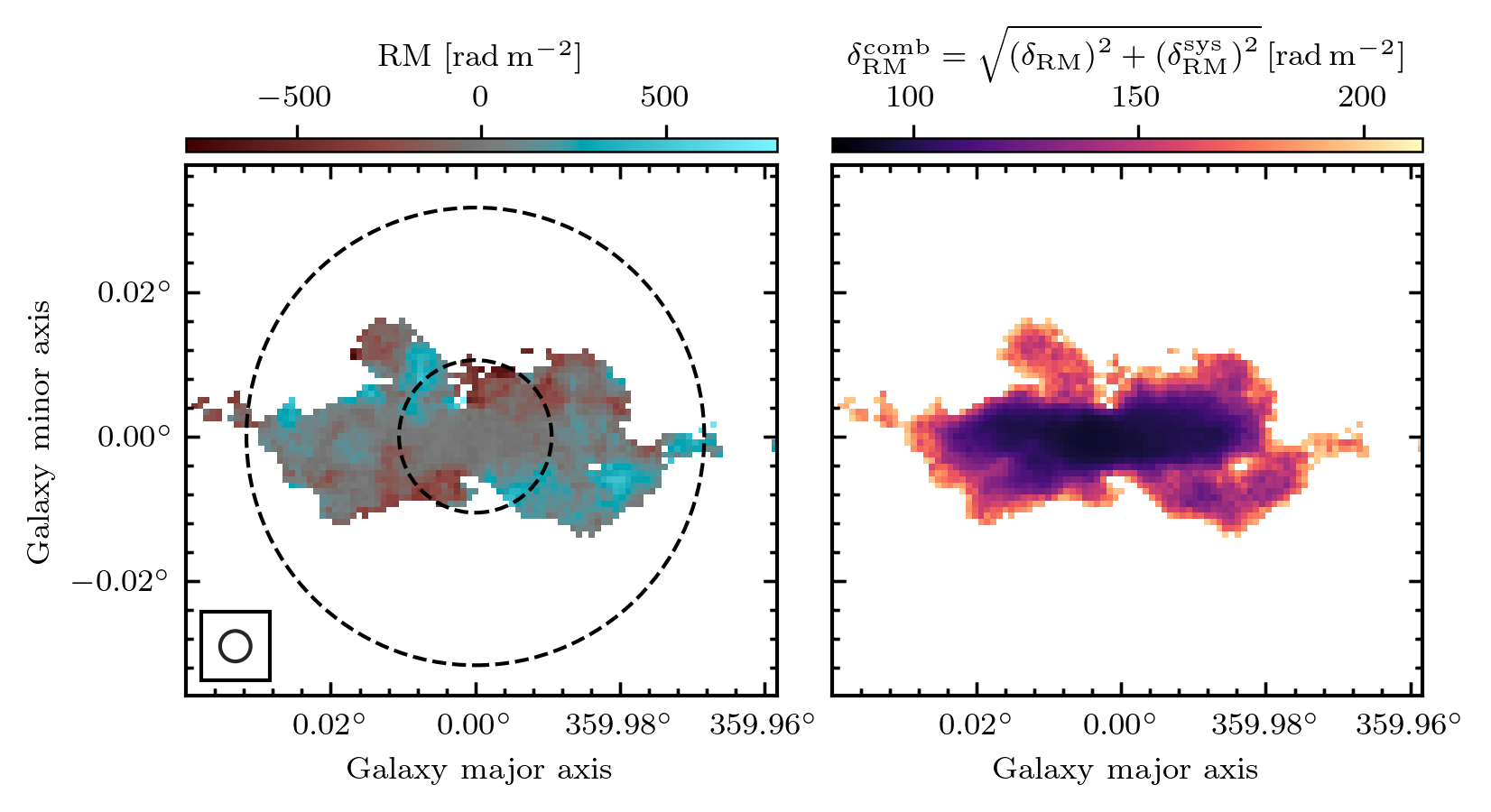}\\
    \includegraphics[width=1\linewidth]{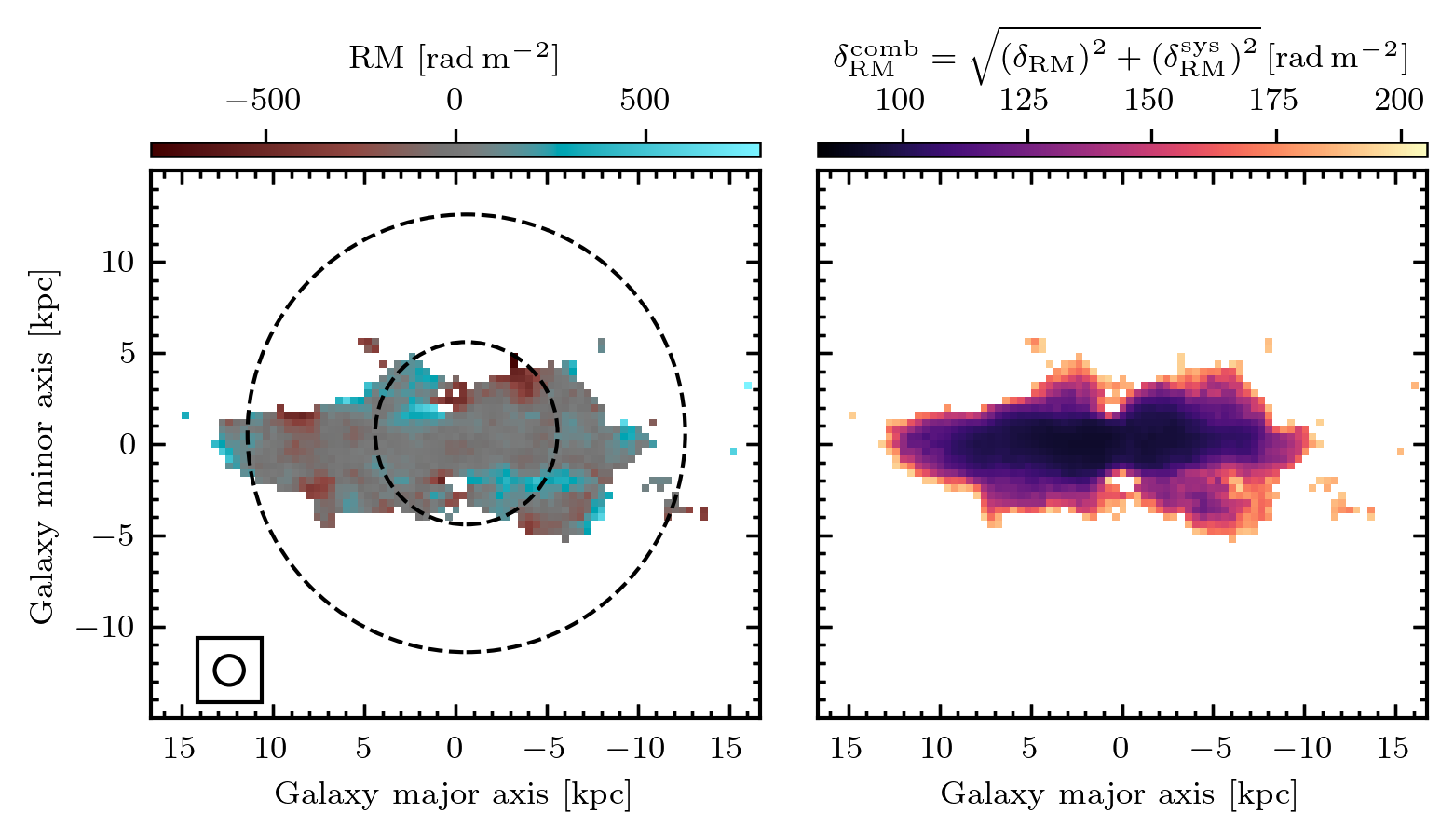}
    \caption{RM map (left column) and $\delta_\mathrm{RM}^\mathrm{comb}$ map (right column) for the PI flux normalised median stack of the rotation alignment strategy using angular scaling (top row) and physical scaling (bottom row). The beam size is indicated in the bottom left corner of the RM map. The dashed black circles indicate the distinction between the central and outer regions that were analysed separately.}
    \label{fig:RM_map_median_piNorm_rot_ang_phy}
\end{figure}

In Fig. \ref{fig:RM_map_median_piNorm_rot_ang_phy} we show the RM and $\delta_\mathrm{RM}^\mathrm{comb}$ maps of the flux normalised median stack, using the rotation alignment strategy for angular size and physical size scaling. In the $\delta_\mathrm{RM}^\mathrm{comb}$ maps we find mostly values of $\sim\!150$\,rad\,m\textsuperscript{-2}, highlighting the combined effect of the introduced stacking systematic and the broad RMSF. To search for a similar RM pattern as reported in \citetalias{2021A&A...649A..94M}, we introduced a sector comparison where we computed the mean RM values when combining all pixels of QI and QIII ($\mu_\mathrm{QI+QIII}$) as well as QII and QIV ($\mu_\mathrm{QII+QIV}$). We further computed the difference of these sector means:
\begin{equation}
    \Delta\mu = \mu_\mathrm{QII+QIV}-\mu_\mathrm{QI+QIII}.
\end{equation}
We estimated the uncertainty of the mean by accounting for the number of independent beams with RM detection ($N_\mathrm{Beam}$), the scatter in each sector ($\sigma_\mathrm{QI+QIII}$ and $\sigma_\mathrm{QII+QIV}$), and further included the systematic RM uncertainty $\delta_{\mathrm{RM}}^\mathrm{sys}$:
\begin{equation}
\delta_{\mu_s}=\frac{\sqrt{\left(\sigma_s\right)^2+\left(\delta_{\mathrm{RM}}^\mathrm{sys}\right)^2}}{\sqrt{N_\mathrm{Beam}}},\ \mathrm{for}\ s \in [\mathrm{QI+QIII,QII+QIV}]. 
\end{equation}

In Table \ref{tab:rm_stat} we present these summarising statistics for multiple alignment, size scaling, and flux normalisation strategies. We checked for differences when including all pixels or splitting central and outer regions, but do not find these analyses to differ significantly. Therefore, the sector statistics we report in Table \ref{tab:rm_stat} are based on combining all pixels from the inner and outer regions (indicated by the two circles in Fig. \ref{fig:RM_map_median_piNorm_rot_ang_phy}).

Similar to \citetalias{2021A&A...649A..94M}, we find predominantly negative RMs in QI and QIII and positive RMs in QII and QIV in case of angular size scaling. However, we do want to highlight that this RM `split' is on the order of the variation of the sectors and the Faraday depth step size that is used in the RM synthesis (see. Sect. \ref{sec:meth_rm}).

To check whether the observed structure predominately comes from individual sources, dominant in RM, we divided the sample of the angular scaling stack (based on the analysis in Fig.~\ref{fig:rm_dist_individual}) into two subsamples: high |RM| and low |RM| galaxies (see Table \ref{tab:stacking_parameters}) and repeated the analysis (see Table \ref{tab:rm_stat}). The detected split is strongest when considering only high |RM| sources, indicating that also the angular
scaling stack using the complete sample might be significantly influenced by individual high |RM| sources.

For the physically scaled stacking, the detected RM split vanishes completely. As we found no significant difference when comparing central and outer regions, the remaining difference between the physical and angular scaling approaches (when considering the quadrant statistics) is the number of galaxies in the stack (as mentioned above: $N_\mathrm{ang}=21$, $N_\mathrm{phy}=25$), which further highlights the possibility that the marginal RM split in the angular scaling stack arises from individual galaxies.

While the analysis in this section yielded inconclusive results, in the next section we perform an analysis that is more similar to the \citetalias{2021A&A...649A..94M} approach.

\subsubsection{RM structures derived from stacking RM maps}
\label{sec:res_rm_struc_rm_maps}

To derive a stacked RM map, we extracted the RM and $\Delta$RM values pixel-wise of each individual galaxy and combined them into a single dataset by computing the median in each pixel. Unlike the procedure in Sect. \ref{sec:res_rm_struc_cubes}, this analysis avoids combining the $Q$ and $U$ signals of individual galaxies. Instead, we performed an RM synthesis on each galaxy independently and subsequently stacked the resulting RM values.

As pointed out before, some galaxies show larger values in their $\mathrm{\overline{|RM|}}$ distribution and might therefore dominate when stacking RM values of multiple galaxies. However, to find a RM structure similar to \citetalias{2021A&A...649A..94M}, we are more interested in the sign of the RM value per quadrant than its magnitude.

Therefore, to mitigate the effect of RM dominant galaxies, we also present the stacking results when normalising the RM maps so that
\begin{equation}
    \mathrm{RM_{norm} = RM\,\frac{50\,rad\,m^{-2}}{\overline{|RM|}}},
\end{equation}
where $\mathrm{\overline{|RM|}}$ indicates the weighted mean (weighted by $1/\Delta\mathrm{RM}^2$) of the |RM| distribution per galaxy. We chose the normalisation value 50\,\radmsquare\ as the representative mean |RM| value for galaxies without an increased |RM| distribution (see Fig. \ref{fig:rm_dist_individual}). In the resulting stacks, we only considered pixels with at least  \rmminnpix\ contributing galaxies.

\begin{figure}
\centering
    \includegraphics[width=1\linewidth]{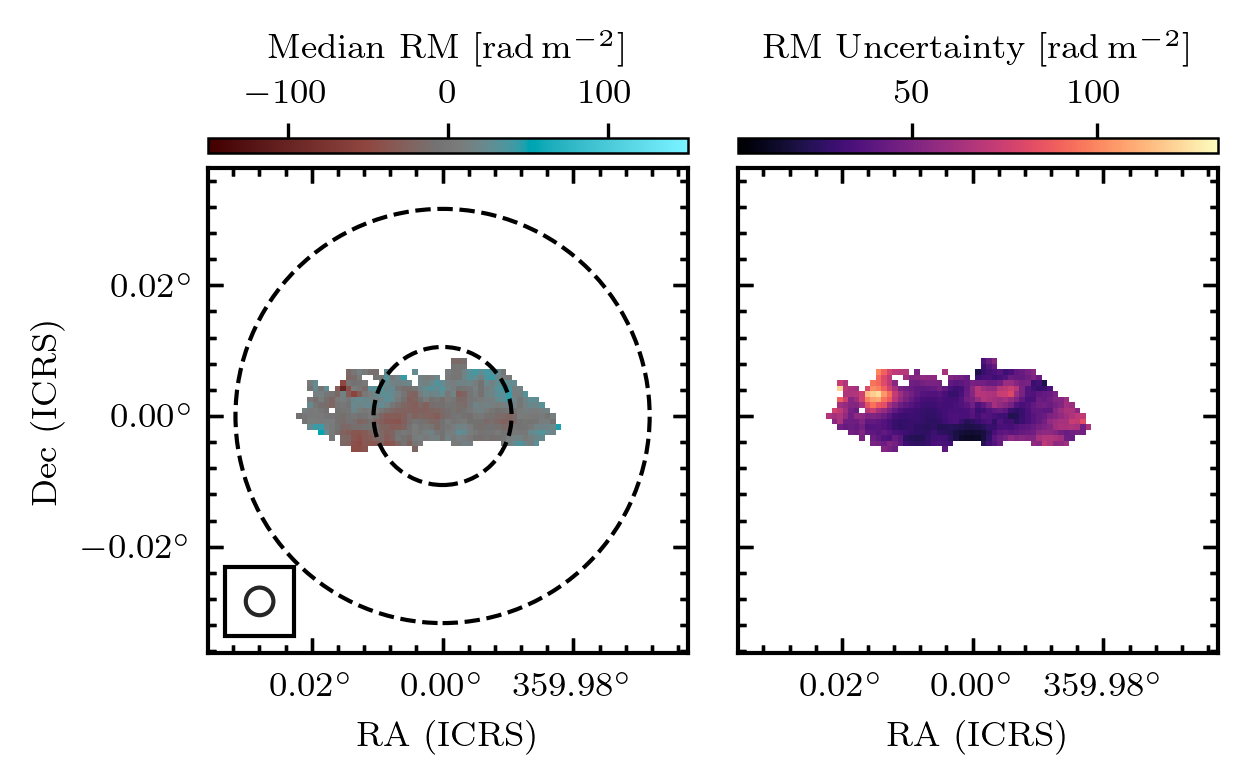}    
    \caption{Median stacked RM map (left panel) and RM uncertainty map (derived as $\left(1/\delta_\mathrm{RM}\right)^2$-weighted standard deviation per pixel, right panel) for the rot-aligned and angular scaled galaxies using RM normalisation. The beam size is indicated in the bottom left corner of the RM map. The dashed black circle indicates the distinction between the central and outer regions that were analysed separately.}
    \label{fig:stacked_rm_rot_RMnorm}
\end{figure}

As an example, we present the stacked RM map for the angular scaled stack using rot alignment and RM normalisation in Fig. \ref{fig:stacked_rm_rot_RMnorm}. Similar to the analysis in Sect. \ref{sec:res_rm_struc_cubes}, we present the summary of the sector comparison in Table \ref{tab:rm_stat}. A quadrupolar structure as found in \citetalias{2021A&A...649A..94M} is not observed.

\begin{table*}
\centering
    \caption{Sector statistics extracted from RM maps of different scaling techniques, flux normalisations, and alignment approaches.}
    \label{tab:rm_stat}
    \begin{tabular}{llllrrrrrrr}
    \hline
    \hline
    scaling & align. & norm. & subsample & $N_{\mathrm{pix}}$ & $\mu_\mathrm{QI+QIII}$ & $\mu_\mathrm{QII+QIV}$ & $\sigma_\mathrm{QI+QIII}$ & $\sigma_\mathrm{QII+QIV}$ & $\Delta\mu$ &$\mathrm{S/N}_{\Delta\mu}$   \\
     & & & & & [\radmsquare] & [\radmsquare] & [\radmsquare] & [\radmsquare] & [\radmsquare] \\
     \hline
\multicolumn{11}{c}{RM derived from stacked Q and $U$ cubes}\\
\hline
ang & rot & noNorm & all & 1243 & $-26\pm26$ & $51\pm27$ & 88 & 102 & $77\pm37$ & 2.1 \\
ang & rot & piNorm & all & 1426 & $-26\pm25$ & $53\pm27$ & 95 & 119 & $79\pm37$ & 2.1 \\
ang & rot & piNorm & highrm & 880 & $-101\pm34$ & $55\pm37$ & 114 & 120 & $156\pm50$ & 3.1 \\
ang & rot & piNorm & lowrm & 1103 & $-12\pm25$ & $27\pm23$ & 72 & 65 & $39\pm34$ & 1.1 \\
ang & std & noNorm & all & 1138 & $-44\pm29$ & $32\pm31$ & 106 & 114 & $76\pm42$ & 1.8 \\
ang & std & piNorm & all & 1203 & $-49\pm29$ & $30\pm31$ & 113 & 122 & $79\pm43$ & 1.8 \\
phy & rot & distNorm & all & 734 & $-1\pm28$ & $24\pm36$ & 64 & 101 & $25\pm46$ & 0.6 \\
phy & rot & piNorm & all & 953 & $-0\pm27$ & $37\pm34$ & 75 & 118 & $37\pm43$ & 0.9 \\
phy & rot & piNorm & ang sample & 1006 & $-10\pm28$ & $45\pm36$ & 84 & 135 & $55\pm45$ & 1.2 \\
phy & std & distNorm & all & 520 & $1\pm35$ & $10\pm45$ & 76 & 107 & $9\pm57$ & 0.2 \\
phy & std & piNorm & all & 799 & $6\pm31$ & $26\pm39$ & 90 & 123 & $20\pm50$ & 0.4 \\
    \hline
\multicolumn{11}{c}{RM derived from stacking individual RM maps}\\
\hline
ang & std & noNorm & all & 540 & $16 \pm 10$ & $14 \pm 10$ & $31$ & $33$ & $2 \pm 14$ &0.1\\
ang & std & RMNorm & all & 540 & $12 \pm 7$ & $5 \pm 6$ & $23$ & $20$ & $8 \pm 9$ & 0.9\\
ang & rot & noNorm & all & 537 & $15 \pm 6$ & $14 \pm 12$ & $21$ & $39$ & $1 \pm 14$ & 0.1\\
ang & rot & RMNorm & all & 537 & $16 \pm 5$ & $7 \pm 7$ & $17$ & $22$ & $9 \pm 8$ & 1.1\\

\hline
\end{tabular}
\tablefoot{Size scaling, alignment strategy, flux or RM normalisation approach, number of pixels (combined for all four quadrants), sector mean of RM values (including uncertainty of the mean) for the two regions (QI+QIII and QII+QIV), RM, standard deviation of the RM values per region, derived RM split between the two regions, and signal to noise ratio of the detected split. The uncertainty for $\Delta\mu$ is derived using Gaussian error propagation.}

\end{table*}

\section{Discussion}
\label{sec:dis}
\subsection{Measuring diffuse polarised emission}
\label{sec:dis_depol}
\begin{figure}
    \centering
    \includegraphics[width=0.75\linewidth]{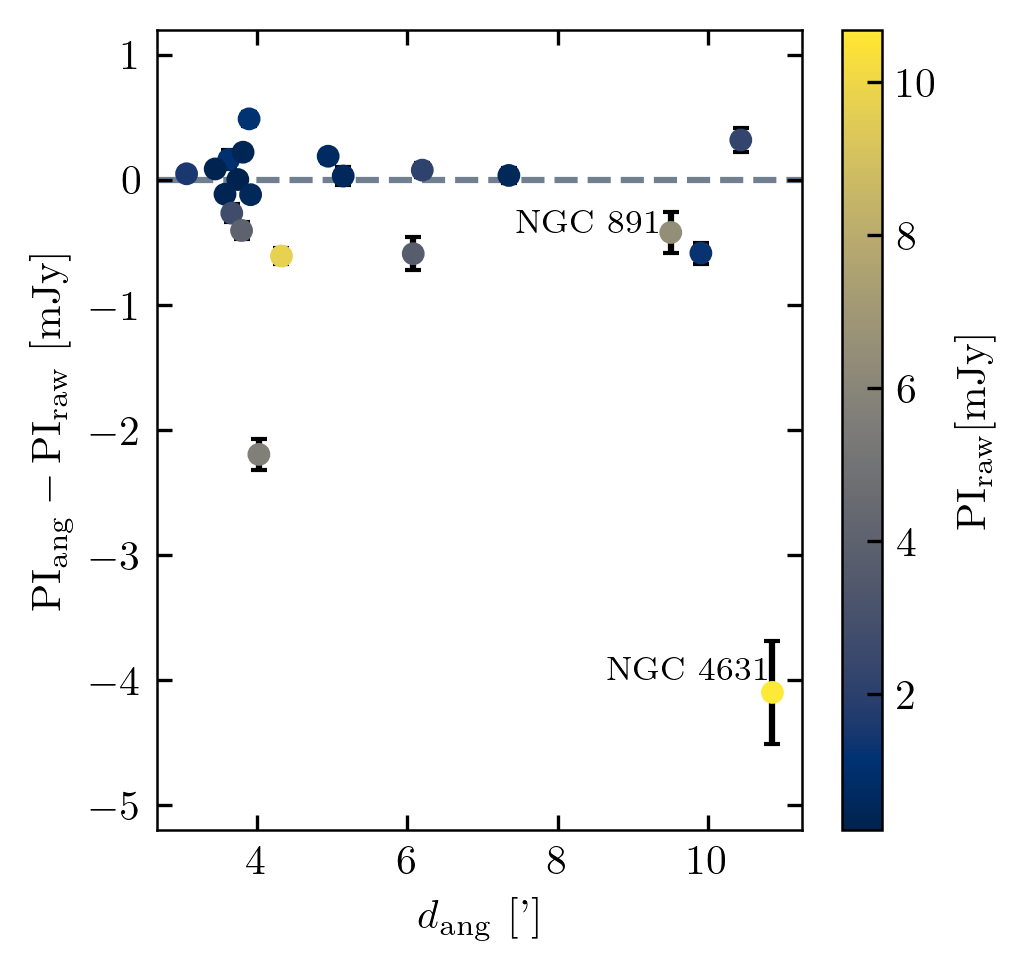}
    \caption{Comparison of polarised flux values measured on the raw data and after smoothing and scaling for galaxies used in the angular size scaling stack plotted against the angular extent.}
    \label{fig:depol}
\end{figure}

Comparing the PI fluxes in Table~\ref{tab:stacking_parameters} for the individual galaxies after the initial data reduction (PI\textsubscript{raw}) and after applying all scaling, convolution, and alignment operations (PI\textsubscript{ang}, PI\textsubscript{phy}) we find some deviation in the derived flux values. To illustrate these deviations, we display the derived flux differences for the ang-scaled datasets and compare them to the angular extent of the galaxies in Fig~\ref{fig:depol}. Here, we can observe two effects. First,  galaxies that show only a low level of polarised emission after the initial data reduction seem to have a higher polarised flux after applying the alignment and scaling operations. Secondly, for galaxies with a large angular extent, the polarised emission is reduced in the data processing.

The increased flux of faint galaxies after the data processing can most likely be attributed to an increase of the polarisation bias on the measurement. For bright galaxies with a large angular extent ($d_{\mathrm{ang}}>4\,\arcmin$), we typically recover a lower polarised flux after the completed data processing. Generally, one can expect this effect. Decreasing the spatial resolution of a dataset will cause distinct regions of polarised emission with differing polarisation angles to be mixed. This will reduce the amount of detected polarised emission. However, the strength of the effect can vary strongly and depends on the original characteristic (small patches of polarised emission vs. large-scale structure) of the polarised emission.

To demonstrate this effect, we explicitly label NGC~891 and NGC~4631, two relatively bright galaxies with a large angular extent, in Fig~\ref{fig:depol}. While the polarised emission of NGC~891 after the complete data processing is only slightly reduced, the polarised emission of NGC~4631 is much more strongly affected by our data processing routine. \citet [][Fig.~8 and Fig.~9]{1991A&A...248...23H} present the polarised emission (including pol. ang. information) of NGC~891 and NGC~4631. Here, the different nature of the observed emission is clearly visible (see also Fig.~6 in \citetalias{2020A&A...639A.112K}). While NGC~891 shows relatively large islands of coherent polarised emission, the polarised emission of NGC~4631 is much more chaotic. The in-depth analysis of NGC~4631 by \citet[][Fig.~1]{2019A&A...632A..11M} also shows this complex structure of the polarised emission with many distinct patches of polarised emission neighbouring each other. Therefore, reducing the spatial resolution in the case of NGC~4631 will mix these distinct regions and thereby reduce the detected polarised emission more strongly compared to NGC~891.

Concluding this section, we note that our analysis shows the difficulties of accurately quantifying the diffuse polarised emission, especially in low signal-to-noise regions. However, we only use the values for the flux normalisation and therefore do not expect the presented results to change significantly due to the systematic uncertainties described above.

\subsection{Correcting for the Milky Way foreground}
As pointed out in Sect.~\ref{sec:meth_MWfg}, we used the RM map provided by \citet{2022A&A...657A..43H} to correct the RM contribution from the MW. \citet{2022A&A...657A..43H} report a \texttt{HEALPix} pixel size of 46.8\,arcmin\textsuperscript{2}. Thus, even the sources of largest angular extent in our study are roughly covered by a single pixel in the MW-RM map. However, based on simulations, \citet{2009A&A...507.1087S} predict Galactic RM variations on much smaller scales and \citet{2025ApJ...982..146P} also find significant Galactic RM substructures on scales below $\sim\!1\,\mathrm{deg}^2$. An inaccurate estimation of the MW RM foreground could significantly influence the large-scale RM patterns in our stacks. However, as the analysis of the RM distribution of individual galaxies (Fig.~\ref{fig:rm_dist_individual}) resulted in mean RM values close to 0\,\radmsquare, we do not expect the RM foreground estimation to strongly influence our results, especially considering our broad RMSF. Nevertheless, future studies using high-accuracy RM measurements of sources with small projected distances to our target galaxies are needed to refine the RM foreground estimation for the CHANG-ES galaxies.

\subsection{PI morphology}
\label{sec:dis_pi_asym}
First, we address the systematic underestimation of the PI identified in App. \ref{app:stack_sys}. Here, the tests of synthetic data show that the recovered PI is underestimated by up to 70\%. In App. \ref{app:stack_sys} we argued that this effect arises from Faraday dispersion \citep{2011MNRAS.418.2336A}, caused by averaging the $Q$ and $U$ values across multiple sources. However, we wish to emphasise that the results described in Sect.~\ref{sec:res_pi_asym} and \ref{sec:res_minor_axis} rely solely on relative measurements. Consequently, we do not expect these findings to be significantly influenced by the aforementioned systematic underestimation.

The detected asymmetry in polarised emission, presented in Sect. \ref{sec:res_pi_asym}, confirms the findings of \citetalias{2020A&A...639A.112K} that galaxies show larger polarised intensity on the approaching side than on the receding side. We report $q$ factors of $\sim\!5\%$, which is in agreement with the mean of $q$ factors reported by \citetalias[][Table 5]{2020A&A...639A.112K}: $\overline{q_{\mathrm{K20}}}=0.07$.

Generally, one can distinguish two scenarios that can explain the detected PI asymmetry. First, galaxies intrinsically have a symmetric emission structure and absorption or depolarisation mechanisms cause the observed polarised emission to appear asymmetric. Such scenarios have been discussed in the context of polarised radio continuum emission by \citet{2010A&A...514A..42B} and \citet{2020A&A...639A.111S}. Alternatively, one might argue that the detected asymmetry is caused by an intrinsically asymmetric distribution of star formation. As an example, such an asymmetry has been found in NGC~891. Deep H$\alpha$ data of NGC~891 \citep{1990A&A...232L..15D, 1990ApJ...352L...1R} show a strong asymmetry where the disc on the approaching side is brighter by a factor of $\sim\!2.5$ compared to the receding side \citep{2007A&A...471L...1K}. \citet{2007A&A...471L...1K} attribute this asymmetry largely to dust absorption but after correcting for this effect, an intrinsic asymmetry of $\sim\!30\%$ remains. However, the fact that the PI asymmetry is observed constantly in multiple galaxies individually (\citet{2010A&A...514A..42B}, \citetalias{2020A&A...639A.112K}) as well as in the stacking results of this study, makes it unlikely that this detected  PI asymmetry results from intrinsic asymmetries that randomly line up but rather points towards depolarisation effects to cause the observed asymmetry. 

To further limit the impact of an intrinsic asymmetry in polarised emission, instead of comparing polarised intensity measurements as performed by \citet{2010A&A...514A..42B}, \citetalias{2020A&A...639A.112K}, and this study, one can also perform a similar study using polarisation fractions (PF). \citet{Skeggs2025}\footnote{\citet{Skeggs2025} can be openly accessed via the library of Queen's university: \url{https://hdl.handle.net/1974/34661}.} performs an in-depth analysis of the asymmetry in polarised emission in CHANG-ES galaxies but computes the $q$ factor (Eq. \ref{eq:q_fac}) using weighted PFs instead of PI measurements. While there are differences for individual galaxies, overall the results of \citet{Skeggs2025} are in agreement with the analysis of \citetalias{2020A&A...639A.112K}. While the PI asymmetry is larger at lower frequencies (cf. \citet{2010A&A...514A..42B}, \citetalias{2020A&A...639A.112K}), \citet{Skeggs2025} report a stronger PF asymmetry in $C$ band compared to $L$ band (1.5\,GHz), highlighting the impact of IFD, which causes the low frequency measurements to be very noisy.

\citet{2010A&A...514A..42B} argue that the observed PI asymmetry is caused by the superposition of a large-scale spiral disc field, which is axisymmetric, and a quadrupolar halo field and the fact that PI observations are near-side biased. In contrast, \citet{2020A&A...639A.111S} and \citet{Skeggs2025} both attribute the observed trends to the impact of spiral arms. \citet{Skeggs2025} argues that the shock at the front of the spiral arm temporarily compresses the B field, causing an increase of total and polarised radio continuum emission \citep[see][]{2005A&A...444..739B,2011MNRAS.412.2396F, 2016A&A...585A..21F}. Furthermore, \citet{Skeggs2025} argues, following the theoretical work of \citet{2017MNRAS.469.4806H}, that the magnetic spiral arms can also extend into the halo of galaxies, which results in a qualitative model that can explain the observed patterns.

In addition to the detected PI asymmetry, the detected minor-axis PI-deficit, presented in Sect. \ref{sec:res_minor_axis} opens another opportunity to infer about the magnetic field structure of star-forming galaxies. First of all, we do not attribute this effect to a decrease in B field strength or CRE density. Typically, the total magnetic field strength \citep[e.g.][]{2013MNRAS.433.1675B,2023A&A...670A.158S} as well as the strength of the coherent field \citep[]{2024ApJ...970...95U} increase towards the centre of star-forming galaxies.

Therefore, we suspect this to be an observational bias due to the geometry of the ordered B field. If the B field in the galactic halo has an X field component \citep{2022A&A...658A.101H}, the B field vectors that are in projection close to the galaxy's minor axis are partly aligned with the LoS. This alignment then causes a decrease in the observed polarised intensity as the PI only traces B field components that are perpendicular to the LoS. This interpretation is supported by the fact that the decrease in flux is stronger for profiles that are offset by 2\,kpc of the disc compared to the central profile.

\begin{figure}
    \centering
    \includegraphics[width=0.75\linewidth]{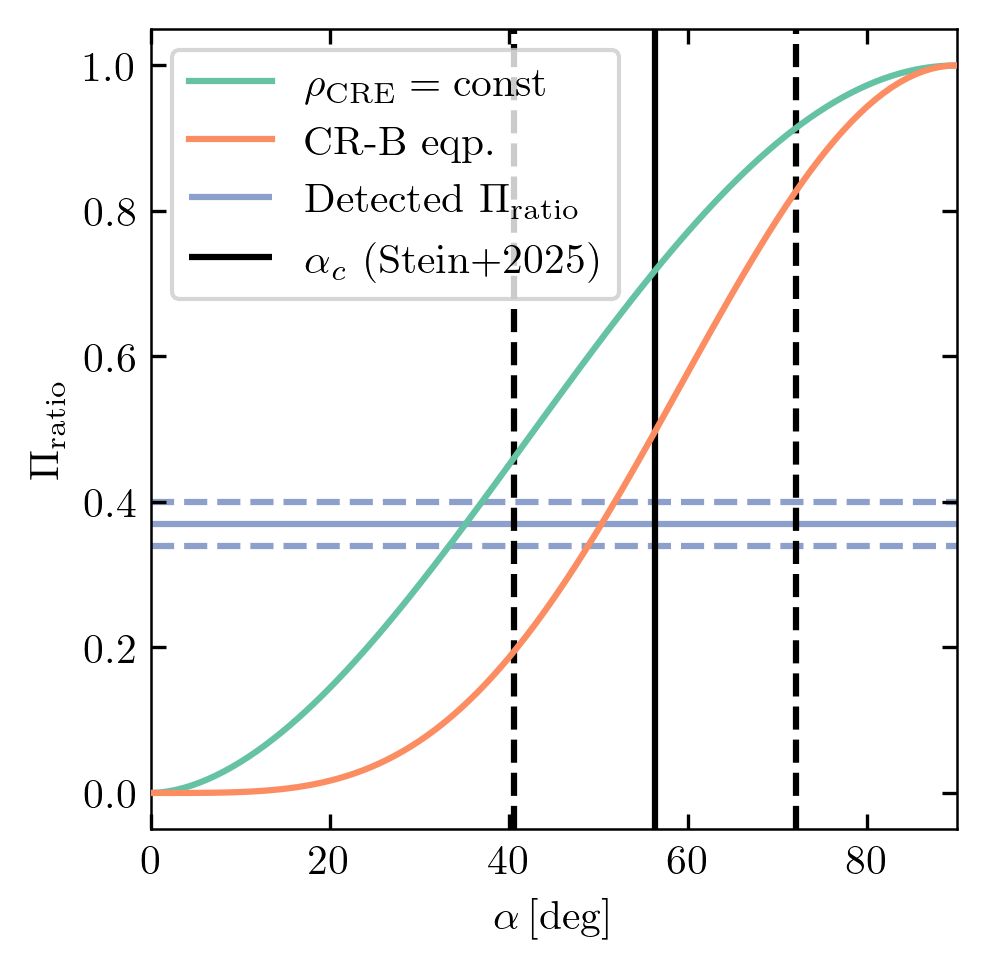}
    \caption{Expected $\Pi_\mathrm{ratio}$ plotted against the opening angle $\alpha$ of X-shaped B field in the halo. Curves for two scenarios are shown: constant CR density (green) and B-field-CR-equipartition  (orange). The detected $\Pi_\mathrm{ratio}$ for the halo profile (blue horizontal line) and the distribution of X shape opening angles as reported by \citet{2025A&A...696A.112S} (black vertical line) are also highlighted. Dashed lines indicate $1\sigma$-uncertainties.}
    \label{fig:pi_ratio_alpha}
\end{figure}

When considering an X-shaped halo B field, the ratio of the B field perpendicular to the LoS scales with the opening angle $\alpha$ of the X field:  $B_\mathrm{min}/B_\mathrm{max}\propto\sin(\alpha)$.
This ratio in magnetic field strengths perpendicular to the LoS results in an observed ratio of PI where the PI reaches its minimum at $x=0$ (the radial B field component is parallel to the LoS, therefore not observed), and its maximum if both B field components (radial and $z$-component) of the X-shaped B field are oriented perpendicular to the LoS. We define this ratio as:
\begin{equation}
   \Pi_\mathrm{ratio} = \frac{\mathrm{PI_{min}}}{\mathrm{PI_{max}}}
\end{equation}
Different physical conditions in the ISM result in different scaling relations of the observed $\Pi_\mathrm{ratio}$ and the ratio $B_\mathrm{min}/B_\mathrm{max}$. In a scenario where the CR density does not rely on the B field strength (constant CR density) $\Pi_\mathrm{ratio}$ scales as
\begin{equation}
    \Pi_\mathrm{ratio}\propto (B_\mathrm{min}/B_\mathrm{max})^{1+\mathrm{SPIX}},
\end{equation}
where SPIX describes the radio synchrotron spectral index ($I_\nu\propto\nu^{-\mathrm{SPIX}}$). In a scenario where the energy densities of CRs and the B field are in equipartition, see \citet{2005AN....326..414B}, the scaling relation between $(B_\mathrm{min}/B_\mathrm{max})^{1+\mathrm{SPIX}}$ and $\Pi_\mathrm{ratio}$ is much stronger: 
\begin{equation}
    \Pi_\mathrm{ratio}\propto (B_\mathrm{min}/B_\mathrm{max})^{3+\mathrm{SPIX}}.
\end{equation}
It is important to note that we do not assume a priori that energy equipartition of CRs and the B field is valid, but that the observed minor-axis PI-deficit allows us to test these scenarios.

Fig.~\ref{fig:pi_ratio_alpha} displays the detected $\Pi_\mathrm{ratio}$ alongside theoretical curves for a range of X-field opening angles ($\alpha$), providing a direct comparison between constant CR density and energy equipartition (eqp) scenarios. Here, we assumed a radio spectral index of SPIX=0.8 for a galactic halo at a height of 2\,kpc \citep{2023A&A...670A.158S}. To distinguish between the two scenarios, an estimate of the X-shape opening angle is needed. Therefore, we further compare the detected $\Pi_\mathrm{ratio}$ with the distribution of the measured opening angles of the X shape, also measured for CHANG-ES galaxies \citep{2025A&A...696A.112S}. As can be seen in Fig.~\ref{fig:pi_ratio_alpha}, the constant CR density assumption predicts a much higher PI ratio for the distribution of opening angles of the CHANG-ES galaxies (expecting a smaller PI decrease close to the galaxy's minor axis). The predicted PI ratio in the case of the equipartition assumption is in much better agreement with the data.

Alternatively, one can consider the minor-axis PI-deficit to be caused by a more turbulent B field at small galactocentric radii, caused by a higher velocity dispersion in the interstellar medium \citep{2014AJ....148..127Y} in this region. This scenario is less probable, however, because any sight-line intersecting the inner galaxy must also traverse a significant path length through the outer regions. Therefore, we consider the described geometric effect to be the cause of the observed minor-axis PI-deficit.

No minor-axis deficit was reported by \citet{2015AJ....150...81W} when stacking the total intensity emission of CHANG-ES galaxies. This is, indeed, also not expected, as the total B-field in star-forming galaxies is dominated by its turbulent (mostly isotropic) component.

To conclude this section, we want to highlight that the PI asymmetry has now been observed across multiple radio wavelengths in multiple studies that made use of different analysis techniques.  However, a detailed understanding of what causes the observed asymmetry is still missing. Therefore, more refined theoretical frameworks with quantitative predictions are needed to further constrain the underlying B field morphology that causes the PI and PF asymmetry. Here, especially magnetohydrodynamical simulations that include CR electrons (either on the fly or in post processing) \citep[e.g.][]{2021MNRAS.508.4072W,2024ApJ...976..136C,2025ApJ...987..204S,2025ApJ...988..214L} will play a key role in explaining these findings.

As presented, the PI decrease at the galaxy's minor axis offers additional insight into the structure of large-scale galactic magnetic fields. Future investigations of individual galaxies, using data with higher resolution in Faraday depth, might search for an increase in |RM| close to the galaxy's minor axis (this is not expected for a stack of multiple galaxies), which would strengthen the idea of an X-shaped B field in galactic haloes.

\subsection{RM structures}
\label{sec:dis_rm_struc}
To search for an RM structure similar to that reported by \citetalias{2021A&A...649A..94M}, we employed two different approaches. First, in Sect. \ref{sec:res_rm_struc_cubes} we presented results from stacking $Q$ and $U$ cubes of multiple galaxies. Other studies that stack the polarised signal of astrophysical objects typically rely on stacking polarised intensities \citep[e.g.][]{2014ApJ...787...99S, 2023SciA....9E7233V}, which requires careful treatment of the non-Gaussian noise properties of the data. This approach is typically needed as the underlying B field geometry of the stack constituents is unknown. In this study, however, we have prior knowledge on the B field geometry in our target galaxies. Aligning them using their galactic disc allows us to stack in the $Q$ and $U$ domain, which prevents the increased background noise found in approaches that involve PI stacking. Furthermore, by applying RM synthesis on the resulting  $Q$ and $U$ cubes, this approach allows us to search for global RM patterns.

From Table \ref{tab:rm_stat}, we can draw several conclusions. In the case of ang-scaling, we detected a marginal split of the RM mean in the combined regions (QI+QIII and QII+QIV). Additionally, we detected no increase in the strength of the RM split when accounting for the galaxy rotation sense in the alignment process, indicating no relation between the galactic rotation sense and the orientation of the B field. This is in agreement with expectations from dynamo theory, which predicts no relation between the rotation sense of the disc and the sign of the B field \citep{2008A&A...487..197M}.

Additionally, we do want to highlight that the detected RM split is very marginal (with varying significance levels $1.1\sigma-3.1\sigma$) and therefore not conclusive. Furthermore, we want to stress that the detected larger RM split in the high |RM| indicates that the stack might be influenced by individual high |RM| sources. In the case of the phy scaling, we do not find any split. As pointed out before, for the sector comparison, the main difference between the ang-scaling and phy-scaling approach is the slightly larger sample size in the case of phy scaling\footnote{The phy-scaling stack contains all galaxies from the ang scaling stack and further includes NGC~2820, NGC~3448, NGC~4388, and NGC~5792}. This further supports the idea that the observed RM pattern in the ang-scaling might arise from the relatively small sample size, where a couple of galaxies randomly align to produce the observed pattern.

To further investigate the impact of the different samples, we reran the physical scaling approach (median stacking, PI flux normalisation), including only galaxies that are in the angular scaling sample. The detected split is stronger compared to the full phy-sample (see Table \ref{tab:rm_sample_stats}), but it does not reach a similar level as in the ang-scaling approach. Therefore, the sample difference contributes to the detected difference of the RM splits, but it cannot fully explain it. However, we want to stress again that all samples and techniques yielded insignificant results.

\begin{figure}
\centering
    \includegraphics[width=1\linewidth]{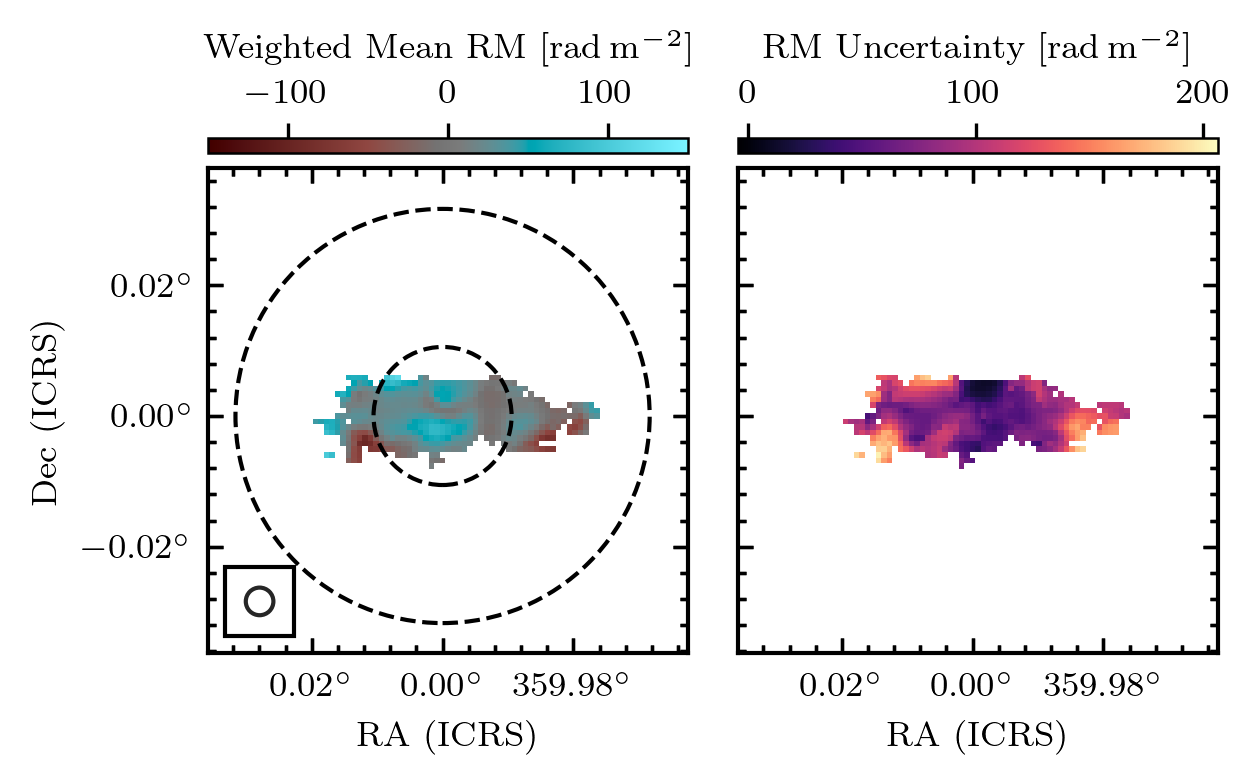}    
    \caption{RM map derived from weighted mean stacking (left panel) and RM uncertainty maps (derived as $\left(1/\delta_\mathrm{RM}^2\right)$-weighted standard deviation per pixel, right panel) for the std-aligned and angular scaled galaxies without RM normalisation. The beam size is indicated in the bottom left corner of the RM map. The black dashed circles indicate the distinction between the central and outer regions that were analysed separately.}
    \label{fig:stacked_rm_standard_wmean}
\end{figure}

Also, our second approach, the stacking of RM values of individual galaxies (Sect. \ref{sec:res_rm_struc_rm_maps}), did not result in an RM split as reported by \citetalias{2021A&A...649A..94M}. While we find throughout this paper median stacking to be more effective than mean stacking, in Fig. \ref{fig:stacked_rm_standard_wmean} we present a RM map that mimics the \citetalias{2021A&A...649A..94M} approach as closely as possible. Here we show the ang-scaled weighted mean stack using standard alignment and no RM normalisation. A RM structure as described by \citetalias{2021A&A...649A..94M} does not appear.

Even though the ang-scaled stacking results described in Sect. \ref{sec:res_rm_struc_cubes} show a similar geometry as reported by \citetalias{2021A&A...649A..94M} (negative RMs in QI and QIII, positive RMs in QII and QIV), the strength of the split strongly differs.  \citetalias{2021A&A...649A..94M} find a split of $\sim\!30$\,\radmsquare, we find a split of $\sim\!100$\,\radmsquare. No difference in RM pattern is observed between central and outer regions, providing no support for models predicting strong central source influence (e.g., supermassive black hole (SMBH)) on galactic B fields. As our analysis relies solely on $C$-band data, resulting in an RMSF of $\sim2000$\,\radmsquare, therefore, detecting (or excluding to detect) a split of $\sim\!30$\,\radmsquare, as reported by \citetalias{2021A&A...649A..94M}, is ultimately not possible. However, theoretical frameworks describing galactic B fields also do not predict to find such a structure in a stack of galaxies.

Simplified models of the mean-field dynamo \citep[e.g.][]{1988ASSL..133.....R} predict that large-scale magnetic fields generated in galaxy discs are symmetric with respect to the equatorial plane (quadrupolar poloidal fields), whereas fields generated in quasi-spherical haloes are antisymmetric (dipolar poloidal fields). Furthermore, a strong connection between the dynamo in the disc and in the halo is anticipated \citep{2008A&A...487..197M}.

However, RM studies of nearby galaxies \citep[e.g.][]{2019A&A...632A..11M} reveal a much more complex structure in galactic haloes than predicted by simplified mean-field dynamo theory. To explain these observations, advanced non-linear models of the mean-field dynamo, taking into account the magnetic buoyancy instability \citep{2023MNRAS.525.2972T,2023MNRAS.525.5597T,2024MNRAS.527.7994Q} are necessary. These studies predict that the initial field structure is wiped away, and the halo field shows a complicated structure with reversals on scales of about 1 kpc \citep{2024MNRAS.527.7994Q}, consistent with observation in the radio haloes of several edge-on galaxies (e.g. \citetalias{2020A&A...639A.112K}). In a stack of many galaxies, no large-scale field pattern remains. In summary, mean-field dynamo models do not predict a universal RM pattern. 

\citetalias{2021A&A...649A..94M} propose a relation between the polarity of the field and the direction of its winding by differential rotation in the accretion disc around the galaxy's supermassive black hole. In this scenario, a universal pattern should be evident in the stacked RM map using rotation alignment.

Also, in a scenario where the halo B field is not governed by a large-scale dynamo but is dominated by the small-scale dynamo and the disc B field that is transported by galactic outflows \citep{2020MNRAS.498.3125P,2021MNRAS.508.4072W}, a general RM pattern, observable in stacks of galaxies, is not to be expected.

To summarise, we do not find an RM pattern as reported by \citetalias{2021A&A...649A..94M}. However, if the \citetalias{2021A&A...649A..94M} RM pattern were to be confirmed in future studies, many theoretical frameworks would be challenged. Solutions that attribute this effect to the SMBH in the centre of galaxies, as discussed in \citetalias{2021A&A...649A..94M}, seem unlikely, as we do not find a change of the RM pattern when comparing the central to the outer region. An alternative approach to explaining such an RM structure has been proposed by \citet{2022A&A...658A.101H}. Here, the discussed RM structure is attributed to the interaction of a large-scale dynamo and the galactic wind.

\section{Summary and outlook}
\label{sec:SandO}
In this paper, we have explored a new technique to stack the polarised signal of star-forming edge-on galaxies and presented the following key findings:
\begin{enumerate}
    \item Compared to the derived PI stack of \citetalias{2020A&A...639A.112K}, the newly derived stacking products show a larger halo extent and a more symmetric structure. Additionally, the derived median stacks seem much less affected by features from individual galaxies. All stacks show diffuse halo emission with an X-shaped polarisation pattern. In the physical size stack, we can trace polarised emission up to 7\,kpc above the galactic disc.
    \item The PI maps that were derived from the stacked Q and $U$ cubes show a PI asymmetry with an asymmetry factor $q\approx5\%$, which is in agreement with the sample-averaged $q$ factors that were reported by \citetalias{2020A&A...639A.112K}.
    \item The detected PI decrease within the halo near the galaxy's minor axis is consistent with the X-shape opening angles derived by \citet{2025A&A...696A.112S}. The magnitude of this decrease suggests that energy equipartition between CRs and B fields holds within galactic haloes, rather than a model of constant CR density.
    \item In both RM stacking approaches, we do not find a significantly large-scale RM structure when analysing. Therefore, we cannot confirm the results reported by \citetalias{2021A&A...649A..94M}.
\end{enumerate}

As we have expanded the analysis of \citetalias{2020A&A...639A.112K} in this paper, it is worth pointing out the differences between our approach and that of \citetalias{2020A&A...639A.112K}, summarised here:
\begin{enumerate}
    \item We excluded one additional galaxy (NGC~660) compared to \citetalias{2020A&A...639A.112K}.
    \item We carried out RM synthesis on the data whereas \citetalias{2020A&A...639A.112K} did not.
    \item We explored multiple flux normalisation approaches, whereas \citetalias{2020A&A...639A.112K} solely scaled by the galaxy distance.
    \item We computed both medians and means whereas \citetalias{2020A&A...639A.112K} computed only the mean.
    \item We provided various options for the rotations, whereas \citetalias{2020A&A...639A.112K} only used the std alignment.
\end{enumerate}
In spite of these differences, our results agree with the findings presented in \citetalias{2020A&A...639A.112K}, as described above.

Extending a similar analysis to slightly lower frequencies might enable us to trace the polarised emission further into the halo, as slightly lower frequency electrons would be traced. Therefore, combining the $Q$ and $U$ data used in this study with the upcoming CHANG-ES $S$ band (3\,GHz) data \citep[see][for the first $S$ band result in total intensity and polarisation]{2024AJ....168..138I,2025ApJ...978....5X,2025A&A...699A.243H} will greatly increase our RM resolution\footnote{Combining $S$ band and $C$ band data, would result in a RMSF of $\sim\!280$\,\radmsquare.} as well as the ability to trace CR electrons in the halo of galaxies. Furthermore, robust predictions of RM maps based on theory or simulations are needed to enable a direct comparison of the observational products and theoretical models.

\begin{acknowledgements} We thank the anonymous referee for a very constructive report that helped to improve our paper. We also gratefully acknowledge Yik Ki (Jackie) Ma for his deep insights and fruitful discussions, which significantly improved this work. Further, we gratefully acknowledge financial support by the German Research Foundation (Deutsche Forschungsgemeinschaft, DFG) through the Collaborative Research Center (CRC; i.e., Sonderforschungsbereich, SFB) 1491.PK acknowledges the support of the BMBF project 05A23PC1 for D-MeerKAT. This research has made use of the CIRADA cutout service at URL \url{cutouts.cirada.ca}, operated by the Canadian Initiative for Radio Astronomy Data Analysis (CIRADA). CIRADA is funded by a grant from the Canada Foundation for Innovation 2017 Innovation Fund (Project 35999), as well as by the Provinces of Ontario, British Columbia, Alberta, Manitoba and Quebec, in collaboration with the National Research Council of Canada, the US National Radio Astronomy Observatory and Australia’s Commonwealth Scientific and Industrial Research Organisation. TW acknowledges financial support from the grant CEX2021-001131-S  funded by MICIU/AEI/ 10.13039/501100011033, from the coordination of the participation in SKA-SPAIN, funded by the Ministry of Science, Innovation and Universities (MICIU).

 In addition to the already acknowledged software packages, this study used the following packages: \texttt{APLpy} \citep{2012ascl.soft08017R}, \texttt{Astropy} \citep{astropy:2013, astropy:2018, astropy:2022}, \texttt{CARTA} \citep{2021zndo...3377984C}, \texttt{CosmosCanvas} \citep{2024ascl.soft01005E}, \texttt{RM-Tools} \citep{2020ascl.soft05003P, 2026arXiv260120092V}.
\end{acknowledgements}

\bibliographystyle{aa}
\bibliography{export-bibtex,additional_refs}

\begin{appendix}
\onecolumn
\FloatBarrier
\section{Properties of the galaxy sample}
\begin{table*}[h!]
    \centering
    \caption{Fundamental information about the initial galaxy sample.}
    \label{tab:fundamental_parameters}
    \begin{tabular}{lccrrrrrrrrr}
    \hline \hline
    Galaxy  & RA                & Dec          & PA    & Q\textsubscript{AS} &$D$     & $d_{\mathrm{ang}}$ & $d_{\mathrm{phy}}$  & B\textsubscript{maj} &B\textsubscript{maj} & $N^{d}_{\mathrm{Beam}}$ & RM\textsubscript{MW} \\
            & [H:M:S]           & [D:M:S]      & [deg] & &[Mpc] & [\SI{}{\arcmin}]&[kpc]   & [\SI{}{\arcsec}] & [kpc]  & & [$\mathrm{rad}\,\mathrm{m}^{-2}$]\\
    \hline
NGC 891 & 02h22m33.4s & +42d20m56.9s & 22.0 & NE & 9.1 & 9.5 & 25.1 & 12.0 & 0.5 & 47.5 & $-72.9\pm11.0$ \\
NGC 2613 & 08h33m22.8s & -22d58m25.2s & 113.5 & NW & 23.4 & 5.0 & 33.7 & 12.0 & 1.4 & 24.7 & $248.0\pm74.0$ \\
NGC 2683 & 08h52m41.3s & +33d25m18.3s & 43.6 & NE & 6.3 & 5.2 & 9.4 & 12.0 & 0.4 & 25.7 & $14.7\pm3.8$ \\
NGC 2820 & 09h21m45.6s & +64d15m28.6s & 61.1 & SW & 26.5 & 1.8 & 13.8 & 12.0 & 1.5 & 8.9 & $-15.4\pm2.6$ \\
NGC 3003 & 09h48m36.2s & +33d25m17.7s & 78.8 & NE & 25.4 & 3.8 & 27.7 & 12.0 & 1.5 & 18.7 & $18.8\pm4.2$ \\
NGC 3044 & 09h53m40.9s & +01d34m46.7s & 114.3 & NW & 20.3 & 3.1 & 18.1 & 12.0 & 1.2 & 15.3 & $-8.2\pm6.6$ \\
NGC 3079 & 10h01m57.8s & +55d40m47.2s & 166.2 & NW & 20.6 & 4.3 & 26.0 & 12.0 & 1.2 & 21.6 & $2.8\pm1.7$ \\
NGC 3432 & 10h52m31.1s & +36d37m07.6s & 33.0 & NE & 9.4 & 3.6 & 10.0 & 12.0 & 0.5 & 18.2 & $11.5\pm2.9$ \\
NGC 3448 & 10h54m39.2s & +54d18m17.5s & 64.8 & SW & 24.5 & 1.8 & 12.5 & 12.0 & 1.4 & 8.7 & $9.3\pm1.8$ \\
NGC 3556 & 11h11m31.0s & +55d40m26.8s & 79.0 & NE & 14.1 & 6.1 & 24.9 & 12.0 & 0.8 & 30.4 & $11.0\pm2.2$ \\
NGC 3628 & 11h20m17.0s & +13d35m22.9s & 103.6 & NW & 8.5 & 9.9 & 24.5 & 12.0 & 0.5 & 49.5 & $10.6\pm6.3$ \\
NGC 3735 & 11h35m57.3s & +70d32m08.1s & 129.7 & SE & 42.0 & 2.8 & 34.4 & 12.0 & 2.4 & 14.1 & $-7.4\pm6.1$ \\
NGC 3877 & 11h46m07.7s & +47d29m40.4s & 36.2 & SW & 17.7 & 3.6 & 18.5 & 12.0 & 1.0 & 17.9 & $22.0\pm6.0$ \\
NGC 4013 & 11h58m31.3s & +43d56m50.7s & 58.6 & NE & 16.0 & 3.5 & 16.1 & 12.0 & 0.9 & 17.2 & $4.5\pm2.1$ \\
NGC 4096 & 12h06m01.1s & +47d28m42.8s & 20.0 & SW & 10.3 & 3.9 & 11.8 & 12.0 & 0.6 & 19.6 & $7.6\pm1.8$ \\
NGC 4157 & 12h11m04.4s & +50d29m04.8s & 64.7 & SW & 15.6 & 3.7 & 16.6 & 12.0 & 0.9 & 18.3 & $24.6\pm5.4$ \\
NGC 4192 & 12h13m48.3s & +14d54m01.2s & 152.4 & NW & 13.6 & 6.2 & 24.4 & 12.0 & 0.8 & 31.0 & $-6.6\pm5.7$ \\
NGC 4217 & 12h15m50.9s & +47d05m30.4s & 49.8 & NE & 20.6 & 3.9 & 23.4 & 12.0 & 1.2 & 19.5 & $3.5\pm1.3$ \\
NGC 4302 & 12h21m42.3s & +14d35m51.1s & 178.5 & N & 19.4 & 3.8 & 21.6 & 12.0 & 1.1 & 19.1 & $-7.1\pm6.4$ \\
NGC 4388 & 12h25m46.8s & +12d39m43.8s & 91.1 & W & 16.6 & 2.3 & 11.1 & 12.0 & 1.0 & 11.5 & $-2.9\pm5.3$ \\
NGC 4565 & 12h36m20.8s & +25d59m15.6s & 135.2 & SE & 11.9 & 10.4 & 36.1 & 12.0 & 0.7 & 52.1 & $4.6\pm3.2$ \\
NGC 4631 & 12h42m08.0s & +32d32m29.4s & 85.7 & SW & 7.4 & 10.8 & 23.4 & 12.0 & 0.4 & 54.2 & $-5.0\pm2.0$ \\
NGC 4666 & 12h45m08.6s & -00d27m42.8s & 40.6 & NE & 27.5 & 4.0 & 32.3 & 12.0 & 1.6 & 20.1 & $-5.7\pm7.9$ \\
NGC 5297 & 13h46m23.7s & +43d52m20.7s & 146.5 & SE & 40.4 & 2.2 & 25.9 & 12.0 & 2.4 & 11.0 & $7.9\pm2.1$ \\
NGC 5775 & 14h53m57.6s & +03d32m40.1s & 148.4 & NW & 28.9 & 3.8 & 32.0 & 12.0 & 1.7 & 19.0 & $13.5\pm8.1$ \\
NGC 5792 & 14h58m22.7s & -01d05m28.3s & 87.2 & NE & 31.7 & 2.6 & 23.7 & 12.0 & 1.8 & 12.9 & $0.2\pm7.0$ \\
NGC 5907 & 15h15m53.2s & +56d19m47.6s & 155.6 & SE & 16.8 & 7.3 & 35.9 & 12.5 & 1.0 & 35.3 & $0.4\pm4.2$ \\
    
   \\

    \hline
    \end{tabular} 
    \tablefoot{Celestial coordinates at epoch J2000.0\textsuperscript{(a)}, major axis position angle (from north, eastwards)\textsuperscript{(b)}, quadrant of the galaxy's approaching side (derived from CHANG-ES \HI-data), distance\textsuperscript{(c)}, on sky and physical diameter of the star-forming disc measured from 22\,$\mu$m data\textsuperscript{(c)}, axis of the circular synthesised beam (on-sky and physical), number of beams that cover the diameter of the star-forming dics, MW foreground RM.\textsuperscript{(d)} \\
    \tablefoottext{a}{Taken from the NASA NED (\url{https://ned.ipac.caltech.edu/}).} \tablefoottext{b}{Taken from the HyperLeda Database \citep[\url{http://leda.univ-lyon1.fr/}][]{2014A&A...570A..13M}.}\tablefoottext{c}{Taken from \citet{2015AJ....150...81W}. 
    \tablefoottext{d}{Taken from \citet{2022A&A...657A..43H}}.\\}
    }
\end{table*}

\begin{table*}[]
      \centering
    \caption{Stacking information about the initial galaxy sample.}
    \label{tab:stacking_parameters}
    \begin{tabular}{lrrccccrrr}
    \hline \hline
    Galaxy  & $\delta$PA & std. rot. & Extra rot. & ang. stack & phy. stack & |RM| sample  & $\mathrm{PI_{ang}}$ & $\mathrm{PI_{phy}}$& $\mathrm{PI_{raw}}$ \\
     & [deg] & [deg] & & & & & [mJy] & [mJy] & [mJy] \\
    \hline
NGC 891 & -2.0 & 66.0 & no & True & True & high     & 6.02 $\pm$ 0.13 & 7.97 $\pm$ 0.17 & 6.44 $\pm$ 0.10 \\
NGC 2613 & -0.5 & -24.0 & yes & True & True & low   & 0.90 $\pm$ 0.03 & 0.93 $\pm$ 0.03 & 0.72 $\pm$ 0.04 \\
NGC 2683 & 2.6 & 49.0 & no & True & True & high     & 0.61 $\pm$ 0.06 & 0.98 $\pm$ 0.16 & 0.58 $\pm$ 0.03 \\
NGC 2820 & 0.0 & 28.9 & yes & False & True &–       & – & 1.23 $\pm$ 0.04 & 1.14 $\pm$ 0.05 \\
NGC 3003 & 0.8 & 12.0 & no & True & True & high     & 0.22 $\pm$ 0.02 & 0.19 $\pm$ 0.03 & 0.22 $\pm$ 0.03 \\
NGC 3044 & -0.7 & -25.0 & yes & True & True & low   & 1.61 $\pm$ 0.02 & 1.49 $\pm$ 0.02 & 1.57 $\pm$ 0.02 \\
NGC 3079 & -2.8 & -79.0 & yes & True & True & low   & 9.12 $\pm$ 0.04 & 8.67 $\pm$ 0.07 & 9.73 $\pm$ 0.05 \\
NGC 3432 & -9.0 & 48.0 & no & True & True & high    & 1.21 $\pm$ 0.06 & 1.22 $\pm$ 0.08 & 1.05 $\pm$ 0.04 \\
NGC 3448 & -4.0 & 21.2 & yes & False & True &–      & – & 0.74 $\pm$ 0.03 & 0.83 $\pm$ 0.04 \\
NGC 3556 & -2.0 & 9.0 & no & True & True & low      & 3.14 $\pm$ 0.09 & 3.17 $\pm$ 0.11 & 3.73 $\pm$ 0.09 \\
NGC 3628 & -0.4 & -14.0 & yes & True & True & high  & 0.66 $\pm$ 0.07 & 2.03 $\pm$ 0.14 & 1.25 $\pm$ 0.04 \\
NGC 3735 & -0.3 & -40.0 & no & False & False &–     & – & – & 1.31 $\pm$ 0.02 \\
NGC 3877 & 2.2 & 56.0 & yes & True & True & high    & 0.26 $\pm$ 0.03 & 0.35 $\pm$ 0.03 & 0.37 $\pm$ 0.03 \\
NGC 4013 & -7.4 & 24.0 & yes & True & True & high   & 0.39 $\pm$ 0.01 & 0.37 $\pm$ 0.03 & 0.30 $\pm$ 0.02 \\
NGC 4096 & 3.0 & 73.0 & yes & True & True & high    & 0.39 $\pm$ 0.03 & 0.61 $\pm$ 0.05 & 0.51 $\pm$ 0.04 \\
NGC 4157 & 1.7 & 27.0 & yes & True & True & low     & 2.41 $\pm$ 0.05 & 2.32 $\pm$ 0.08 & 2.67 $\pm$ 0.05 \\
NGC 4192 & 0.4 & -62.0 & yes & True & True & low    & 2.18 $\pm$ 0.04 & 2.15 $\pm$ 0.08 & 2.10 $\pm$ 0.03 \\
NGC 4217 & 0.8 & 41.0 & no & True & True & low      & 1.68 $\pm$ 0.03 & 1.58 $\pm$ 0.06 & 1.19 $\pm$ 0.05 \\
NGC 4302 & -1.0 & -89.5 & yes & True & True & high  & 0.68 $\pm$ 0.02 & 0.90 $\pm$ 0.04 & 0.46 $\pm$ 0.04 \\
NGC 4388 & 2.1 & 1.0 & yes & False & True & -       & – & 1.55 $\pm$ 0.04 & 2.25 $\pm$ 0.03 \\
NGC 4565 & 0.0 & -45.2 & no & True & True & low     & 2.57 $\pm$ 0.05 & 3.31 $\pm$ 0.11 & 2.25 $\pm$ 0.08 \\
NGC 4631 & 1.7 & 6.0 & yes & True & True & low      & 6.58 $\pm$ 0.38 & 6.25 $\pm$ 0.32 & 10.68 $\pm$ 0.16 \\\
NGC 4666 & 0.0 & 49.4 & no & True & True & high     & 3.45 $\pm$ 0.05 & 3.95 $\pm$ 0.05 & 5.64 $\pm$ 0.11 \\
NGC 5297 & -2.5 & -59.0 & no & False & False &–     & – & – & 0.38 $\pm$ 0.02 \\
NGC 5775 & 2.4 & -56.0 & yes & True & True & low    & 3.59 $\pm$ 0.05 & 3.57 $\pm$ 0.04 & 4.00 $\pm$ 0.05 \\
NGC 5792 & 1.2 & 4.0 & no & False & True &–         & – & 0.28 $\pm$ 0.03 & 0.44 $\pm$ 0.02 \\
NGC 5907 & 0.6 & -65.0 & no & True & True & low     & 0.57 $\pm$ 0.04 & 0.89 $\pm$ 0.03 & 0.53 $\pm$ 0.04 \\
    \hline
    \end{tabular}
    \tablefoot{Galaxy name, required fine tuning of the Hyperleda PA to align galaxies horizontally, rotation angle applied used for galaxy alignment in the standard case (positive values indicate counter-clockwise rotation), indicator if an extra $180^\circ$ rotation is required so that the galaxy's approaching side is to the left; indicators for inclusion in the angular or physical stack and high or low RM subsample, and measured polarised flux of individual galaxies after applying all data processing steps (Sect. \ref{sec:meth_smr}) of the angular or physical scaling procedure.}

\end{table*}
\FloatBarrier
\clearpage
\section{Data processing flow chart}
\begin{figure*}[h!]
    \includegraphics[width=0.95\linewidth]{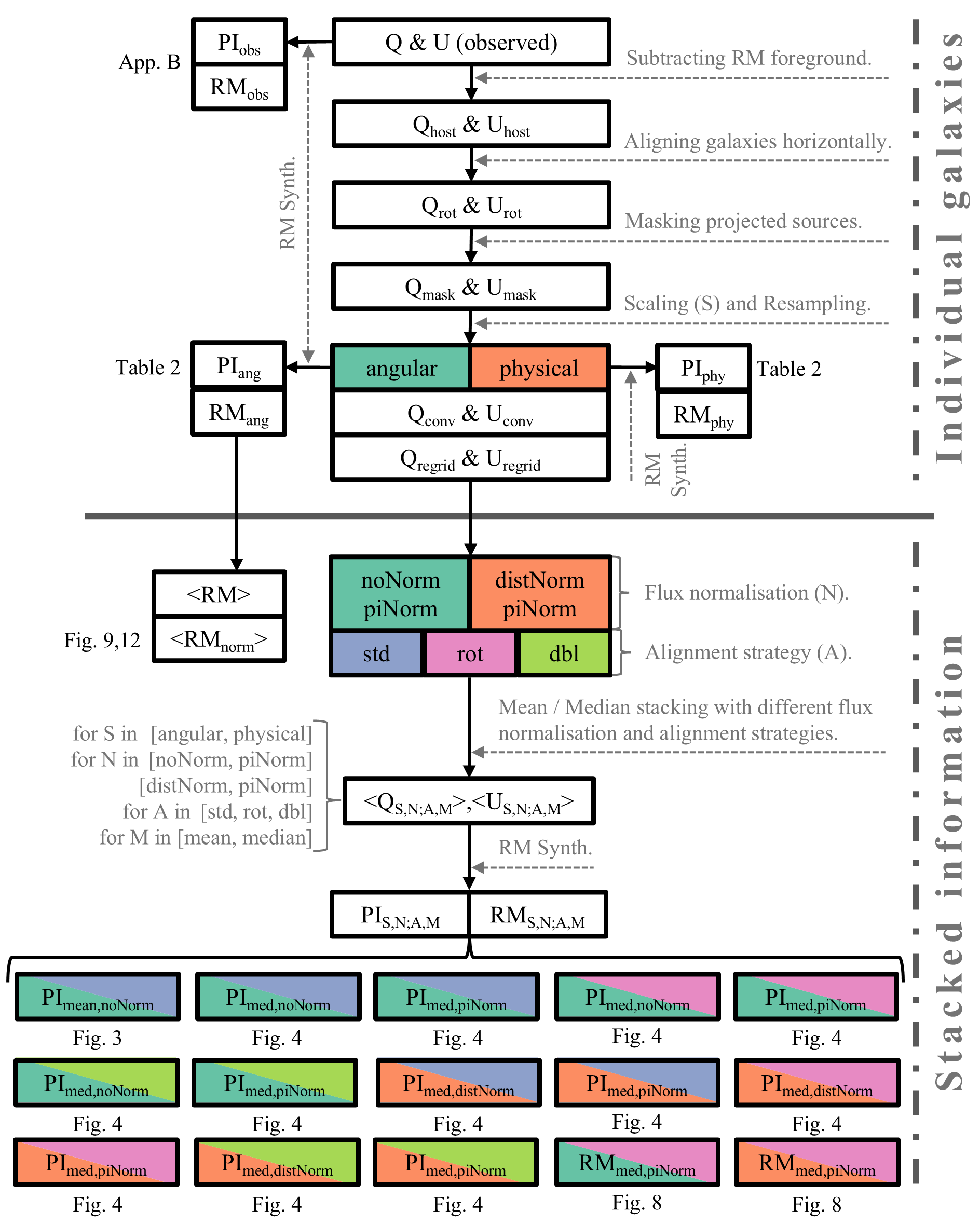}
    \caption{Overall workflow of the data processing displayed as a flowchart.}
    \label{fig:meth_flowchart}
\end{figure*}
\FloatBarrier
\clearpage
\section{Image atlas initial data reduction}
\label{sec:app_img_atals}
\begin{figure*}[h!]
    \centering
    \includegraphics[width=0.9\linewidth]{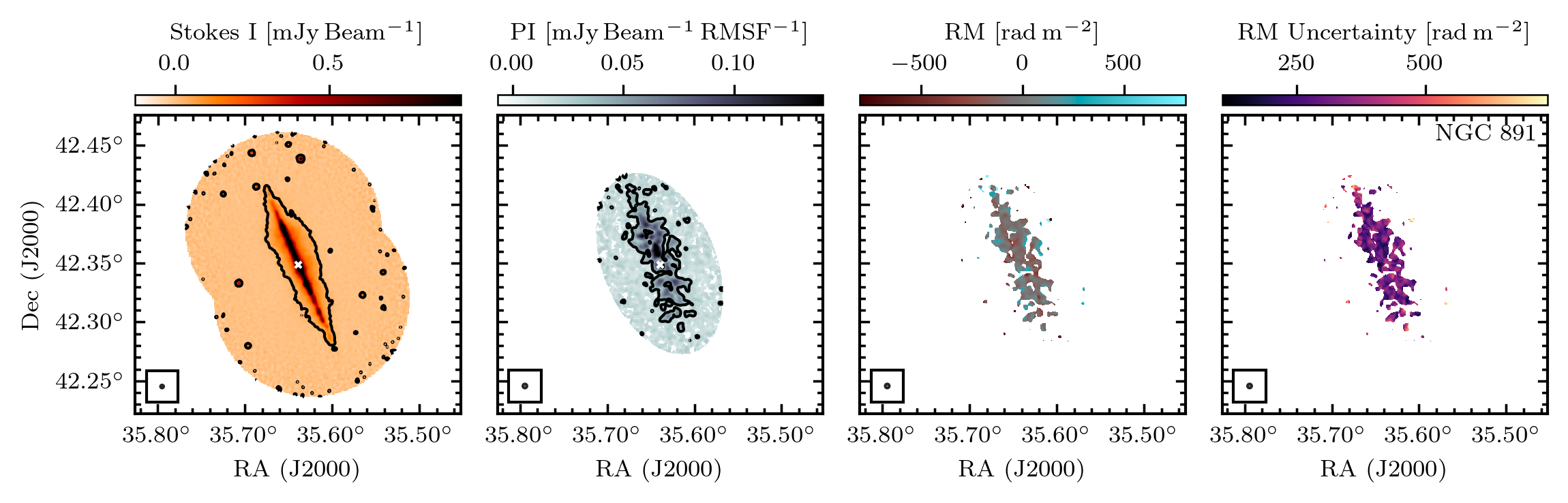}\\
    \includegraphics[width=0.9\linewidth]{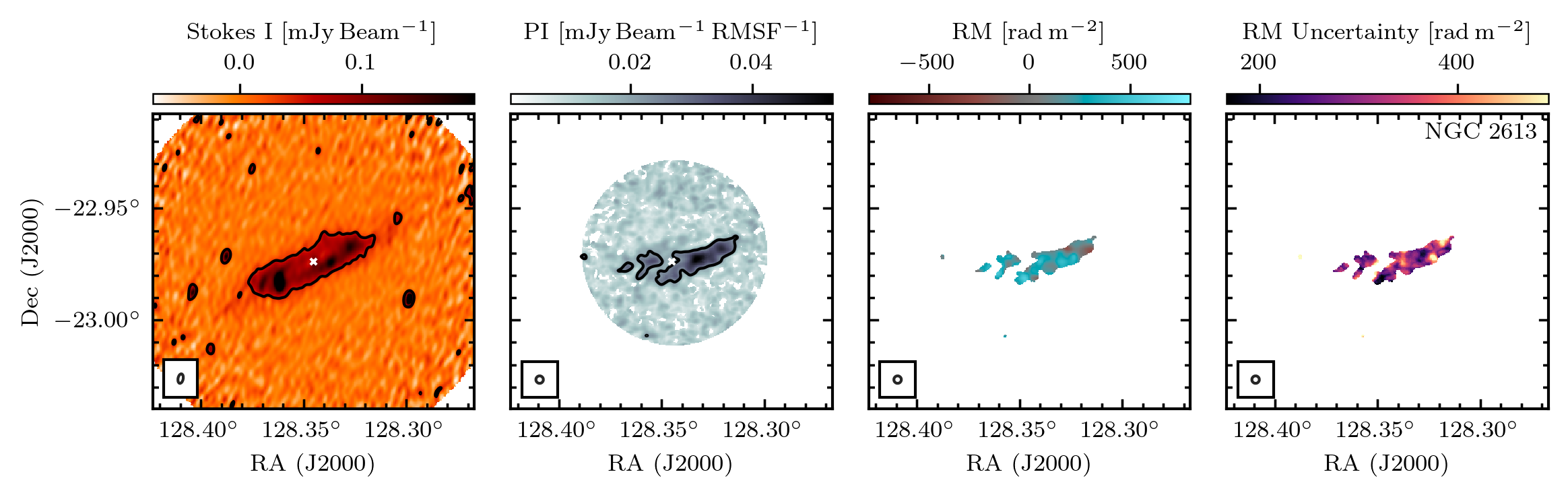}\\
    \includegraphics[width=0.9\linewidth]{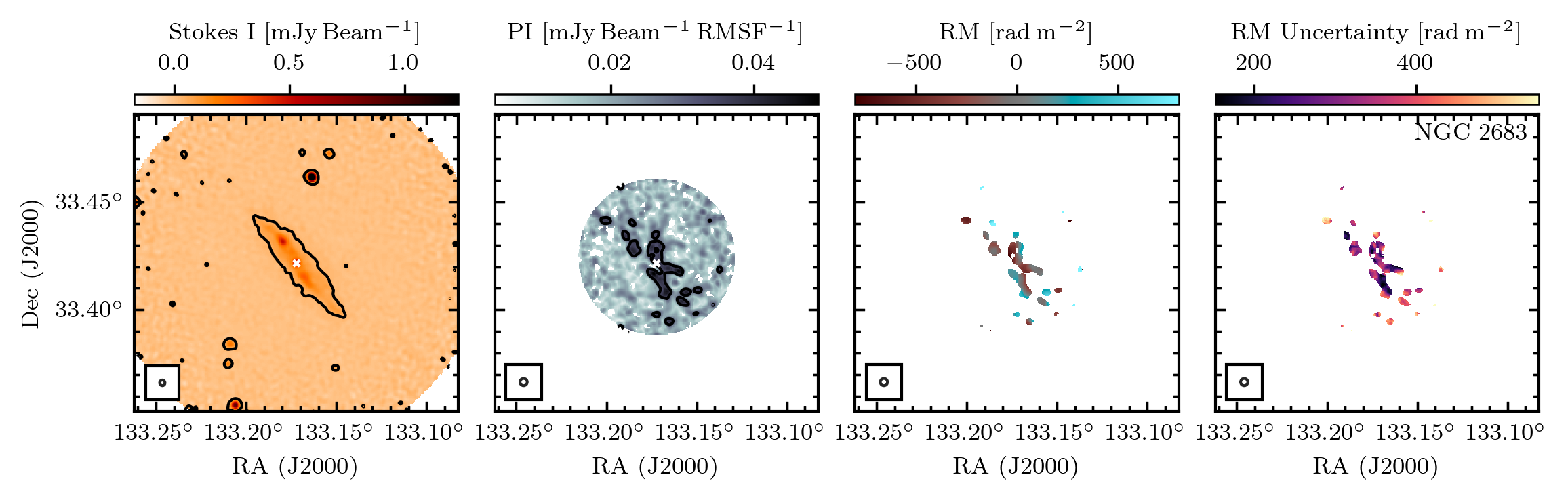}\\
    \includegraphics[width=0.9\linewidth]{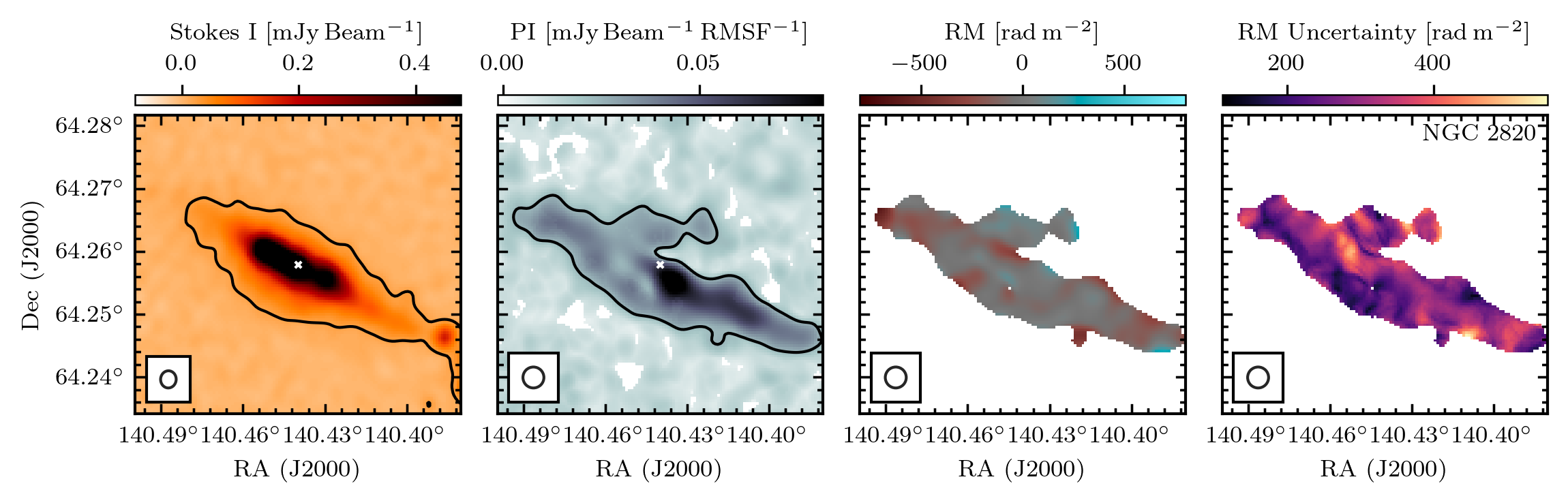}\\
    \caption{Image atlas after applying RM synthesis to the datasets after initial data reduction. From left to right, the panels show the D array $C$ band total intensity map (+$3\sigma$ contour) taken from \citet{2015AJ....150...81W}, peak PI map (+$3\sigma$ contour), RM map, RM uncertainty map. The shape of the radio beam is indicated in the bottom left corner of each panel. Displayed galaxies: NGC~891, NGC~2613, NGC~2683, and NGC~2820.}
    \label{fig:app_img_atlas_891_2613_2683_2820}
\end{figure*}

\begin{figure*}
    \centering
    \includegraphics[width=0.9\linewidth]{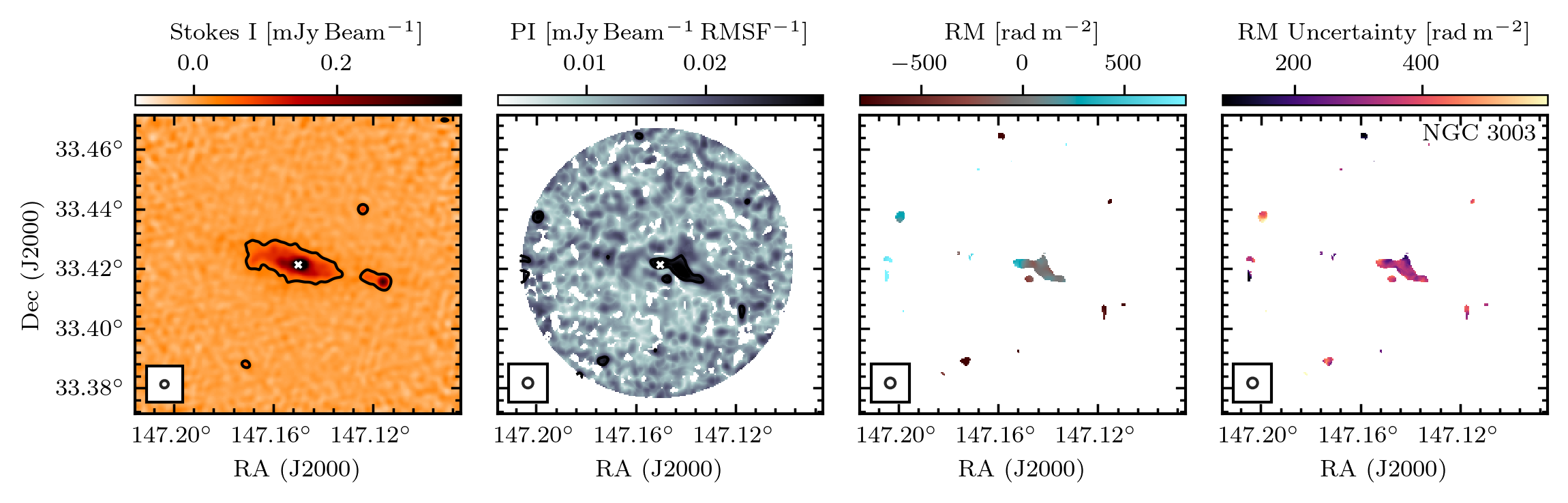}\\
    \includegraphics[width=0.9\linewidth]{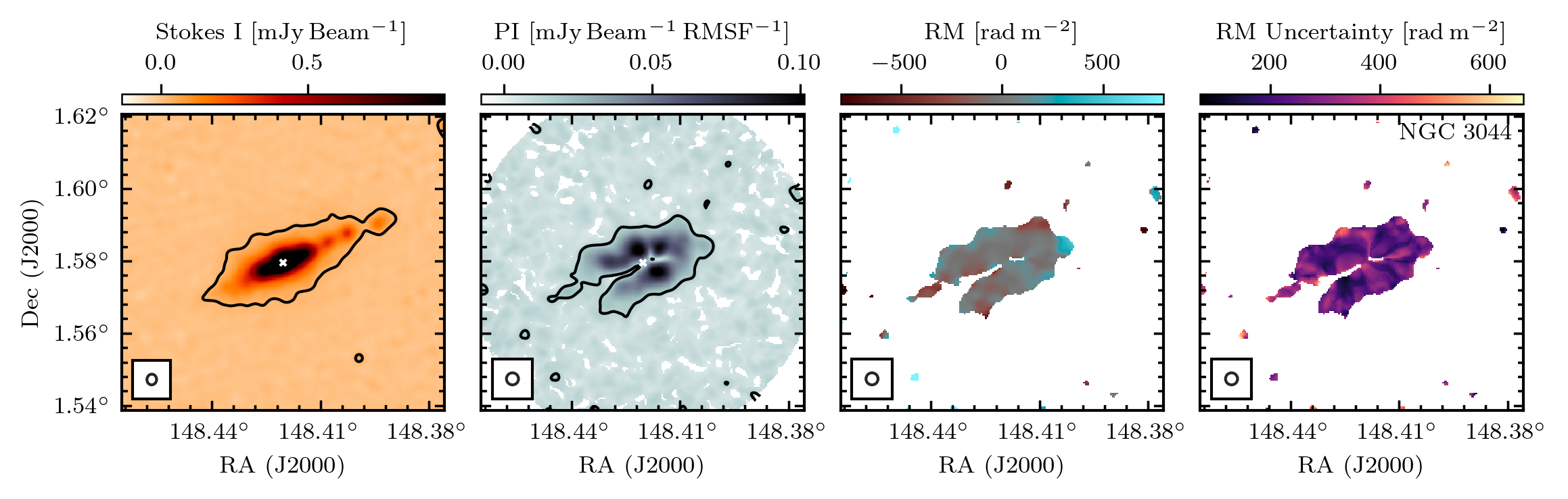}\\
    \includegraphics[width=0.9\linewidth]{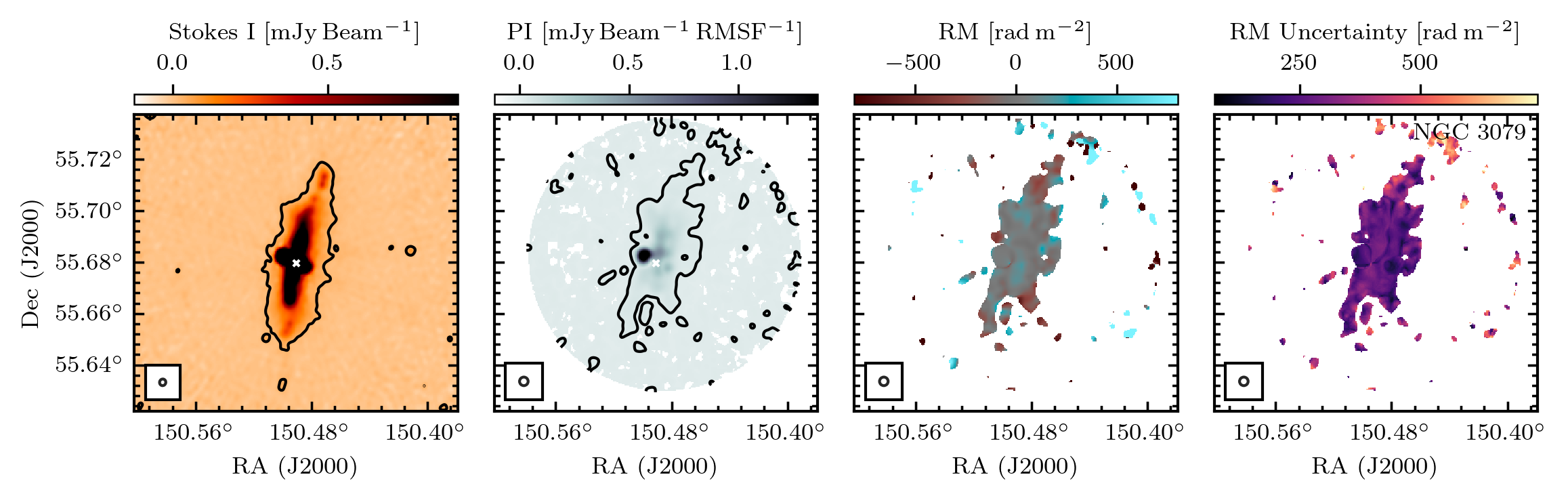}\\
    \includegraphics[width=0.9\linewidth]{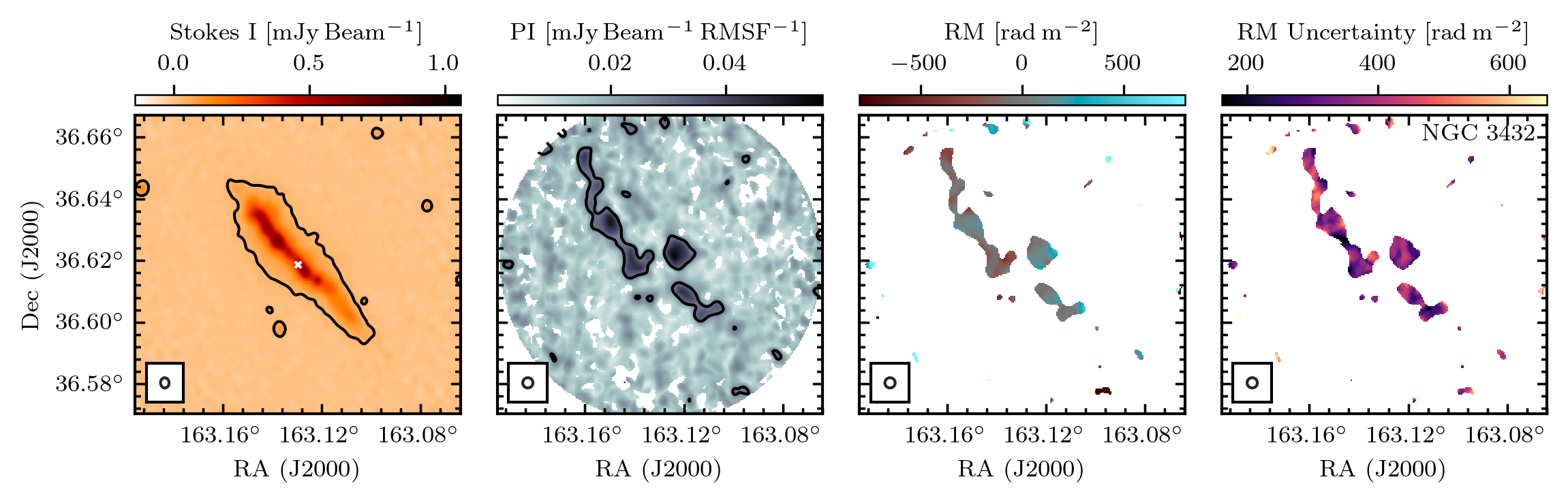}\\
    \caption{Continuation of Fig. \ref{fig:app_img_atlas_891_2613_2683_2820}. Displayed galaxies: NGC 3003, NGC 3044, NGC 3079, and NGC 3432.}
    \label{fig:app_img_atlas_3003_3044_3079_3432}
\end{figure*}

\begin{figure*}
    \centering
    \includegraphics[width=0.9\linewidth]{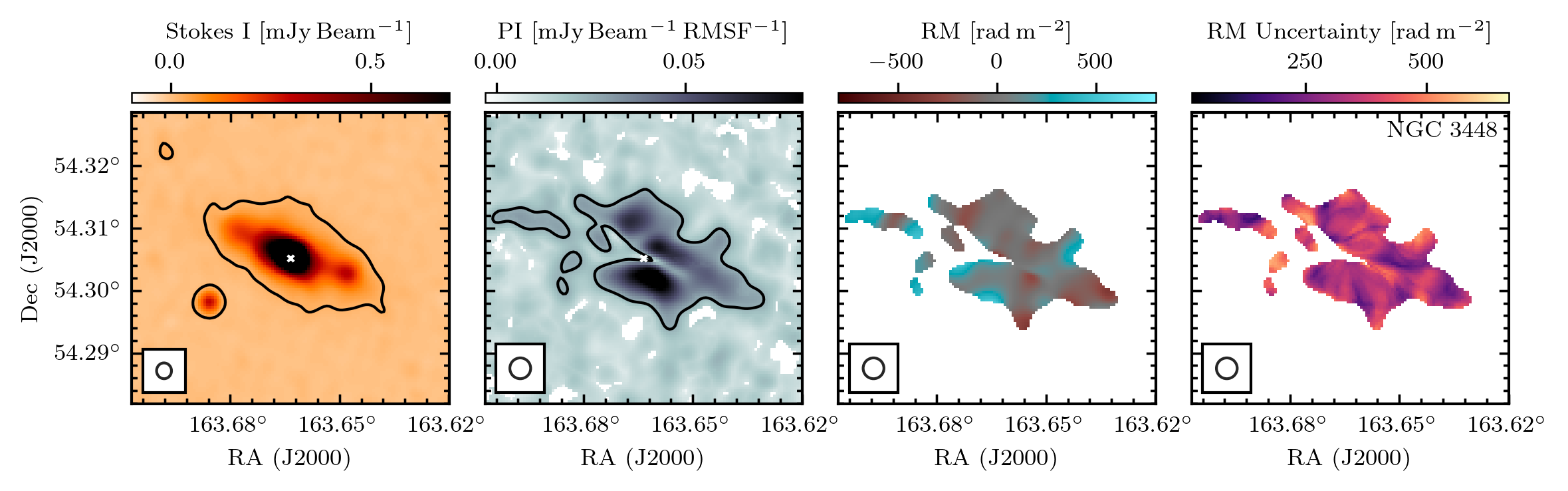}\\
    \includegraphics[width=0.9\linewidth]{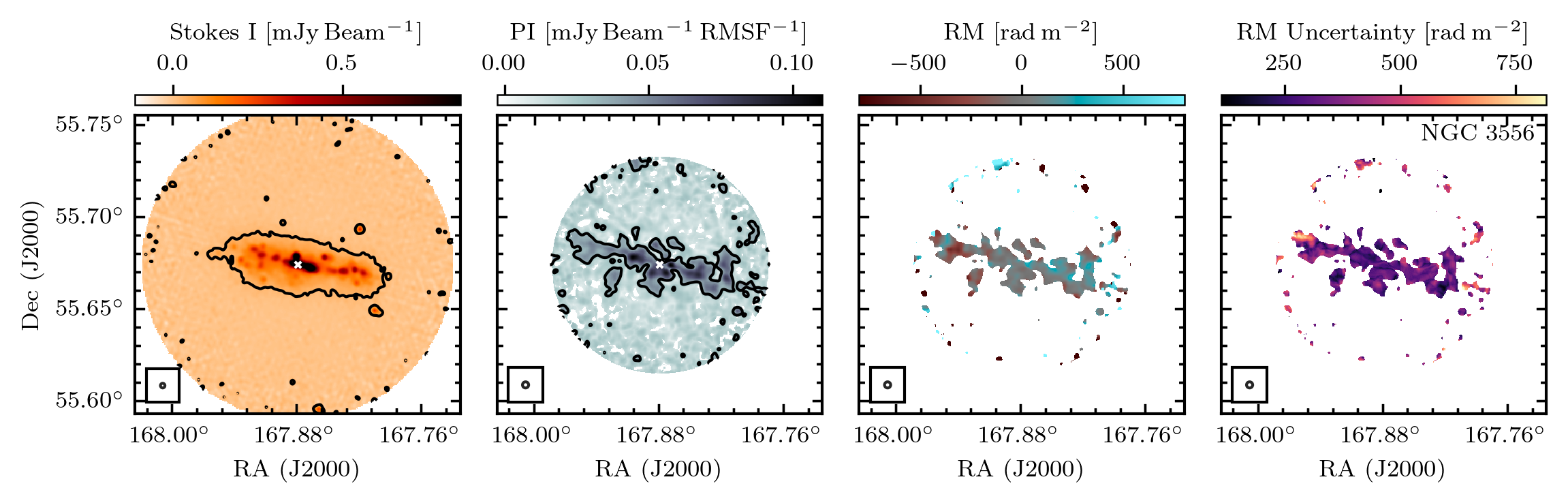}\\
    \includegraphics[width=0.9\linewidth]{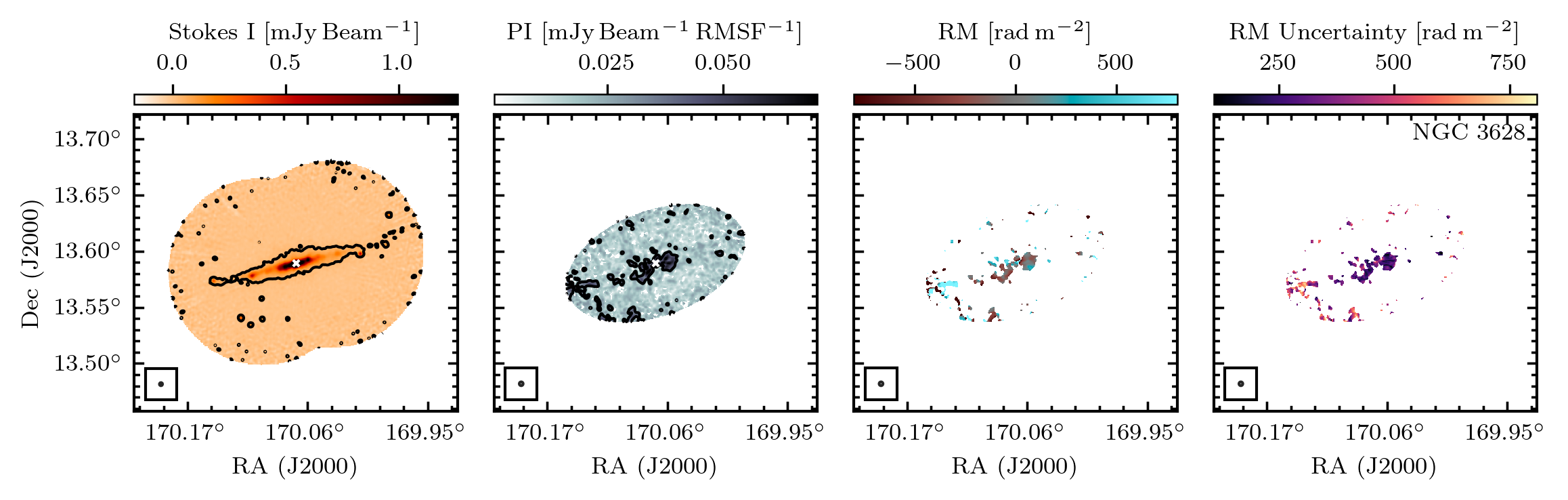}\\
    \includegraphics[width=0.9\linewidth]{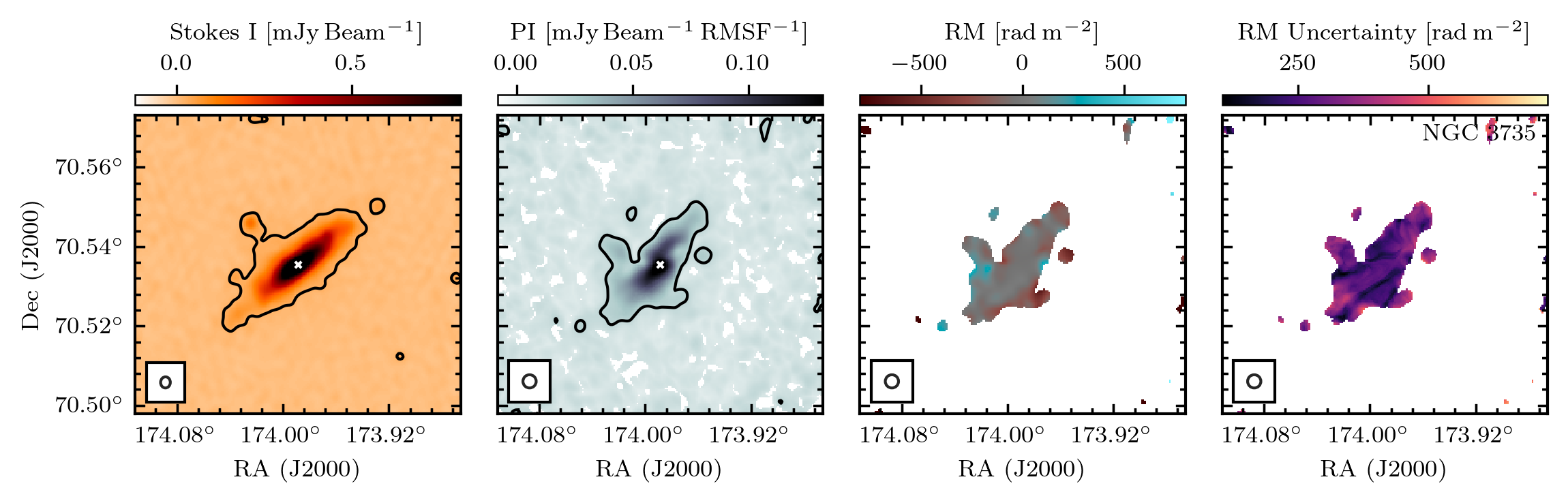}\\
    \caption{Continuation of Fig. \ref{fig:app_img_atlas_891_2613_2683_2820}. Displayed galaxies: NGC 3448, NGC 3556, NGC 3628, and NGC 3735.}
    \label{fig:app_img_atlas_3448_3556_3628_3735}
\end{figure*}

\begin{figure*}
    \centering
    \includegraphics[width=0.9\linewidth]{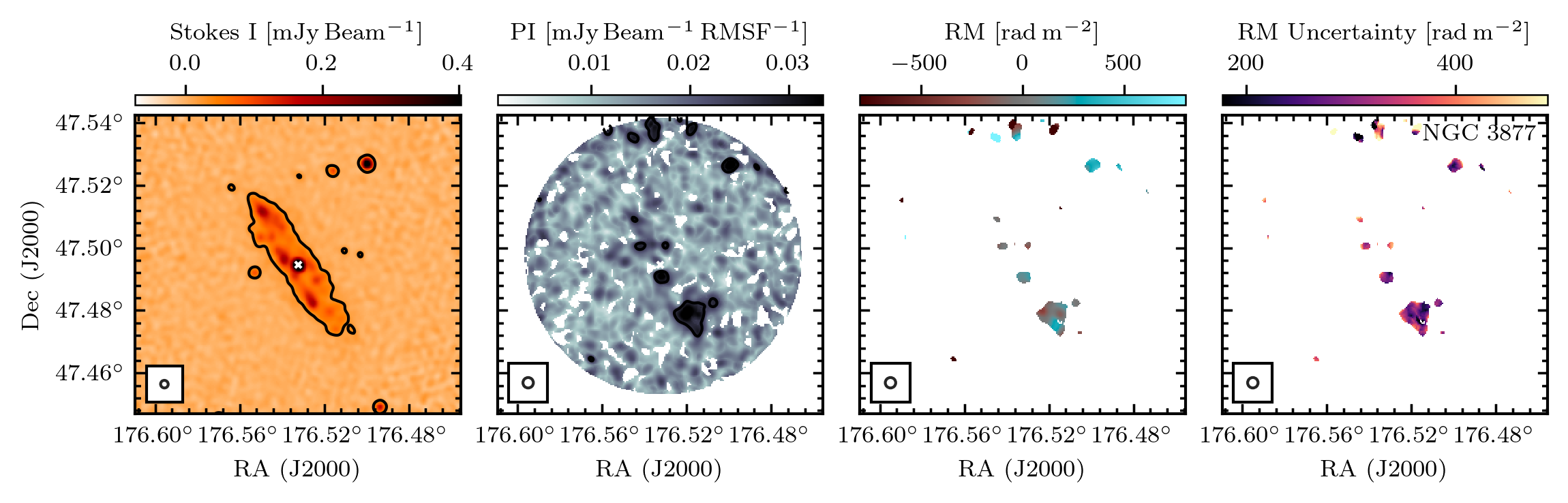}\\
    \includegraphics[width=0.9\linewidth]{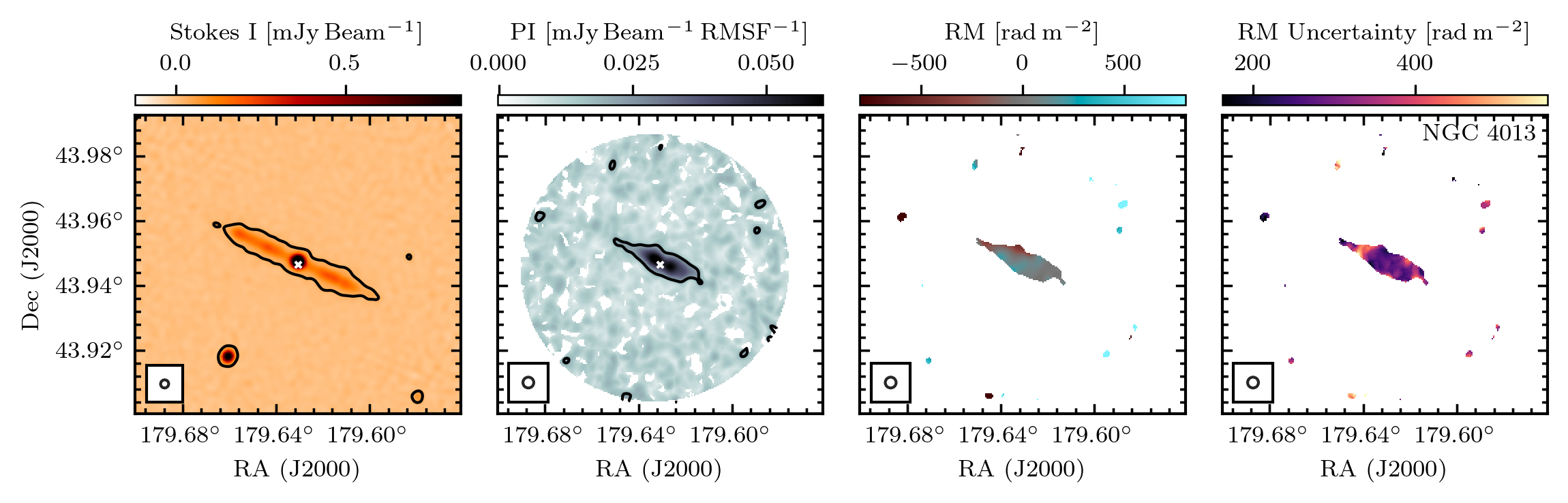}\\
    \includegraphics[width=0.9\linewidth]{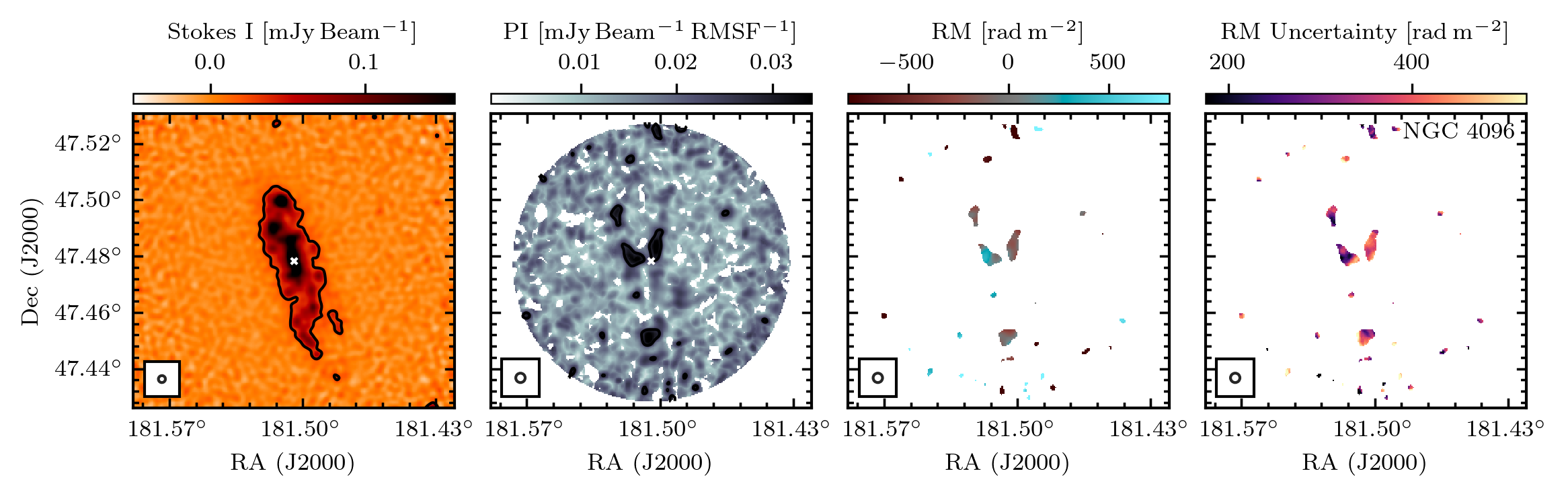}\\
    \includegraphics[width=0.9\linewidth]{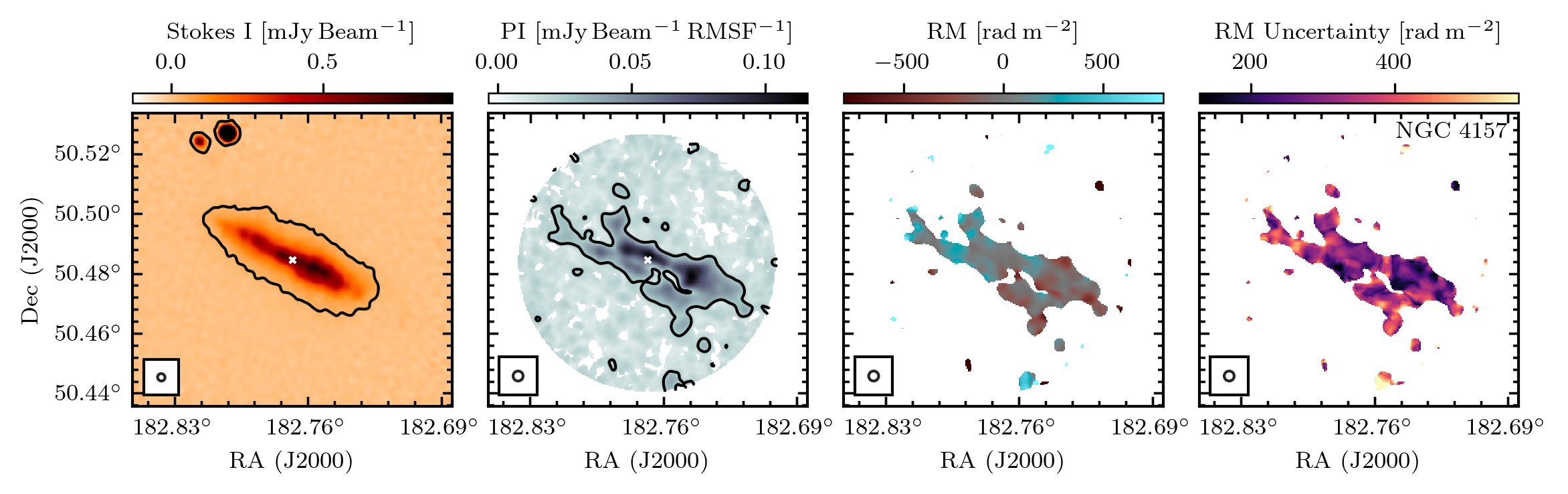}\\
    \caption{Continuation of Fig. \ref{fig:app_img_atlas_891_2613_2683_2820}. Displayed galaxies: NGC 3877, NGC 4013, NGC 4096, and NGC 4157.}
    \label{fig:app_img_atlas_3877_4013_4096_4157}
\end{figure*}

\begin{figure*}
    \centering
    \includegraphics[width=0.9\linewidth]{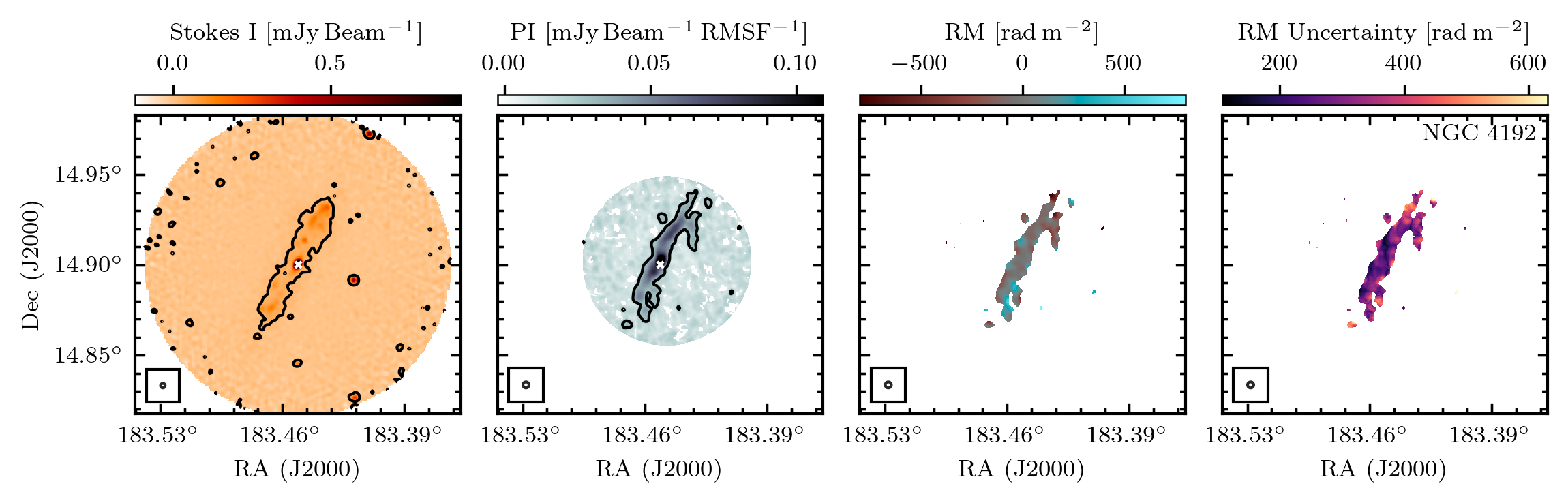}\\
    \includegraphics[width=0.9\linewidth]{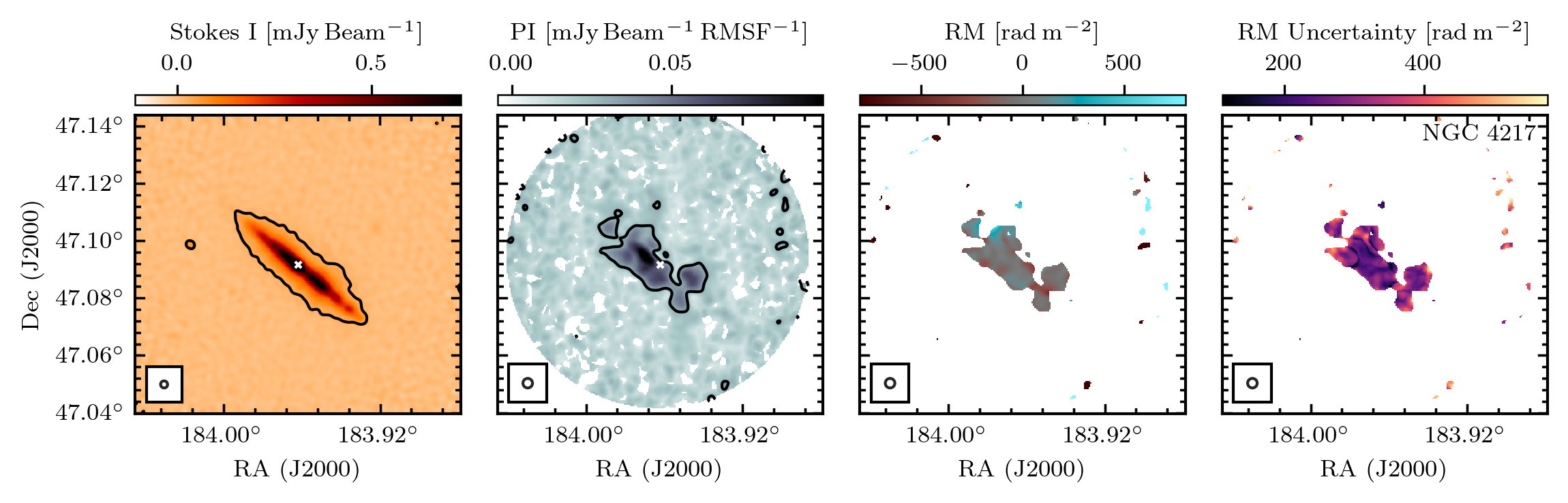}\\
    \includegraphics[width=0.9\linewidth]{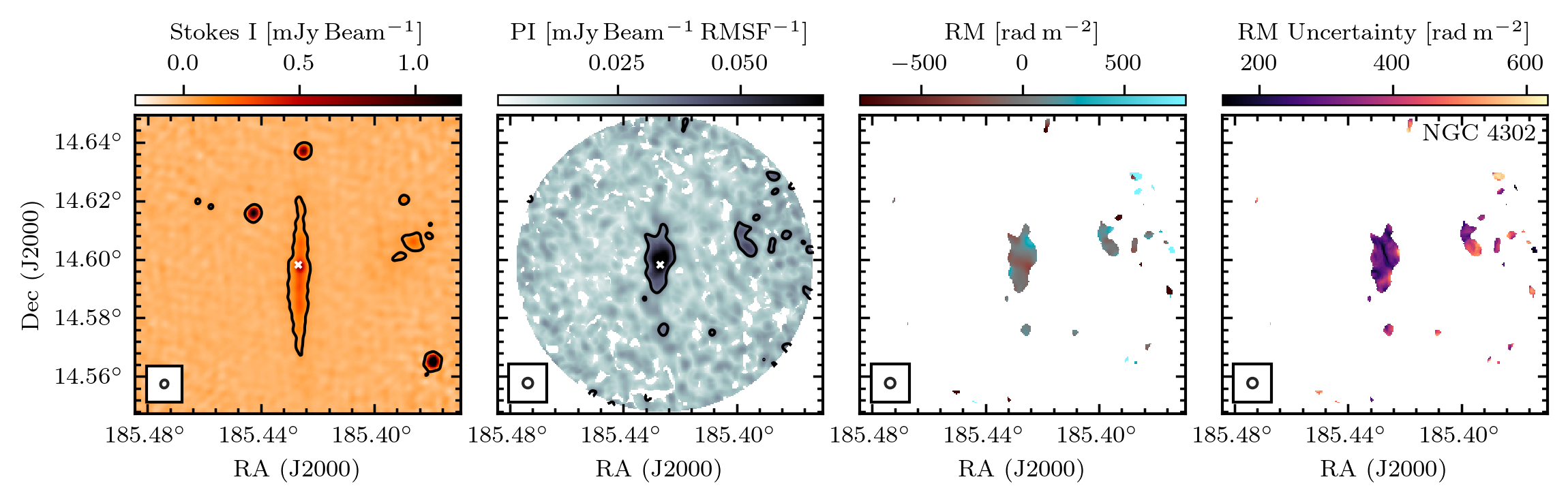}\\
    \includegraphics[width=0.9\linewidth]{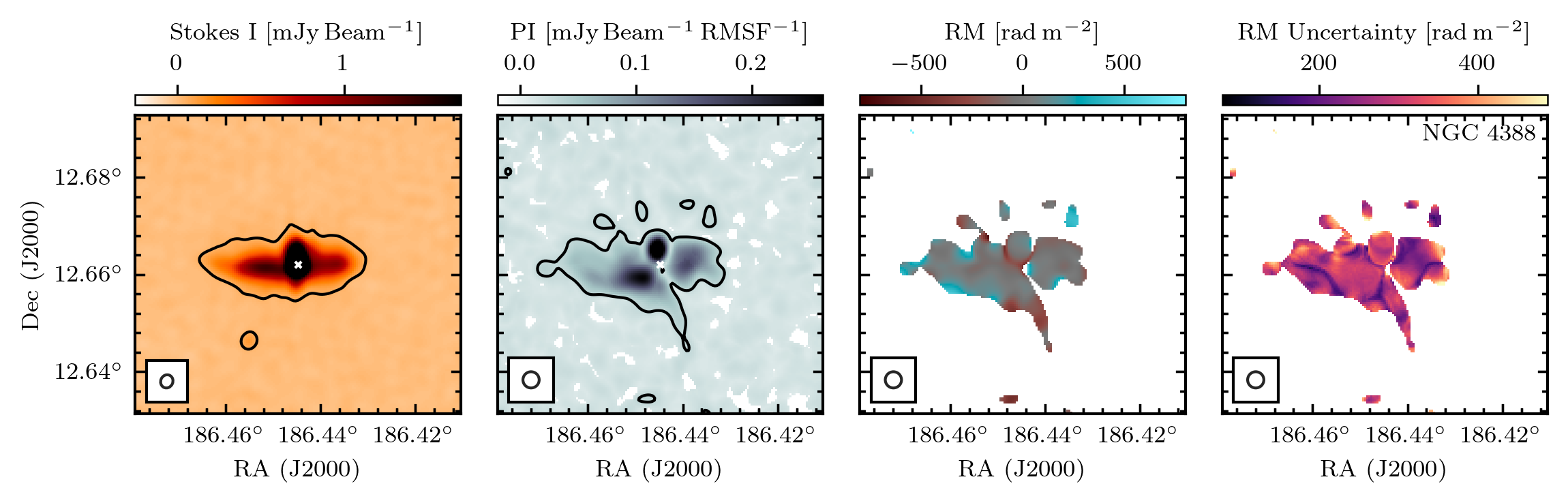}\\
    \caption{Continuation of Fig. \ref{fig:app_img_atlas_891_2613_2683_2820}. Displayed galaxies: NGC 4192, NGC 4217, NGC 4302, and NGC 4388.}
    \label{fig:app_img_atlas_4192_4217_4302_4388}
\end{figure*}

\begin{figure*}
    \centering
    \includegraphics[width=0.9\linewidth]{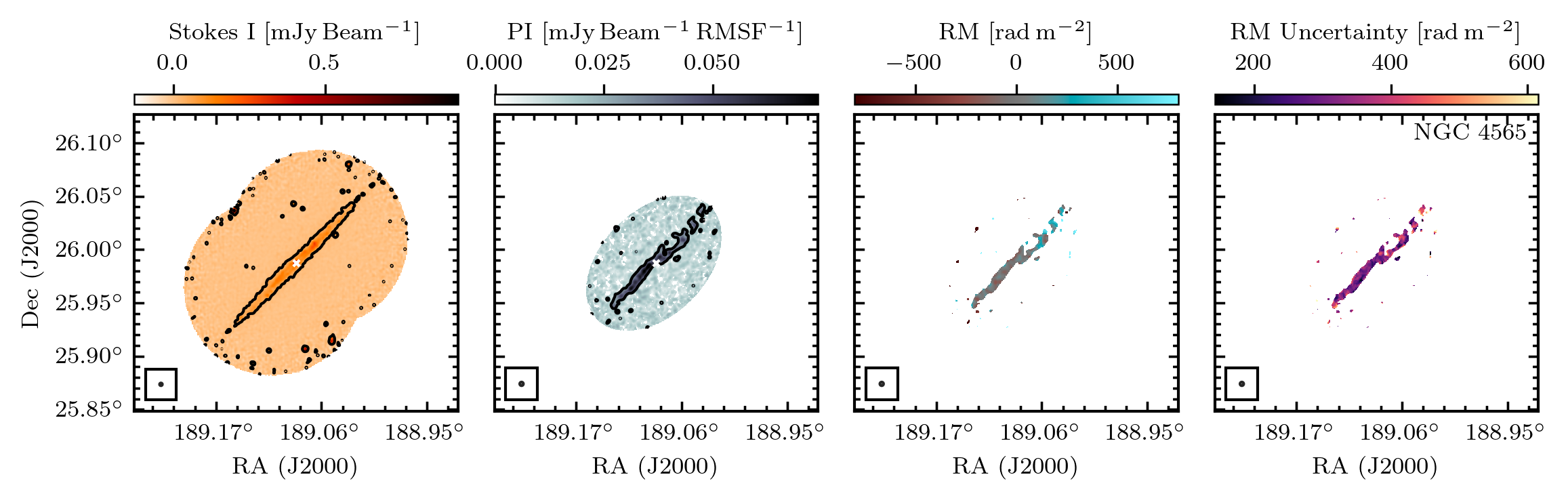}\\
    \includegraphics[width=0.9\linewidth]{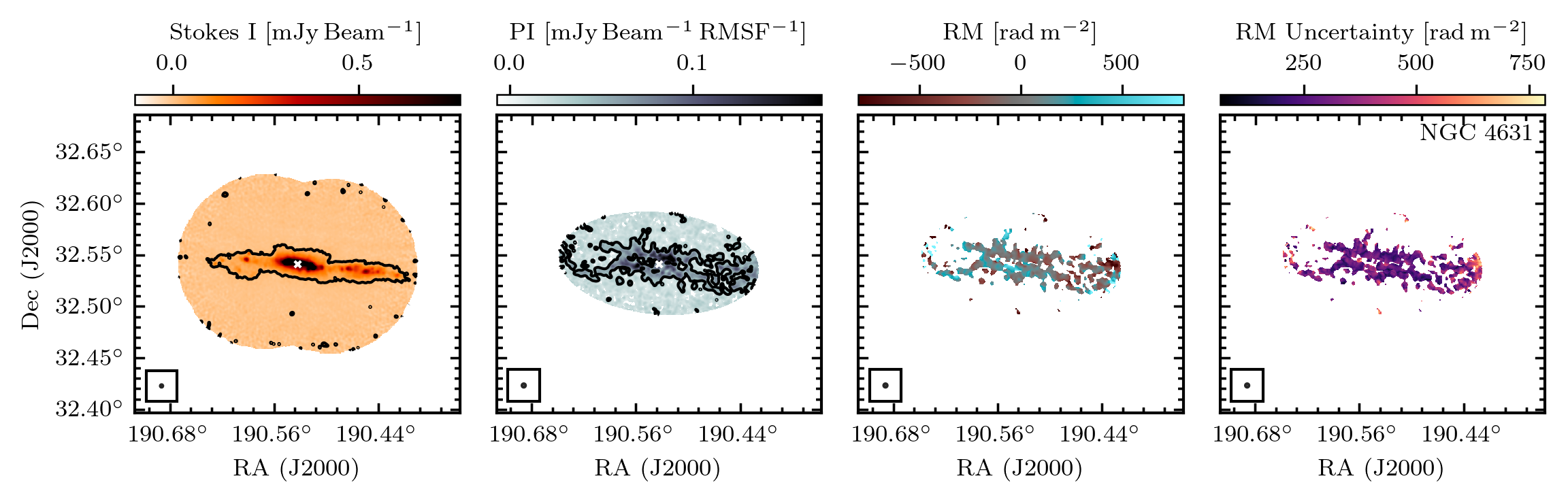}\\
    \includegraphics[width=0.9\linewidth]{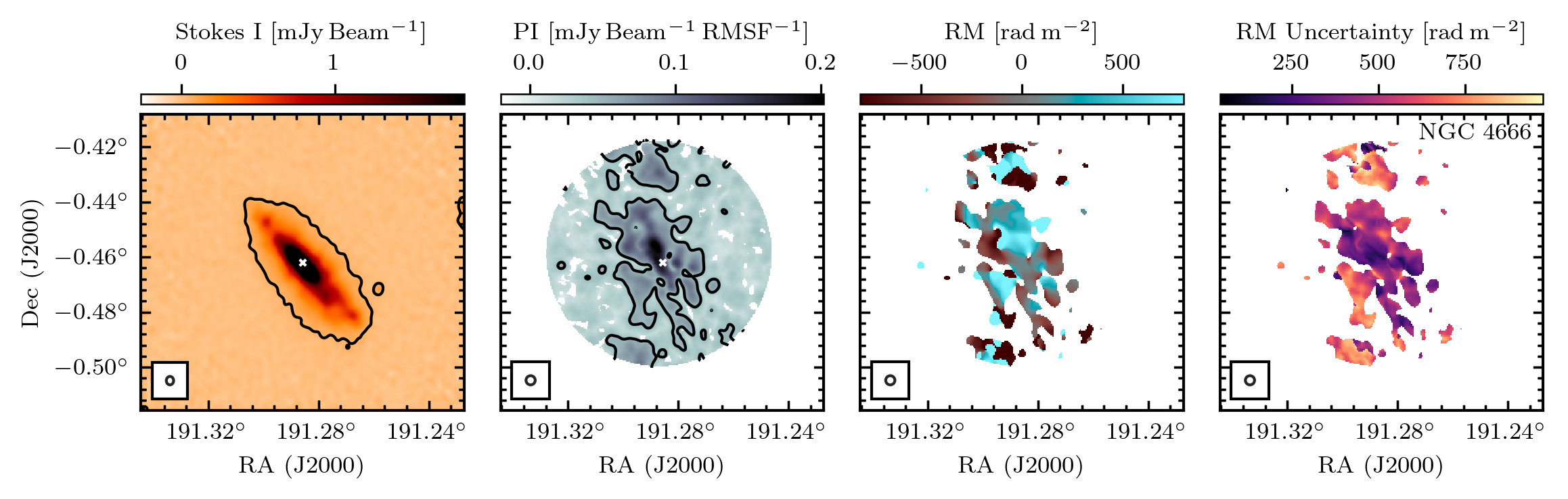}\\
    \includegraphics[width=0.9\linewidth]{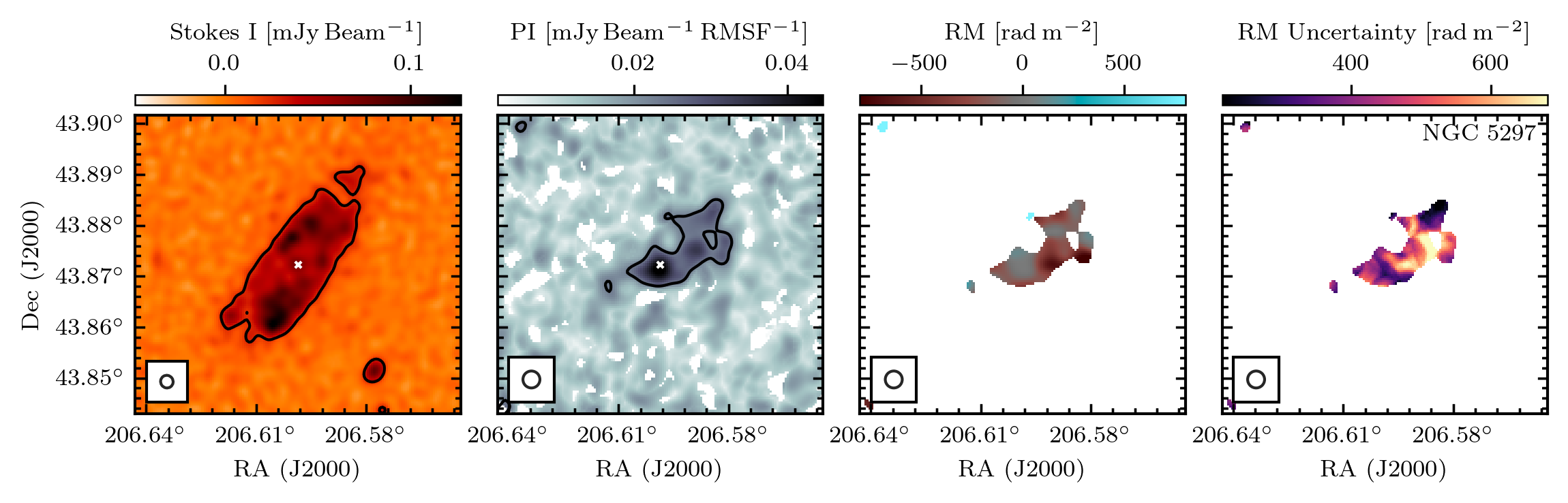}\\
    \caption{Continuation of Fig. \ref{fig:app_img_atlas_891_2613_2683_2820}. Displayed galaxies: NGC 4565, NGC 4631, NGC 4666, and NGC 5297.}
    \label{fig:app_img_atlas_4565_4631_4666_5297}
\end{figure*}

\begin{figure*}
    \centering
    \includegraphics[width=0.9\linewidth]{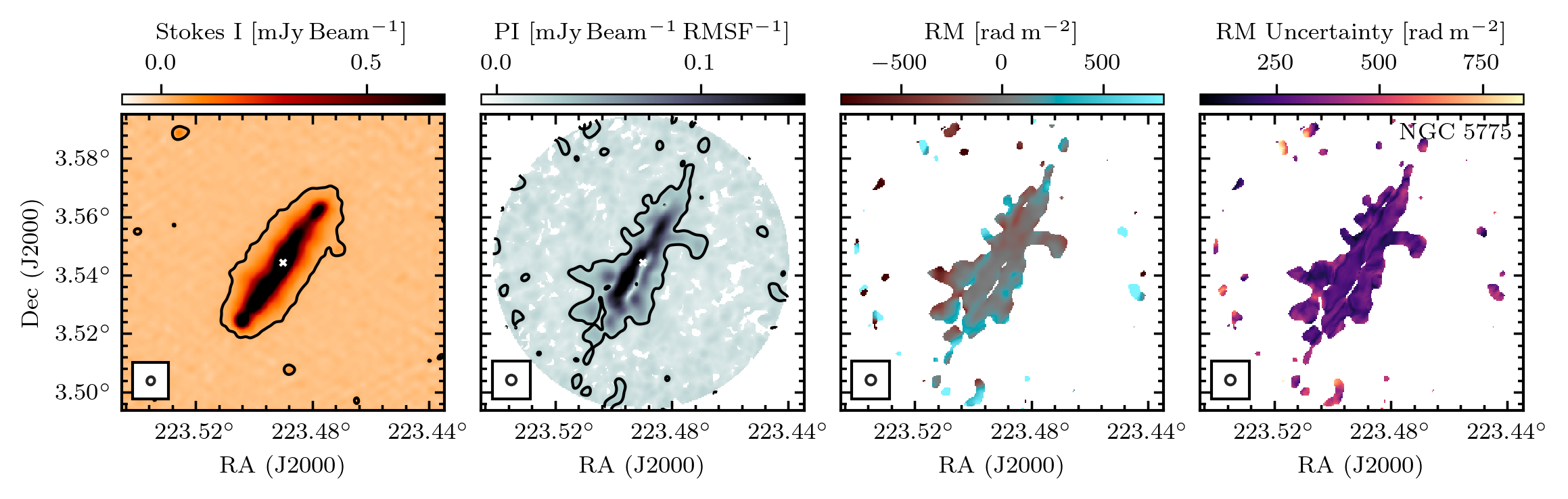}\\
    \includegraphics[width=0.9\linewidth]{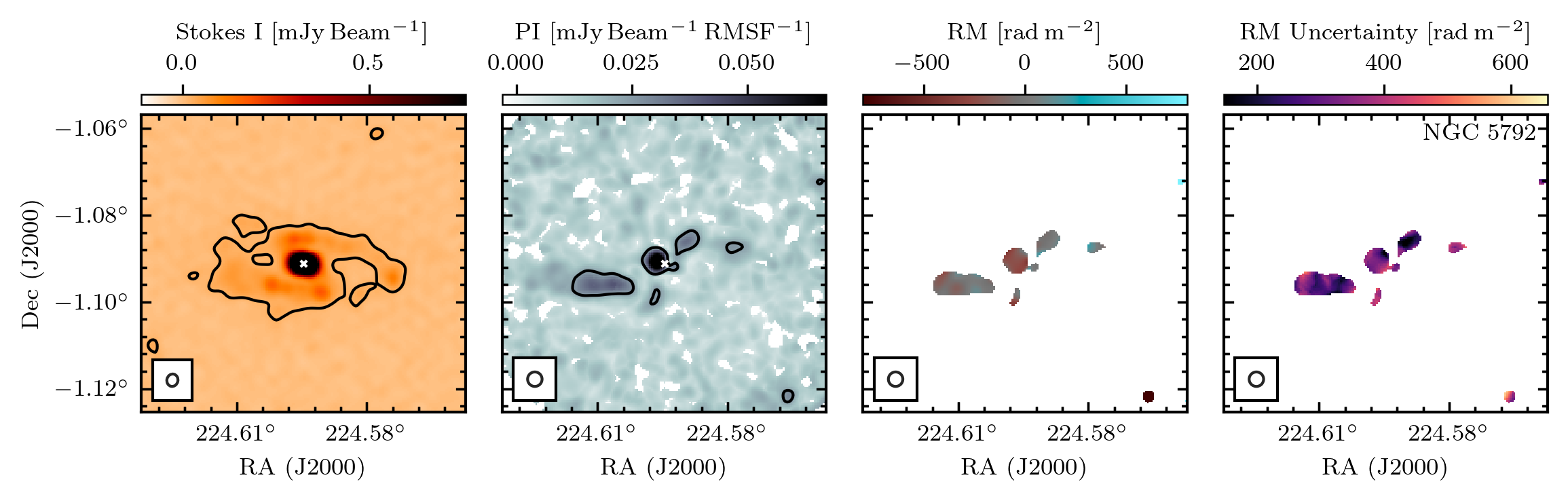}\\
    \includegraphics[width=0.9\linewidth]{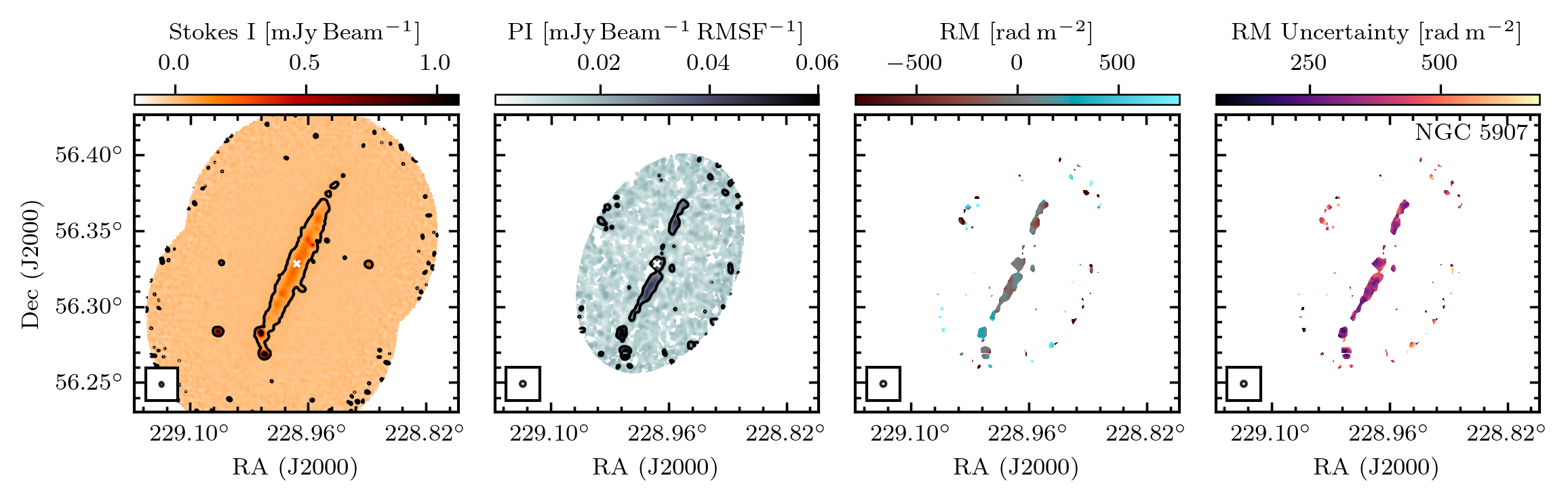}\\
    \caption{Continuation of Fig. \ref{fig:app_img_atlas_891_2613_2683_2820}. Displayed galaxies: NGC 5775, NGC 5792, and NGC 5907.}
    \label{fig:app_img_atlas_5775_5792_5907}
\end{figure*}
\FloatBarrier
\section{Stacking systematics}
\label{app:stack_sys}

When performing RM synthesis using the RM-Tools implementation, the Faraday depth peak-fitting algorithm \citep{2026arXiv260120092V} provides uncertainty estimates for the derived values of RM, $\chi_0$, and PI. We indicate  these uncertainties as $\delta_\mathrm{RM}$, $\delta_{\chi_0}$, and $\delta_{\mathrm{PI}}$. These strongly depend on the signal-to-noise ratio (S/N) and the frequency coverage of the input data. In this section, we derive estimates for additional systematic uncertainties that do not arise from the S/N of the data but are introduced through the stacking of multiple individual sources with varying RM, $\chi_0$, and PI. We  refer to these additional systematic uncertainties as  $\delta_\mathrm{RM}^{\mathrm{sys}}$, $\delta_{\chi_0}^{\mathrm{sys}}$, and $\delta_{\mathrm{PI}}^{\mathrm{sys}}$.

To quantify the impact of our stacking routine, we performed an experiment using synthetic data that represents a single pixel. We first generated a set of `sources', each defined by three input parameters: $\mathrm{RM^{inp}}$, $\mathrm{PI^{inp}}$, and $\chi_0^{\mathrm{inp}}$. The parameters were drawn from underlying distributions, which are described and motivated below. Using these generated parameter sets, we compiled synthetic Stokes $Q$ and Stokes $U$ spectra for each source, using the frequency spacing of this study. We then averaged the $Q$ and $U$ spectra of the individual sources:
\begin{eqnarray*}
    \langle Q\rangle = \frac{\sum_{i=1}^{n}Q_i}{n}, \langle U\rangle = \frac{\sum_{i=1}^{n}U_i}{n}, 
\end{eqnarray*}
where $n$ is the number of sources. In the following experiments, we set $n=25$, similar to the number of galaxies in the physical scaling stacks presented in this paper. The averaged spectra  $\langle Q\rangle$ and  $\langle U\rangle$ were then used as input parameters for RM synthesis. With RM synthesis, we derived three output (recovered) values: $\mathrm{RM^{recov}}$, $\mathrm{PI^{recov}}$, and $\chi_0^{\mathrm{recov}}$. $\mathrm{RM^{recov}}$ is the Faraday depth at the peak in the Faraday spectrum. To estimate the systematic uncertainties, we then compared $\mathrm{RM^{recov}}$, $\mathrm{PI^{recov}}$, and $\chi_0^{\mathrm{recov}}$ to $\langle\mathrm{RM^{inp}}\rangle$, $\langle\mathrm{PI^{inp}}\rangle$, and $\langle \chi_0^{\mathrm{inp}}\rangle$, where $\langle\mathrm{PI^{inp}}\rangle$ describes the mean PI of the generated sources, $\langle\mathrm{RM^{inp}}\rangle$ describes the PI-weighted mean of the input RM values, and $\langle \chi_0^{\mathrm{inp}}\rangle$ describes the PI-weighted mean of the input $\chi_0$ values:
\begin{eqnarray*} 
\langle\mathrm{PI^{inp}}\rangle = \frac{\sum_{i=1}^{n}\mathrm{PI}_i^{\mathrm{inp}}}{n}, 
\langle\mathrm{RM^{inp}}\rangle = \frac{\sum_{i=1}^{n}\mathrm{RM}_i^{\mathrm{inp}} \mathrm{PI}_i^{\mathrm{inp}}}{\sum_{i=1}^{n}\mathrm{PI}_i^{\mathrm{inp}}},
\langle \chi_0^{\mathrm{inp}}\rangle = \frac{\sum_{i=1}^{n}\chi_{0i}^{\mathrm{inp}} \mathrm{PI}_i^{\mathrm{inp}}}{\sum_{i=1}^{n}\mathrm{PI}_i^{\mathrm{inp}}}.
\end{eqnarray*}

The PI values were drawn from a normal distribution (ND), with a mean of  $\mu_{\mathrm{PI}}=2.4$ and a standard deviation of $\sigma_{\mathrm{PI}}=2.3$, restricting the selection to positive values. These values were derived by computing the mean and standard deviation of the PI\textsubscript{phy} values in Table~\ref{tab:stacking_parameters}\footnote{Further, we tested if drawing from a Laplacian distribution, which is in better agreement with the data, affects our test results, but did not find a significant effect.}. We also explored the effect of flux normalisation, where we normalised the $Q$ and $U$ spectra of each source by its PI value before stacking.

For $\chi_0$, values were drawn from a uniform distribution $\chi_0\in[0,\pi/2]$\,\radmsquare. This restriction of the $\chi_0$ values is motivated by the applied alignment strategy. In the disc, most galaxies show a polarisation pattern that indicates the B field to be aligned with the disc. In the halo, prior studies \citep[e.g.][]{2025A&A...696A.112S} have shown that most galaxies in the analysed sample show an X-shaped polarisation pattern. Therefore, we did not assume a completely random underlying $\chi_0$ distribution but restricted the distribution to the quoted range\footnote{We note that the exact location of this interval does not affect our tests; only its width is relevant.}.

For the distribution of RM values, we considered three scenarios. First, we assumed that the RM values in a galaxy follow a ND ($\mu_{\mathrm{RM}}$=0\,\radmsquare, $\sigma_{\mathrm{RM}}=$150\,\radmsquare, Scenario A \& B). These values are representative of the analysed sample based on the derived RM maps of individual galaxies. Secondly, we assumed the RM values to follow a different ND ($\mu_{\mathrm{RM}}$=100\,\radmsquare, $\sigma_{\mathrm{RM}}=$75\,\radmsquare, Scenario C \& D). This assumption represents a scenario where there is a global preference for positive or negative RMs in a specific region (similar to a scenario as proposed by \citetalias{2021A&A...649A..94M})\footnote{The values for the mean and the standard deviation of the distribution were chosen to be representative for regions that show coherent RM structures in individual galaxies.}. As a final scenario, we considered the same RM distribution as in the previous example, but did not assume a global preference of RM signs in a specific quadrant. Therefore, we randomly assigned positive and negative RM signs with equal probabilities. This results in a bimodal distribution with peaks at $\pm100$\,\radmsquare where each peak has a width of 75\,\radmsquare (Scenario E \& F). In Table \ref{tab:app_stack_sys} we list all tested scenarios.

\begin{figure}
    \centering
    \includegraphics[width=0.75\linewidth]{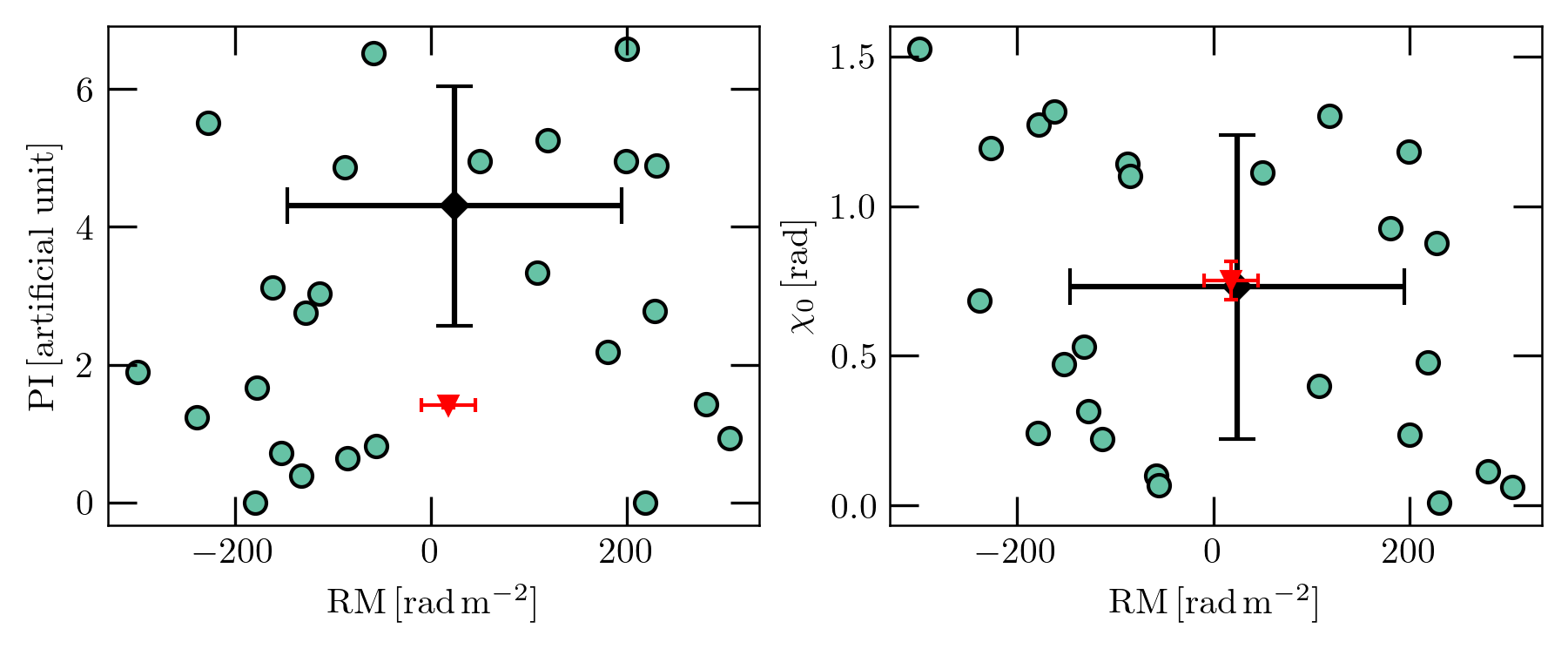}\\
    \caption{RM synthesis results based on simulated sources. For a set of 25 simulated sources (scenario E, Table \ref{tab:app_stack_sys}), synthetic $Q$ and $U$ spectra were averaged. The panels show the individual simulated sources (turquoise circles), the PI-weighted means of the distributions ($\langle\mathrm{RM^{inp}}\rangle$, $\langle\mathrm{PI^{inp}}\rangle$, and $\langle \chi_0^{\mathrm{inp}}\rangle$) as black diamonds and the recovered measurements ($\mathrm{RM^{recov}}$, $\mathrm{PI^{recov}}$, and $\chi_0^{\mathrm{recov}}$) when performing RM synthesis on $\langle Q \rangle$ and $\langle U \rangle$ (red triangle).  Black error bars reflect the standard deviation of the input sources. Red error bars represent the uncertainties derived in the RM synthesis ($\delta_\mathrm{RM}$, $\delta_{\chi_0}$, and $\delta_{\mathrm{PI}}$). Displayed are the source distributions in RM-PI space (left) and RM-$\chi_0$ space (right). }
    \label{fig:app_stack_sys_1relasiation}
\end{figure}

\begin{table}[]
    \centering
    \caption{Set of tested underlying RM and PI distributions. }
    \label{tab:app_stack_sys}
    \begin{tabular}{lllrrrr}
    \hline 
    \hline
    Scenario & RM distribution & PI distribution & $\Delta\mathrm{RM}$ & $\Delta\mathrm{RM_{clipped}}$ & $\Delta\mathrm{PI_{rel}}$  & $\Delta\chi_0$\\
    &$[\mathrm{rad\,m^{-2}}]$& [artificial units] &  $[\mathrm{rad\,m^{-2}}]$ & $[\mathrm{rad\,m^{-2}}]$  & & [rad] \\
    \hline 
    A & ND $(\mu=0,\, \sigma=150)$ & ND $(\mu=2.3,\, \sigma=2.4)$ & $0\pm90$ &$0\pm60 $ & $0.67\pm0.10$ & $-0.02\pm0.18$ \\
    B & ND $(\mu=0,\, \sigma=150)$ & normalised                   & $0\pm70$ &$0\pm50 $ & $0.53\pm0.11$ & $0.00\pm0.12$ \\
    C & ND $(\mu=100,\, \sigma=75)$ & ND $(\mu=2.3,\, \sigma=2.4)$& $0\pm18$ &$0\pm13 $ & $0.58\pm0.08$ & $0.00\pm0.03$ \\
    D & ND $(\mu=100,\, \sigma=75)$ & normalised                  & $0\pm13$ &$0\pm12 $ & $0.41\pm0.10$ & $0.00\pm0.03$ \\
    E & bimodal & ND $(\mu=2.3,\, \sigma=2.4)$                    & $0\pm60$ &$0\pm30 $ & $0.64\pm0.09$ &  $0\pm0.13$\\  
    F & bimodal & normalised                                      & $0\pm40$ &$0\pm28 $ & $0.47\pm0.10$ &  $0\pm0.06$\\
    \hline
    \end{tabular}
\tablefoot{For each parameter set, the distributions of the differences between the PI-weighted underlying source distribution and the RM synthesis measurement (RM, PI\textsubscript{rel} (normalised by the mean of the input PI values), and $\chi_0$) are listed. RM distributions: The magnitudes of RM values are drawn from a normal distribution (ND) and two scenarios account for a random RM sign (bimodal distributions). PI distribution: We either consider a ND that is representative of the flux distribution listed in Table \ref{tab:stacking_parameters} (PI\textsubscript{phy}) or apply a flux normalisation.}    
\end{table}

Fig. \ref{fig:app_stack_sys_1relasiation} displays this comparison for one random set of 25 sources, without PI normalisation and a bimodal RM distribution (Scenario E in Table \ref{tab:app_stack_sys}). Here, RM and $\chi_0$ are correctly recovered while the PI is underestimated. Furthermore, the uncertainty estimates derived by the RM synthesis ($\delta_\mathrm{RM}$, $\delta_{\chi_0}$, and $\delta_{\mathrm{PI}}$) underestimate the scatter of the underlying distributions.

To further quantify the systematic uncertainties of the individual scenarios, we repeated the experiment displayed in Fig.~\ref{fig:app_stack_sys_1relasiation} 1000 times and compared the PI-weighted input to the recovered values. For this, we introduce the following quantities:
\begin{eqnarray*}
    \Delta\mathrm{PI_{rel}}=\frac{\langle\mathrm{PI^{inp}}\rangle -\mathrm{PI^{recov}} }{\langle\mathrm{PI^{inp}}\rangle},\ \Delta\mathrm{RM}=  \langle\mathrm{RM^{inp}}\rangle - \mathrm{RM^{recov}},\ \Delta\chi_0=  \langle \chi_0^{\mathrm{inp}}\rangle - \chi_0^{\mathrm{recov}}.
\end{eqnarray*}

\begin{figure}
    \centering
    \includegraphics[width=0.8\linewidth]{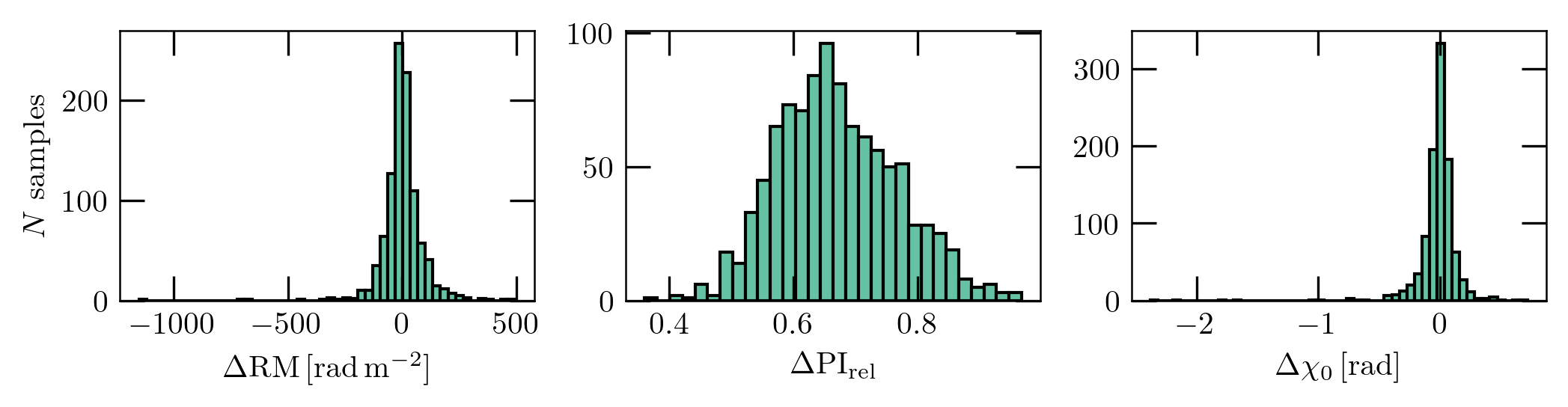}\\
    \includegraphics[width=0.8\linewidth]{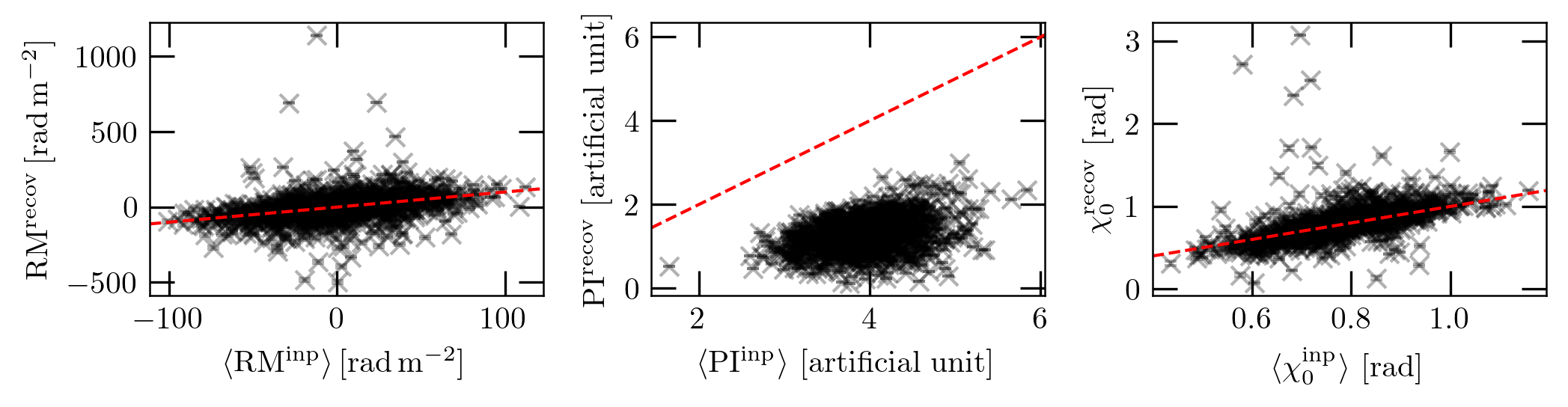}\\
    \includegraphics[width=0.8\linewidth]{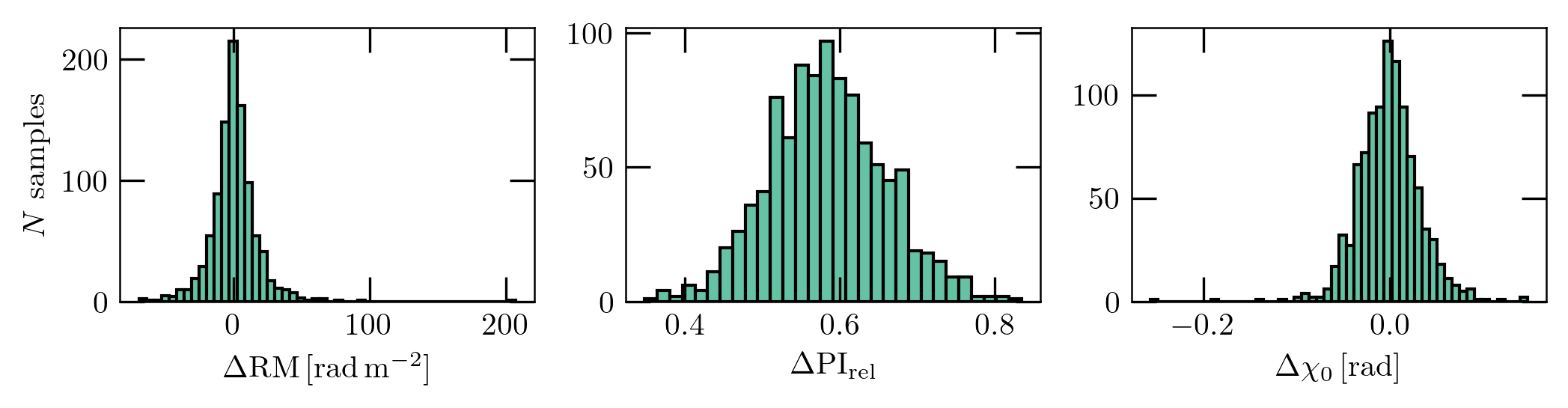}\\
    \includegraphics[width=0.8\linewidth]{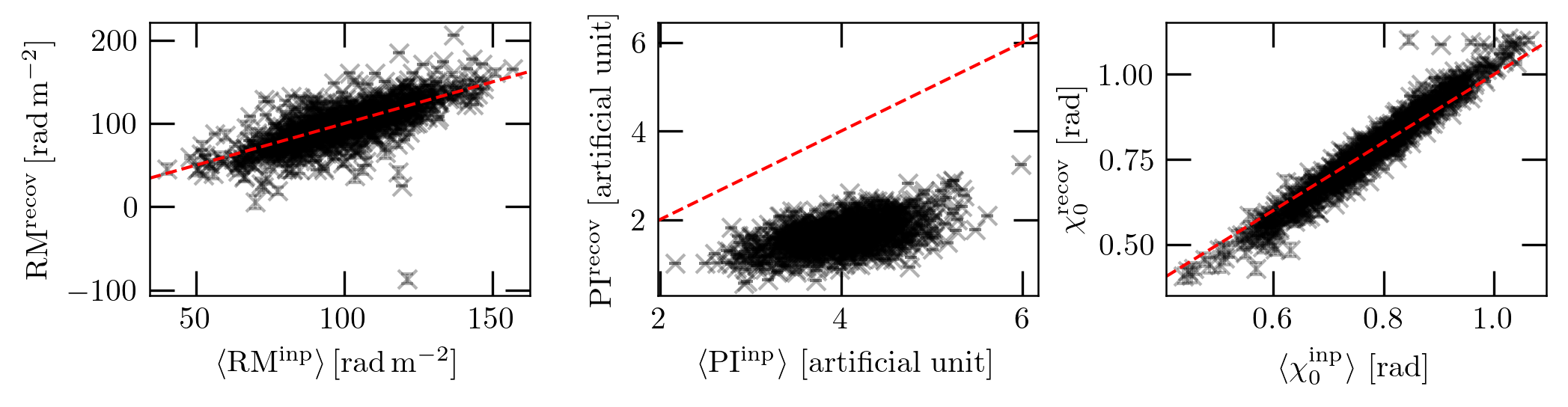}\\
    \caption{First and third row: distributions of $\Delta\mathrm{RM}$ (left panel), $\Delta\mathrm{PI_{rel}}$  (middle panel), and $\Delta\chi_0$ (right panel) for 1000 samples of sources; second and fourth row: scatter plots comparing $\mathrm{RM^{recov}}$, $\mathrm{PI^{recov}}$, and $\chi_0^{\mathrm{recov}}$ to  
    $\langle\mathrm{RM^{inp}}\rangle$, $\langle\mathrm{PI^{inp}}\rangle$, and $\langle \chi_0^{\mathrm{inp}}\rangle$. In the scatter plots, the red line indicate the one-to-one relation. First and second row show Scenario A, third and fourth row show Scenario C.}
    \label{fig:app_stack_sys_hist_scatter}
\end{figure}

As an example of our procedure, we display the comparison of PI-weighted simulation input with the recovered quantities in Fig. \ref{fig:app_stack_sys_hist_scatter}, for scenarios A and C. The first and third row of Fig. \ref{fig:app_stack_sys_hist_scatter} show the distributions of the $\Delta\mathrm{RM}$, $\Delta\mathrm{PI_{rel}}$, and $\Delta\chi_0$ for both scenarios. If the peak of the distribution is centred on zero, the method correctly recovers the PI-weighted input parameters. In this case, we define the standard deviation of the three distributions as the systematic uncertainties ($\delta_\mathrm{RM}^{\mathrm{sys}}$, $\delta_{\chi_0}^{\mathrm{sys}}$, and $\delta_{\mathrm{PI}}^{\mathrm{sys}}$) that are introduced by the stacking routine. If the distribution does not peak at zero (as for $\Delta\mathrm{PI_{rel}}$), a systematic offset is introduced by the stacking routine. In Table~\ref{tab:app_stack_sys}, we list the mean and the standard deviation ($\mu\pm\sigma$) of $\Delta\mathrm{RM}$-, $\Delta{\chi_0}$-, and $\Delta{\mathrm{PI_{rel}}}$-distributions for all tested scenarios. As can be seen in the scatter plots for scenario A (second row Fig.~\ref{fig:app_stack_sys_hist_scatter}), the method can produce catastrophic outliers, where especially the RM estimate is off. Therefore, we also list the sigma-clipped\footnote{We use the \texttt{astropy sigma\_clipped\_stats} module with a $\sigma$-threshold of 3 and a maximum number of 5 sigma-clipping iterations.} mean and standard deviation of the $\Delta\mathrm{RM}$ distribution ($\Delta\mathrm{RM_{clipped}}$) in Table~\ref{tab:app_stack_sys}.

From Table \ref{tab:app_stack_sys} we deduce the following results. Firstly, in all analysed scenarios, the method correctly recovers the PI-weighted values for $\chi_0$ and RM. A systematic uncertainty is introduced for both quantities. Especially for the recovered RM value, this uncertainty scales with the width of the underlying RM distribution. Secondly, for all scenarios, the recovered PI is systematically underestimating the underlying distribution of PI values. We attribute this depolarisation effect to Faraday dispersion \citep[see e.g.][]{2011MNRAS.418.2336A} that is introduced by averaging over multiple sources with different sets of PI, RM, and $\chi_0$. This averaging process can cause the individual $Q$ and $U$ spectra to cancel out, thereby systematically reducing the recovered PI. This effect can be significantly reduced by performing a flux normalisation before stacking.

\begin{figure}
    \centering
    \includegraphics[width=0.4\linewidth]{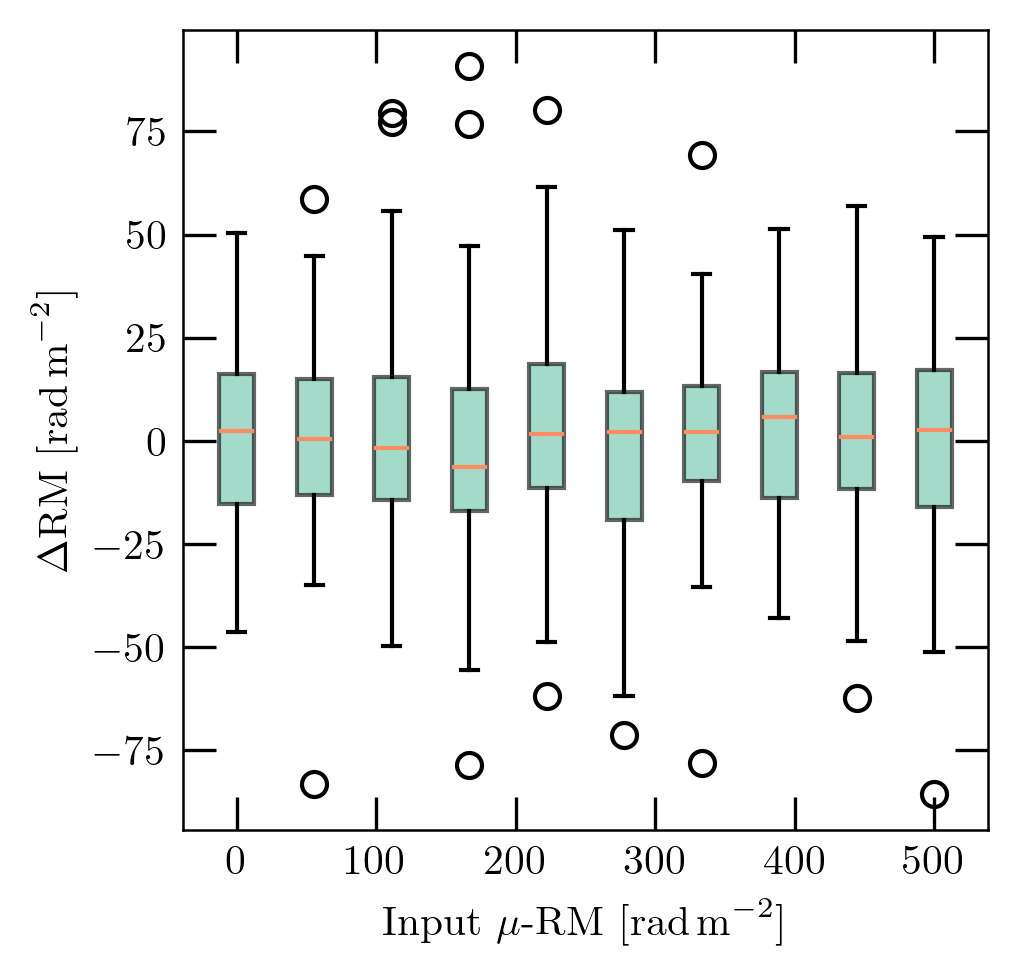}
    \includegraphics[width=0.4\linewidth]{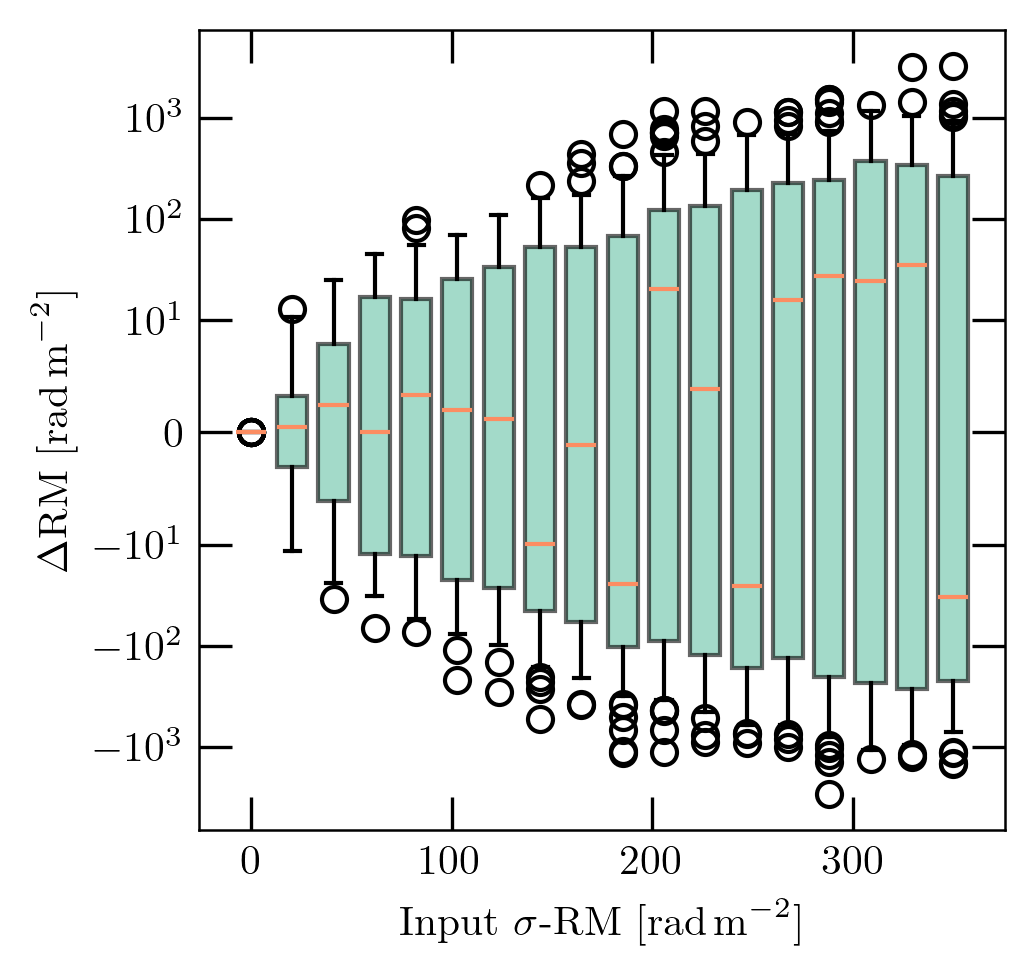}
    \caption{Box plots for $n=100$-samples drawn (with a setup similar to scenario C) but varying the mean (left panel) or standard deviation (right panel) of the underlying RM distribution. The orange line in each box indicates the sample median. The boxes extend from quartile 1 to quartile 3. The whiskers extend from the box to the farthest data point lying within 1.5 times the inter-quartile range from the box. Outlier values are marked as black circles. The right panel employs a symmetric logarithmic scale on the $y$-axis. Within the range of $[-10, 10]$\,\radmsquare, the scale is linear; outside of this interval, it transitions to logarithmic scaling.}
    \label{fig:app_stack_sys_rm_mean_std}
\end{figure}

As a final test case, we compared the impact of the mean and the standard deviation of the underlying RM distribution. Firstly, to analyse the impact of varying the mean of the underlying RM distribution. We held the standard deviation constant at $\sigma_{\mathrm{RM}}=75$\,\radmsquare\ and applied no flux normalisation (similar to scenario C). We varied the $\mu_{\mathrm{RM}}$, covering a range of $[0,500]$\,\radmsquare. For each $\mu_{\mathrm{RM}}$-value, we drew 100 samples and analysed the resulting distribution of $\Delta\mathrm{RM}$. The resulting distributions are visualised in the left panel of Fig.~\ref{fig:app_stack_sys_rm_mean_std}. As the widths of the $\Delta\mathrm{RM}$-distributions as well as the distribution centres (here estimated from the median (orange line)) stay relatively constant for the tested range of the input mean RM, we conclude that the method's uncertainty does not depend on the mean of the underlying RM distribution.

Secondly, we performed the same best but held $\mu_{\mathrm{RM}}$ constant 50\,\radmsquare\ and instead varied $\sigma_{\mathrm{RM}}$. This test is visualised in the right panel of Fig.~\ref{fig:app_stack_sys_rm_mean_std}. Here, the data shows that the resulting $\Delta\mathrm{RM}$-distribution strongly depends on the $\sigma_{\mathrm{RM}}$. The centre of the $\Delta\mathrm{RM}$ distributions starts to deviate from 0\,\radmsquare. Compared to the distribution widths, this offset remains insignificantly small. If $\sigma_{\mathrm{RM}}$ exceeds a value of 180\,\radmsquare\, the introduced systematic uncertainty strongly rises, reaching a value of $\delta_\mathrm{RM}^{\mathrm{sys}}=230$\,\radmsquare. However, we note that this magnitude of scatter in the RM distribution exceeds the level found in the analysed galaxy sample.

In summary, we conclude that the most accurate stacking results can be obtained for the most concentrated distributions of RM and PI. Nevertheless, as the real underlying RM distribution is suspected to vary strongly (similar to scenario A or E), as a conservative estimate, we account for a systematic uncertainty of $\delta_\mathrm{RM}^{\mathrm{sys}}=90$\,\radmsquare\ and $\delta_{\chi_0}^{\mathrm{sys}}=0.18$\,rad (the largest uncertainties of all scenarios) in the analysis of this paper. Nevertheless, we stress that this result is dependent on the frequency coverage, as well as the nature of the studied objects. Therefore, we recommend performing similar tests before applying this technique to a different dataset.
\end{appendix}
\end{document}